\documentclass[prd,twocolumn,nofootinbib,showpacs,10pt]{revtex4-1}
\usepackage{textcomp}
\usepackage{amsmath}
\usepackage{amsfonts}
\usepackage{calc}
\usepackage{mathrsfs}
\usepackage{amssymb}
\usepackage{amsmath}
\usepackage{array}
\usepackage{color}
\usepackage{bm}
\usepackage{graphicx}
\usepackage{hyperref}

\DeclareMathOperator{\STF}{STF}
\newcommand{\qmax}{{q_{\rm max}}}
\newcommand{\lmax}{{\ell_{\rm max}}}
\newcommand{\qSmax}{\mathcal{Q}}
\newcommand{\lSmax}{\mathcal{L}}
\newcommand{\e}{\epsilon}
\newcommand{\nhat}{\hat{n}}
\newcommand{\E}{\mathcal{E}}
\newcommand{\B}{\mathcal{B}}

\begin{document}
\title{Nonlinear gravitational self-force. I. Field outside a small body} 
\author{Adam Pound$^1$} 
\affiliation{$^1$School of Mathematics, University of Southampton, Southampton,
United Kingdom, SO17 1BJ}
\date{\today}

\begin{abstract}
A small extended body moving through an external spacetime $g_{\alpha\beta}$ creates a metric perturbation $h_{\alpha\beta}$, which forces the body away from  geodesic motion in $g_{\alpha\beta}$. The foundations of this effect, called the gravitational self-force, are now well established, but concrete results have mostly been limited to linear order. Accurately modeling the dynamics of compact binaries requires proceeding to nonlinear orders. To that end, I show how to obtain the metric perturbation outside the body at all orders in a class of generalized wave gauges. In a small buffer region surrounding the body, the form of the perturbation can be found analytically as an expansion for small distances $r$ from a representative worldline. 
Given only a specification of the body's multipole moments, the field obtained in the buffer region suffices to find the metric everywhere outside the body via a numerical puncture scheme. Following this procedure at first and second order, I calculate the field in the buffer region around an arbitrarily structured compact body at sufficiently high order in $r$ to numerically implement a second-order puncture scheme, including effects of the body's spin. I also define $n$th-order (local) generalizations of the Detweiler-Whiting singular and regular fields and show that in a certain sense, the body can be viewed as a skeleton of multipole moments.
\end{abstract}
\maketitle 


\section{Introduction}\label{intro}
In general relativity, bodies affect the spacetime geometry around them, and their motion is, in turn, affected by that geometry. Historically, study of this problem of motion has focused on the post-Newtonian limit of weak fields and slow velocities. In the strong-field regime, bodies haves typically been approximated as test particles, which move on geodesics of a background spacetime that is unaffected by them; or, more generally, they have been treated as test \emph{bodies}, which are accelerated only by the coupling of their multipole moments to the background curvature. However, gravitational wave astronomy will soon allow us to observe binaries with strong fields and fast motion, providing fresh impetus to solve the problem of motion in the strong-field regime.

For binaries made up of bodies of comparable mass, the problem has been tackled with numerical relativity~\cite{Centrella:10} or analytically with effective one body (EOB) theory~\cite{Buonanno-Damour:98}. For binaries with extreme mass ratios, in which a small mass $m$ emits gravitational waves and spirals into a large mass $M$, the problem has been approached via the gravitational self-force approximation: the smaller mass is treated as a source of small perturbations of the background spacetime of the larger, and that perturbation forces the small body away from geodesic motion in the background~\cite{Mino:96,Quinn-Wald:97,Poisson-Pound-Vega:11}. Besides modeling these extreme-mass-ratio inspirals (EMRIs), the self-force provides an essential point of comparison with the methods used for intermediate- and equal-mass binaries~\cite{Favata:11,LeTiec-etal:11a,LeTiec-etal:11a,LeTiec-etal:12} and can be used to fix mass-dependent parameters in EOB~\cite{Damour:09,Barack-Damour-Sago:10, Barausse-etal:11}.

The gravitational self-force is now well understood at linear order in $m$~\cite{Poisson-Pound-Vega:11,Gralla-Wald:08,Pound:10a,Pound:10b}. However, to extract orbital parameters from a waveform emitted by an EMRI, one requires a theoretical description accurate to second order, as shown by the following simple scaling argument: Suppose $z(t)$ is the body's trajectory and that we have an equation for its acceleration $a$, but that our equation differs from the true acceleration by an error $\delta a$. Then our result for $z(t)$ will differ from the true position by $\delta z\sim t^2\delta a$. Therefore, to obtain a model accurate at order 1, we require $t^2\delta a \ll M$. An inspiral occurs on the timescale over which energy and angular momentum change substantially, which is the radiation-reaction time $t_{\rm rr}\sim M^2/m$, suggesting that we require $\delta a\ll m^2/M^3$; that is, we need $a$ to be accurate at least through order $m^2$. This conclusion is also borne out by a more rigorous scaling argument~\cite{Hinderer-Flanagan:08}.

Furthermore, comparisons with numerical simulations suggest that the second-order self-force would provide a highly accurate description of intermediate-mass-ratio binaries and even a reasonably accurate description of comparable-mass binaries~\cite{LeTiec-etal:11a,LeTiec-etal:12}, both of which should soon be observed by Advanced LIGO~\cite{Brown:07,Abadie:10}. The second-order force would also fix EOB parameters quadratic in $m$. Finally, second-order results would allow us to determine the error in various first-order approximations, such as the geodesic-force approximation used in Ref.~\cite{Warburton:11}.

Hence, there is a need to determine the self-force at least through second order, and proceeding to still higher order could be useful for studying intermediate- and equal-mass binaries. Harte~\cite{Harte:12} has shown that even outside the realm of perturbation theory, material bodies of sufficiently small size behave as test bodies in a certain effective metric, where the effective metric is defined by subtracting out the body's `self-field', which is bound to the body and has no direct influence on its bulk motion. This split into self-field and effective field overcomes the longstanding obstacle in applying Dixon's treatment of extended bodies to the self-gravitating case (as discussed in Sec. 13 of Ref.~\cite{Dixon:74}). In the context of the self-force, it is a powerful extension of the analogous first-order result of Detweiler and Whiting~\cite{Detweiler-Whiting:02}, who showed that the first-order self-force is equivalent (at order $m$) to geodesic motion in a certain effective metric. However, Harte's method does not apply to the motion of black holes, because it involves integrals over the body's interior. More importantly, it does not provide a concrete means of solving the Einstein field equation (EFE) to obtain the total metric or the effective metric---this must be done in a perturbative context. 

In recent years, two perturbative methods have emerged: one that expands the perturbation due to the body in a strict power series, representing the body's motion using deviation vectors pointing away from a reference geodesic~\cite{Gralla-Wald:08}; and one that uses a general expansion which self-consistently incorporates the corrections to the motion, working with an accelerated worldline rather than deviation vectors~\cite{Pound:10a,Pound:10b}. Both of these methods were devised with sufficient scope to be applied at any order in perturbation theory. The power series method is limited to short timescales over which the deviation vectors remain small, but there is the prospect of combining it with a two-timescale method~\cite{Hinderer-Flanagan:08} to create a valid approximation on the radiation-reaction time. The self-consistent approximation can be expected to be valid on long timescales, and it can always be reduced to the power series method by performing an additional local-in-time expansion.

Gralla has recently applied the power series method to second order~\cite{Gralla:12}. Shortly before that, I reported the second-order results of the self-consistent method~\cite{Pound:12}. In this paper and its sequel~\cite{Pound:12c}, I present the explicit expressions that would not fit within the short letter format of Ref.~\cite{Pound:12}. However, besides recapitulating those results, the present paper serves a larger purpose of showing how the EFE may be solved for an arbitrarily structured small body at any order in a self-consistent expansion. I also show, with slightly less certitude, how the $n$th-order equation of motion will be determined. While most of the necessary ideas were sketched in Ref.~\cite{Pound:10a}, here they are put in considerably more concrete form, while still avoiding detailed calculations. 

Both the power series and self-consistent expansions are based on the method of matched asymptotic expansions from singular perturbation theory~\cite{Kevorkian-Cole:96, Eckhaus:79, Kates:81,Pound:10b}. In this method, one uses two expansions, one valid very near the body and the other valid everywhere else, and information is fed between them by insisting that they agree in a buffer region around the body where they can both be expected to be valid. This method avoids modeling the body as a point particle (except in a well-defined sense discussed later in this paper), since such models are not well defined in the full, nonlinear theory.

A brief review of the method will be worthwhile. Suppose ${\sf g}_{\mu\nu}(\e)$ is an exact solution to the EFE containing the
small body on a manifold $\mathcal{M}$, where $\e$ is an expansion parameter
that counts powers of the body's mass. Now let $r$ be some measure of distance
from the body and $\mathcal{R}$ represent the spacetime's lengthscales,
excluding those of the body itself. For $r\sim\mathcal{R}$, well outside the body, in any global
coordinates $x^\alpha$ in a vacuum region $\Omega$, one can use the \emph{outer
expansion} 
\begin{equation}
{\sf g}_{\mu\nu}(x^\alpha,\e)=g_{\mu\nu}(x^\alpha)+h_{\mu\nu}(x^\alpha,\e)
\end{equation}
on a manifold $\mathcal{M}_E$. $(g_{\mu\nu},\mathcal{M}_E)$ defines an external
background spacetime with no small body in it, and
\begin{equation}
h_{\mu\nu}(x,\e)=\sum_{n\geq1}\e^n h_{\alpha\beta}^{(n)}(x;\e)
\end{equation}
describes perturbations due to the body. Each term $h^{(n)}_{\mu\nu}$ is allowed to depend on $\e$ through a dependence on the $\e$-dependent worldline $\gamma$ that will represent the body's motion in $\mathcal{M}_E$. This is the defining characteristic of the self-consistent approach~\cite{Pound:10a,Pound:10b}; in the method of Gralla and Wald, no such dependence is allowed. As a concrete example of the quantities involved in an outer expansion, in an EMRI, $g_{\mu\nu}$ is the spacetime of the supermassive black hole, the external lengthscale is $\mathcal{R}\sim M$, and $x^\alpha$ might be Boyer-Lindquist coordinates of the supermassive black hole.

Now, for $r\sim\e\mathcal{R}$, very near the body, the
metric varies rapidly, and there, in coordinates $(t,x^a)$ approximately
centered on the body, one can use the \emph{inner expansion} 
\begin{equation}
{\sf g}_{\mu\nu}(t,\tilde x^a,\e)=g_{{\rm I}\mu\nu}(t,\tilde x^a)+\sum_{n\geq1}\e^n H^{(n)}_{\mu\nu}(t,\tilde x^a)
\end{equation}
on a manifold $\mathcal{M}_{\rm I}$, where $(g_{{\rm I}\mu\nu},\mathcal{M}_{\rm I})$ is the body's spacetime were it isolated, 
and $H^{(n)}_{\mu\nu}$ describes perturbations. Here the components in the $(t,x^a)$ coordinates are written in terms of rescaled spatial coordinates $\tilde x^a\equiv x^a/\e$. These scaled coordinates serve to keep distances fixed relative to the body's mass in the limit $\e\ll1$, sending distances much larger than the mass off toward infinity. The use of a single scaling factor makes the approximation most appropriate
for compact bodies, in which the linear dimension is comparable to the mass; I return 
to this issue in Sec.~\ref{discussion}. Scaling only distances, not $t$, is 
equivalent to assuming the body possesses no fast internal dynamics. That is, there is no evolution on the short timescale of the body's mass and size.

In  the buffer region around the body, defined by $\e\ll r/\mathcal{R}\ll1$, the inner expansion can be expressed in unscaled coordinates and
then expanded for $r \gg \e\mathcal{R}$ (i.e., for distances that are large on the scale of the inner expansion) and the outer expansion can be expanded for $r\ll\mathcal{R}$ (i.e., for distances that are small on the scale of the outer expansion). Since the inner and outer expansions are assumed to approximate the same metric, the results of these re-expansions in the buffer region must match order by order in both $r$ and
$\e$. References~\cite{Eckhaus:79,Pound:10b} contain more details about the underlying assumptions in this procedure. But for my purposes here, the essential aspect is that an outer expansion can be obtained with only minimal knowledge of the inner expansion~\cite{Gralla-Wald:08,Pound:10a}. Furthermore, it can be obtained almost entirely from information in the buffer region. Say the $n$th-order perturbation is expanded as $h^{(n)}_{\mu\nu}=\sum_p r^p h^{(n,p)}_{\mu\nu}$ in the buffer region.\footnote{The coefficients $h^{(n,p)}_{\mu\nu}$ are not necessarily independent of $r$: they may have a functional dependence on $\ln r$, as discussed in Sec.~\ref{nth_outer_expansion}. However, the terms are still well ordered for small $r$, since $r^p(\ln r)^q\ll r^{p'}(\ln r)^{q'}$ for $p>p'$.} One must allow negative powers of $r$, since at least part of the field will fall off with distance from the body. But there is a lower bound on what the most negative power might be at a given order. I allow no negative powers of $\e$ in the inner expansion, meaning $\e^n h^{(n)}_{\mu\nu}$ must have no negative powers of $\e$ when written as a function of $\tilde r=r/\e$, from which it follows that 
\begin{equation}
\e^n h^{(n)}_{\mu\nu} = \frac{\e^n}{r^n}h^{(n,-n)}_{\mu\nu}+O(\e^n r^{-n+1}).\label{most_singular_part}
\end{equation}
Also, since $\e^n/r^n$ is independent of $\e$ in $\tilde r$ coordinates, it must be equal to a term in the zeroth-order part of the inner expansion, $g_{{\rm I}\mu\nu}$. Say $g_{{\rm I}\mu\nu}$, written in terms of the unscaled coordinates and expanded, reads\footnote{The fact that the inner background must be asymptotically flat, containing no positive powers of $r$, follows from the assumption that the outer expansion contains no negative powers of $\e$.}
\begin{equation}
g_{{\rm I}\mu\nu}=\sum_{n\geq0}\frac{\e^n}{r^n}g^{(n)}_{{\rm I}\mu\nu}.
\end{equation}
We then have $h^{(n,-n)}_{\mu\nu}=g^{(n)}_{{\rm I}\mu\nu}$. Therefore, at each order in $\e$, the most singular (as a function of $r$) piece of the metric perturbation $h^{(n)}_{\mu\nu}$ in the buffer region is determined by the $r\gg m$ asymptotic behavior of the body's unperturbed metric.

Now, from the perspective of the outer expansion, the buffer region surrounds a region $\B\subset\mathcal{M}_{\rm E}$. If $\gamma$ is in $\B$, then we can take $r$ to be a radial distance from $\gamma$, and the form~\eqref{most_singular_part} will hold true. We then have both the body---in the full spacetime---and $\gamma$---in $\mathcal{M}_{\rm E}$---in the region surrounded by the buffer, even if $\B$ is not diffeomorphic to the region surrounded by the buffer in $\mathcal{M}$ (say, if a black hole or wormhole resides therein). If mass dipole terms, which indicate the position of the center of mass relative to the origin of the coordinates, vanish in this coordinate system, then the body is appropriately centered ``on" $\gamma$; that is, on their respective manifolds both $\gamma$ and the body are at the center of the region surrounded by the buffer.\footnote{This notion of mass-centeredness based on mass dipole terms is not straightforward at high order in $\e$, as discussed in Sec.~\ref{nth_motion}. I assume that some copacetic notion of centeredness can be imposed.} $\gamma$ is then a meaningful and accurate representation of the body's motion.

These ideas were discussed at length in Ref.~\cite{Pound:10a}. In the present paper, I focus on their essential practical consequence: how a solution obtained entirely in the buffer region, without reference to details of the body's internal structure or global boundary conditions, can be used to construct a global solution to the $n$th-order EFE in the external spacetime. By substituting an expansion of the form~\eqref{most_singular_part} into the EFE, one can solve order by order in $r$. Because $r$ is small in the buffer region, spatial derivatives dominate over temporal ones. Therefore, solving the EFE is reduced to a process of solving a sequence of flat-space Poisson equations, and the solutions to corresponding homogeneous equations comprise all the freedom in the general solution. For multipole number $\ell$, these homogeneous solutions behave as $1/r^{\ell+1}$ and $r^\ell$, as is familiar from classical electromagnetism and Newtonian gravity. The $1/r^{\ell+1}$ terms are identified with the body's multipole moments or corrections thereto; the $r^\ell$ terms are identified with free radiation.

Inspired by this, I define a split of the $n$th-order solution of the EFE into a \emph{self-field} $h^{\rm S}_{\mu\nu}$ and an \emph{effective field} $h^{\rm R}_{\mu\nu}$. Roughly speaking, the self-field is constructed from the set of multipole moments and bound to the body. The effective field is a vacuum solution to the EFE that propagates independently of the body, even though the body created it, and in the solution obtained in the buffer region, it consists entirely of unknown functions, which can be determined only from global boundary conditions. My definition of the self-field carries a similar meaning to Harte's~\cite{Harte:12}, and the metric $g_{\mu\nu}+h^{\rm R}_{\mu\nu}$ defines an effective metric similar to Harte's, but because my effective field satisfies the vacuum EFE at all orders while Harte's does not, the definitions differ at second order and beyond.

Using the self-field found in the buffer region, one can obtain a global solution via a puncture scheme (also known as an effective-source scheme), just as at first order~\cite{Barack-Golbourn:07,Vega-Detweiler:07,Dolan-Barack:11,Dolan-Barack-Wardell:11,Vega-Wardell-Diener:11,Wardell-etal:11}. Such a scheme, in the present context of a small but extended body or black hole, begins by allowing the expressions for $h^{\rm R}_{\mu\nu}$ and $h^{\rm S}_{\mu\nu}$ in the buffer region to hold in the region $\B$ as well. Since doing so does not affect the field in the buffer region, it also does not affect the field values outside the buffer, out in the external universe. Continuing the expression for $h^{\rm S}_{\mu\nu}$ into $\B$ makes it into a field that diverges at $r=0$: the true self-field in the interior of the body, whatever it may be, is replaced with this divergent field, and the self-field becomes the \emph{singular field}. But we are not interested in obtaining $h^{\rm S}_{\mu\nu}$ in the body's interior, nor even in the buffer region, since it is already known there. We are interested only in $h^{\rm R}_{\mu\nu}$, which is the unknown part of the field in the buffer region. Extending the buffer-region expression for $h^{\rm R}_{\mu\nu}$ leads to a field that is smooth at $r=0$, which is the \emph{regular field}. Therefore, in a region covering $\B$ we can rewrite the EFE as an equation to be solved for $h^{\rm R}_{\mu\nu}$, by subtracting off the contribution of the singular field. The singular field, or any approximation to it that has the same singularity structure, serves as the titular `puncture' in the scheme. At the same time as the field equation is solved for the regular field, the `position' of the puncture is moved via the equation of motion for $\gamma$.

With such a scheme, the physical problem in the region covering the body, with all its matter fields, singularities (in the case of a black hole), or other oddities (in the case of exotic matter), is replaced with an \emph{effective} problem. One well-behaved problem free of singularities (except in the case of a black hole) in that region is replaced with another, but the variable being solved for is changed from the physical field to the effective field $h^{\rm R}_{\mu\nu}$. By appropriately converting from $h^{\rm R}_{\mu\nu}$ to the full field $h_{\mu\nu}$ outside that region, one can obtain $h_{\mu\nu}$ globally in the external spacetime. One will then have found the physical field outside the body solely from a specification of the body's multipole moments (which determine the singular/self-field) and initial data (which determines the effective/regular field), both of which are freely specifiable.

From this perspective, the various regularization methods that have been used in the self-force problem to remove the `singular part' of the field~\cite{Poisson-Pound-Vega:11} arise only as a practical necessity: we cannot determine the physical metric \emph{inside} the body, nor are we interested in doing so, which prompts us to replace it with the fiction of a singular field solely as a means of calculating the physical metric \emph{outside} the body.

In the paper proper, I make these notions more precise. Section~\ref{wave_gauges} presents the equations to be solved in the outer expansion, generalized beyond Refs.~\cite{Pound:10a,Pound:10b,Pound:12} to a wide class of wave gauges. This leads into the analysis of the $n$th-order outer expansion in Sec.~\ref{nth_outer_expansion}, where I show the form of the solution in the buffer region, define the singular/self- and regular/effective fields, and describe the puncture scheme in greater detail. I also derive a stress-energy tensor that effectively represents the body's composition as a skeleton of (corrected) multipole moments supported on the body's worldline. In addition, I show the relationship between the $n$th-order acceleration of the body's worldline and the gravitational field in its neighbourhood; however, this suffices to determine the motion only once one makes an appropriate choice of mass-centeredness and specifies the evolution of the body's multipole moments.

Section~\ref{explicit} applies this $n$th-order analysis at first and second order to obtain the first of two explicit results reported in Ref.~\cite{Pound:12}: an expression for the second-order singular/self-field through order $r$, which is sufficiently high order to find a global second-order solution numerically via a puncture/effective-source scheme. Previously, in Ref.~\cite{Pound:10a}, the second-order singular field was obtained through order $r^0$, but explicitly acceleration-dependent terms were dropped. The second result reported in Ref.~\cite{Pound:12} was that if the body is spherical, then $\gamma$ satisfies the geodesic equation in the effective metric $g_{\mu\nu}+h^{\rm R}_{\mu\nu}$ through second order in $\e$. In the sequel~\cite{Pound:12c}, I will provide the detailed expressions involved in the derivation of that equation of motion.

I conclude the present paper in Sec.~\ref{discussion} with a discussion of future work and a comparison of my method and results to other second-order analyses: that of Gralla~\cite{Gralla:12}, mentioned above, earlier results of Rosenthal~\cite{Rosenthal:06a,Rosenthal:06b}, and a recent discussion of Detweiler~\cite{Detweiler:12}.

Throughout, I work in units of $G=c=1$. Greek indices range from 0 to 3. Lowercase Latin indices refer to spatial coordinates. Uppercase Latin indices denote multi-indices; for example, $L\equiv i_1\cdots i_\ell$. $T_{\langle L\rangle}=\displaystyle\mathop{\rm STF}_{L}T_L=\hat T_L$ indicate a tensor symmetric and trace-free (STF) with respect to $\delta_{ab}$.

\section{Outer expansion in generalized wave gauges}\label{wave_gauges}
Let $D\subset\mathcal{M}_{\rm E}$ be a vacuum region of the background spacetime (i.e., a region in which $g_{\mu\nu}$ satisfies the vacuum EFE). Furthermore, let it be a globally hyperbolic region with Cauchy surface $\Sigma$, such that $D=D^+(\Sigma)$ and wave equations can be solved in $D$ given only initial data on $\Sigma$. Now let $\Omega\subset\mathcal{M}_{\rm E}$ be $D$ with the region surrounded by the buffer removed; that is, $\Omega=D\backslash\B$. 
 $\Omega$ then corresponds to a vacuum region outside the body in the full spacetime. This will be the region in which I seek a solution to the EFE.

In $\Omega$, the self-consistent method of Ref.~\cite{Pound:10a} is modelled on standard post-Minkowskian expansions in the harmonic gauge, as used in post-Newtonian theory~\cite{Blanchet:06}. It is convenient to use any field variable $\phi^{\mu\nu}$ satisfying
\begin{equation}
\phi^{\mu\nu} = \bar h^{\mu\nu}+O(\bar h^2),
\end{equation}
where an overbar indicates trace-reversal with the background metric $g_{\mu\nu}$, and the $O(\bar h^2)$ indicates anything that becomes of quadratic or higher order when expanded for small $\bar h^{\mu\nu}$. For example, some candidate fields are $\phi^{\mu\nu}\equiv\bar h^{\mu\nu}$, $\phi^{\mu\nu}\equiv h^{\mu\nu}-\frac{1}{2}{\sf g}^{\mu\nu}{\sf g_{\rho\sigma}}h^{\rho\sigma}$, and $\phi^{\mu\nu}=H^{\mu\nu}\equiv g^{\mu\nu}-\frac{\sqrt{-{\sf g}}}{\sqrt{- g}}{\sf g}^{\mu\nu}$. The latter choice is discussed in detail in Ref.~\cite{Harte:11v1}; it is the field used in post-Newtonian theory in the case that $g_{\mu\nu}$ is Minkowski. 

To disentangle the field from the matter degrees of freedom, one may first impose a convenient gauge condition. I adopt a generalized wave gauge defined by 
\begin{equation}
\nabla_{\!\nu}\phi^{\mu\nu}=Z^\mu{}_{\nu\rho}\phi^{\nu\rho}+D^{\mu},\label{generalized_gauge}
\end{equation}
where $Z^\mu{}_{\nu\rho}$ and $D^\mu$ are specified functions of $x^\alpha$ and $\e$.\footnote{One could instead impose ${}^{\sf g}\nabla_{\!\nu}\phi^{\mu\nu}=Z^\mu{}_{\nu\rho}\phi^{\nu\rho}+D^{\mu}$, where ${}^{\sf g}\nabla_{\!\nu}$ is the covariant derivative compatible with ${\sf g}_{\mu\nu}$, but doing so would be equivalent to a redefinition of $\phi^{\mu\nu}$.} I restrict $Z^\mu{}_{\nu\rho}$ to be smooth on $\mathcal{M}_{\rm E}$ and independent of $\e$ (in any coordinates that are independent of $\e$), for reasons to be made clear in a moment. For simplicity, I assume $D^\mu$ to be a smooth function of $\e$, admitting a power series $D^{\mu}(x,\e)=\sum \e^nD^{\mu}_{(n)}(x)$. I allow $D^{\mu}_{(n)}(x)$ to diverge as $1/r^{n+1}$ as $r\to0$, which preserves the singularity structure~\eqref{most_singular_part} of the metric perturbation.

If $\phi^{\mu\nu}=\bar h^{\mu\nu}$, then \eqref{generalized_gauge} is a generalized Lorenz gauge condition. If $\phi^{\mu\nu}=H^{\mu\nu}$, then \eqref{generalized_gauge} reduces to a generalized harmonic gauge condition when $g_{\mu\nu}$ is Minkowski; in the case where $\phi^{\mu\nu}=H^{\mu\nu}$ and $Z^\mu{}_{\nu\rho}=0$, the gauge condition can be written as a condition in the full spacetime rather than in the background~\cite{Harte:11v1}. In all cases, the gauge condition~\eqref{generalized_gauge} makes the linearized Einstein tensor hyperbolic by removing its elliptic part.

Therefore, with Eq.~\eqref{generalized_gauge} imposed, the exact vacuum EFE $G^{\mu\nu}[{\sf g}]=0$ in $\Omega$ is put in relaxed form, as in post-Newtonian theory. It becomes a weakly nonlinear wave equation,
\begin{align}
E^{\mu\nu}{}_{\rho\sigma}\phi^{\rho\sigma} &= C^{\mu\nu}-16\pi\tau^{\mu\nu},\label{wave_equation}
\end{align}
where
\begin{align}
E^{\mu\nu}{}_{\rho\sigma} &\equiv g^{\mu}_{\rho}g^{\nu}_{\sigma}g^{\alpha\beta}\nabla_{\!\alpha}\nabla_{\!\beta}+2R^{\mu}{}_\rho{}^\nu{}_\sigma \nonumber\\
						& -2(Z^{(\mu}{}_{\rho\sigma}{}^{;\nu)}+Z^{(\mu}{}_{\rho\sigma}\nabla^{\nu)})\nonumber\\
						& +g^{\mu\nu}(Z^{\lambda}{}_{\rho\sigma;\lambda}+Z^{\lambda}{}_{\rho\sigma}\nabla_{\!\lambda})
\end{align}
becomes the usual wave operator of linearized gravity when $Z^\mu{}_{\nu\rho}=0$, returning $E^{\mu\nu}{}_{\rho\sigma}\phi^{\rho\sigma}=\Box\phi^{\mu\nu}+2R^\mu{}_\rho{}^\nu{}_\sigma\phi^{\rho\sigma}$ in that case. The smoothness and $\e$-independence of $Z^\mu{}_{\nu\rho}$ on $\mathcal{M}_{\rm E}$ ensures that $E^{\mu\nu}{}_{\rho\sigma}$ is a nicely behaved operator in the background spacetime. The source terms on the right-hand side of the wave equation~\eqref{wave_equation} are
\begin{align}
C^{\mu\nu} &\equiv 2\nabla^{(\mu}D^{\nu)}-g^{\mu\nu}\nabla_{\!\rho}D^{\rho}, 
\end{align}
which is a gauge term, and
\begin{equation}
\tau^{\mu\nu}[\phi]\equiv \frac{1}{8\pi}\left(G^{\mu\nu}[g,\phi]-\delta G^{\mu\nu}[\phi]\right),\label{tau}
\end{equation}
which can be thought of as the stress-energy tensor of the gravitational field, comprising all terms in $G^{\mu\nu}[{\sf g}]$ that are nonlinear in $\phi^{\mu\nu}$. Here $G^{\mu\nu}[g,\phi]$ is the full Einstein tensor $G^{\mu\nu}[g+h]$ after rewriting $h_{\mu\nu}$ in terms of $\phi^{\mu\nu}$, and $\delta G^{\mu\nu}[\phi]$ is the linearized Einstein tensor written in terms of $\phi^{\mu\nu}$. $\tau^{\mu\nu}$ can be written with the gauge condition already imposed or not. Note that unlike the Landau-Lifshitz pseudotensor or its generalization to curved backgrounds in Ref.~\cite{Harte:11v1}, $\tau^{\mu\nu}$ contains second derivatives of the metric perturbation, meaning in addition to its `gravitational energy' terms, it contains terms that should not be so interpreted. 

Equation~\eqref{wave_equation} can be solved without constraining the body's bulk motion or the evolution of its matter degrees of freedom. No stress-energy tensor appears in Eq.~\eqref{wave_equation}, since the small body lies outside $\Omega$, which may make this lack of constraints seem vacuous. But we can say more precisely that in $\Omega$, the wave equation can be solved without constraining the evolution of the multipole moments in the buffer region, including that of mass dipole terms which shall be used to define the bulk motion. 

Given its smooth dependence on $\e$, $C^{\mu\nu}$ can be expanded in an ordinary power series $C^{\mu\nu}(x,\e)=\sum \e^nC^{\mu\nu}_{(n)}(x)$. I next assume an asymptotic expansion
\begin{equation}\label{phi_expansion}
\phi^{\mu\nu}(x,\e)=\sum_{\e\geq1}\e^n\phi^{\mu\nu}_{(n)}(x;\e)
\end{equation}
in which the terms $\phi^{\mu\nu}_{(n)}(x;\e)$ are constructed by splitting the weakly nonlinear wave equation~\eqref{wave_equation} into a sequence of wave equations
\begin{align}
E^{\mu\nu}{}_{\rho\sigma}\phi^{\rho\sigma}_{(n)} &= C^{\mu\nu}_{(n)}-16\pi\tau^{\mu\nu}_{(n)},\label{nth_wave}
\end{align}
where $\tau^{\mu\nu}_{(n)}$ contains all pieces of $\tau^{\mu\nu}$ with an overall factor of $\e^n$: $8\pi\tau^{\mu\nu}_{(n)}=\delta^2 G^{\mu\nu}[\phi_{(n-1)},\phi_{(1)}]+\delta^2 G^{\mu\nu}[\phi_{(n-2)},\phi_{(2)}]+\ldots+\delta^n G^{\mu\nu}[\phi_{(1)}]$, where $\delta^n G^{\mu\nu}[\phi]$ is the $n$th-order Einstein tensor. (The notation is described in Appendix~\ref{second-order_tensors}.) More concretely, at first and second order these equations read
\begin{align}
E^{\mu\nu}{}_{\rho\sigma}\phi^{\rho\sigma}_{(1)} &= C^{\mu\nu}_{(1)},\\
E^{\mu\nu}{}_{\rho\sigma}\phi^{\rho\sigma}_{(2)} &= C^{\mu\nu}_{(2)}-2\delta^2G^{\mu\nu}[\phi_{(1)}].
\end{align}

Equations~\eqref{nth_wave} can be solved in sequence for each $\phi^{\mu\nu}_{(n)}$ without constraining the motion of the small body, and $\phi^{\mu\nu}_{(n)}(x;\e)$ is taken to be the solution even when the motion is faithfully represented (i.e., when the buffer region at a distance $\e\ll r/\mathcal{R}\ll1$ from $\gamma$ always surrounds the body). The dependence on $\e$ in each term thus takes the form of a functional dependence on $\gamma$. I assume that despite the dependence on $\e$, each $\phi^{\mu\nu}_{(n)}(x;\e)$ is uniformly of order 1.\footnote{It may also contain a dependence on $\ln\e$ coming from the perturbation of light cones. Such logarithmic terms are discussed in Appendix~\ref{logarithms}. If they arise, then the statement of uniformity should instead read that the coefficient of each $(\ln\e)^q$ in $\phi^{\mu\nu}_{(n)}(x;\e)$ is uniformly of order 1.} If this is found to be untrue in a particular case, then the asymptotic series \eqref{phi_expansion} has failed.

The motion is determined via the gauge condition, which ensures that a solution $\phi^{\mu\nu}$ to the wave equation~\eqref{wave_equation} is also a solution to the EFE (and therefore the Bianchi identity). However, I do not wish to split the gauge condition \eqref{generalized_gauge} into a sequence of conditions of the type $\nabla_{\!\nu}\phi^{\mu\nu}_{(n)}=Z^\mu{}_{\nu\rho}\phi^{\nu\rho}_{(n)}+D^{\mu}_{(n)}$. Those conditions would imply that $\phi^{\mu\nu}_{(1)}$ satisfies the linearized EFE and therefore the linearized Bianchi identity, which would constrain the acceleration $a^\mu$ of $\gamma$ to vanish~\cite{Pound:10a}. Yet I do wish to split Eq.~\eqref{generalized_gauge} into equations that can be solved for each $\phi^{\mu\nu}_{(n)}$. To do so, I assume the acceleration can be expanded as
\begin{equation}
a^\mu(\tau,\e) = \sum_n\e^na^\mu_{(n)}(\tau,\e),\label{a_expansion}
\end{equation}
where at time $\tau$, each $a^\mu_{(n)}$ is allowed to depend on the past history of the $\e$-dependent worldline $\gamma$, in the same manner as $\phi^{\mu\nu}$ does. When written in a coordinate system centered on $\gamma$, all fields involved in the gauge condition---$g_{\mu\nu}$, $\phi^{\mu\nu}$, $Z^\mu{}_{\rho\sigma}$, and $D^\mu$---will naturally depend on $\gamma$'s acceleration. So in such a coordinate system, one may substitute the expansion for $a^\mu$, together with the expansion of $\phi^{\mu\nu}$, into the gauge condition to arrive at the equation
\begin{align}
\sum_{s=1}^{n}\bar \delta^{n-s}\nabla_{\!\nu}\phi^{\mu\nu}_{(s)}
								&= \sum_{s=1}^{n}\bar\delta^{n-s}Z^{\mu}{}_{\rho\sigma}\phi^{\rho\sigma}_{(s)}\nonumber\\
								&\quad	+\sum_{s=1}^{n}\bar\delta^{n-s}D^{\mu}_{(s)}\label{nth_gauge}
\end{align}
for each $n$, where $\bar\delta^s$ returns the $s$th-order term in the expansion with respect to $a^\mu$ and its derivatives. For example, $\bar\delta^2 a^\mu\dot a^\nu=a^\mu_{(2)}\dot a^\nu_{(0)}+a^\mu_{(0)}\dot a^\nu_{(2)}+2a^\mu_{(1)}\dot a^\nu_{(1)}$. $\bar\delta^0$ indicates evaluation at $a^{\mu}=a^\mu_{(0)}$. Equation \eqref{nth_gauge} can also be written as an equation for $\phi^{\mu\nu}_{(n)}$ in terms of the lower-order fields:
\begin{align}
\bar \delta^0&\left(\nabla_{\!\nu}\phi^{\mu\nu}_{(n)}-Z^{\mu}{}_{\rho\sigma}\phi^{\rho\sigma}_{(n)}-D^{\mu}_{(n)}\right)\nonumber\\
			&= \sum_{s=1}^{n-1}\bar \delta^{n-s}\Bigg(Z^{\mu}{}_{\rho\sigma}\phi^{\rho\sigma}_{(s)}	+D^{\mu}_{(s)}-\nabla_{\!\nu}\phi^{\mu\nu}_{(s)}\Bigg).
\end{align}

There is no obvious way to express the $\bar\delta^s$ operation in a covariant manner, and one could do away with it entirely by instead solving approximate, rather than exact, equations for each $\phi^{\mu\nu}_{(n)}$. In `little o' notation, these equations would read 
\begin{align}
\e\bigg(\nabla_{\!\nu}\phi^{\mu\nu}_{(1)}-Z^\mu{}_{\rho\sigma}\phi_{(1)}^{\rho\sigma}-D^\mu_{(1)}\bigg) &=o(\e),\label{1st_gauge_approx}\\
\sum_{n=1,2}\!\e^n\bigg(\nabla_{\!\nu}\phi^{\mu\nu}_{(n)}-Z^\mu{}_{\rho\sigma}\phi_{(n)}^{\rho\sigma}-D^\mu_{(n)}\bigg) &=o(\e^2),\label{2nd_gauge_approx}
\end{align}
etc., where $o(\e^n)$ means terms that approach zero faster than $\e^n$ for small $\e$. Equations~\eqref{1st_gauge_approx} and \eqref{2nd_gauge_approx} follow directly from Eq.~\eqref{generalized_gauge}, given the assumed uniformity of each $\phi^{\mu\nu}_{(n)}$. However, the expansion of the acceleration helps to systematize the calculation, and it trivially leads to order-reduced equations in cases where derivatives of the acceleration would appear on the right-hand side of the equation of motion.

As in traditional post-Minkowskian expansions, these gauge conditions are closely related to the conservation of the source $\tau^{\mu\nu}$. Taking the divergence of the wave equation~\eqref{wave_equation} and using the identities $\nabla_{\!\nu}C^{\mu\nu}=\Box D^\mu$ and $\nabla_{\!\nu}E^{\mu\nu}{}_{\rho\sigma}\phi^{\rho\sigma}=\Box(\nabla_{\!\nu}\phi^{\mu\nu}-Z^\mu{}_{\nu\rho}\phi^{\nu\rho})$, we find
\begin{equation}
\Box H^\mu = -16\pi\nabla_{\!\nu}\tau^{\mu\nu},
\end{equation}
where I have defined
\begin{equation}
H^\mu\equiv \nabla_{\!\nu}\phi^{\mu\nu}-Z^\mu{}_{\nu\rho}\phi^{\nu\rho}-D^\mu,
\end{equation}
in terms of which the gauge condition reads $H^\mu=0$. From this it follows that if (i) $\nabla_{\!\nu}\tau^{\mu\nu}=0$ in $\Omega$ and (ii) $H^\mu=\nabla_{n}H^\mu=0$ on the boundary $\partial\Omega$, where $n^\alpha$ is a vector normal to $\partial\Omega$, then $H^\mu=0$ everywhere in $\Omega$. Equation~\eqref{tau} together with the Bianchi identity shows that if $G^{\mu\nu}[g,\phi]=o(\e^n)$, then $\nabla_{\!\nu}\tau^{\mu\nu}=o(\e^{n+1})$. Combining these results, we can show by induction over $n$ that if (i) the data on the part of $\Sigma$ outside the buffer region satisfies the gauge condition, (ii) the gauge condition is satisfied in the buffer region to an accuracy $H^\mu=o(\e^n)$, and (iii) the first $n$ wave equations are satisfied, then the gauge condition is satisfied up to $o(\e^n)$ errors everywhere in $\Omega$ (putting aside issues of secularly growing errors in $H^\mu$). Usually, one has that the conservation equation $\nabla_{\!\nu}(T^{\mu\nu}+\tau^{\mu\nu})=0$ together with gauge-satisfying initial data suffices to enforce the gauge condition everywhere; here, since we exclude the region containing matter, enforcement of the gauge condition in the buffer region takes over the role of conservation of material stress-energy.

Note that in previous work~\cite{Pound:10a,Pound:10b,Pound:12}, I used the metric perturbation $h_{\mu\nu}$ as the field variable, rather than something based on its trace-reverse. Had I used a field $\phi_{\mu\nu} = h_{\mu\nu}+O(h^2)$ as the variable in this section, everything would have carried through almost identically. The gauge condition would have involved the divergence of the trace-reversed field, $(g_\mu^\lambda g^{\nu\sigma}-\frac{1}{2}g_{\mu}^\nu g^{\lambda\sigma})\nabla_{\!\nu}\phi_{\lambda\sigma}=Z_{\mu}{}^{\nu\rho}\phi_{\nu\rho}+D_\mu$, and the sequence of wave equations for $\phi^{(n)}_{\rho\sigma}$ would have followed from an expansion of the exact EFE $R_{\mu\nu}[g+h]=0$, since the linearized Ricci tensor in a wave gauge takes the form of a wave operator acting on $h_{\mu\nu}$ in the same way the linearized Einstein tensor takes the form of a wave operator acting on $\bar h_{\mu\nu}$. In the present paper, I favor the trace-reversed perturbation for the sake of the simpler gauge condition.

Also note that throughout this section, I have used the language of post-Newtonian theory, in which the exact EFE is said to be relaxed by adopting a gauge in which the EFE can be solved without constraining the source. One then seeks a field $\phi^{\mu\nu}$ that solves both the relaxed EFE and the gauge condition as accurately as possible. In self-force literature~\cite{Quinn-Wald:97,Gralla-Wald:08,Gralla:12}, there is a notion of relaxing the gauge condition, meaning one imposes $\nabla_{\!\nu}\phi^{\mu\nu}_{(1)}=0$ but then allows $\nabla_{\!\nu}\e\phi^{\mu\nu}_{(1)}=O(\e^2)$, which avoids constraining the motion at order $\e$. These two notions of relaxation are similar but distinct, and there is no sense in which I relax the gauge condition on a perturbation: $\nabla_{\!\nu}\phi^{\mu\nu}_{(1)}=0$ is simply never imposed; only within the context of a Taylor series (i.e., regular perturbation theory) would the condition $\nabla_{\!\nu}\phi^{\mu\nu}=0$ imply the condition $\nabla_{\!\nu}\phi^{\mu\nu}_{(1)}=0$, which would then have to be relaxed in order to self-consistently evolve the position of the body and the field it produces.


\section{$n$th-order outer expansion}\label{nth_outer_expansion}


Before constructing explicit solutions to the field equations for a particular choice of $\phi^{\mu\nu}$, $Z^\mu{}_{\rho\sigma}$, and $D^\mu$, I discuss the general form of the solution at any order. This involves two steps: first, deriving a local expression for a certain self-field in the buffer region; second, using that local expression to obtain the global solution via a puncture/effective-source scheme. I first derive the general solution to the $n$th-order wave equation~\eqref{nth_wave} in the buffer region, subject only to the matching condition, with neither the body's composition nor other boundary conditions specified.

From the form of the field, I also derive an effective form for the body's $n$th-order stress-energy tensor in terms of the body's multipole moments and the corrections to them. Finally, from the gauge condition~\eqref{nth_gauge} in the buffer region, I derive the relationship between the $n$th-order acceleration of $\gamma$ and the metric perturbation in its neighbourhood.

\subsection{General solution in buffer region}\label{nth_buffer_solution}
In the buffer region, I work in local coordinates $(t,x^a)$ centered on $\gamma$ and use $\delta_{ab}$ to raise and lower lowercase Latin indices. Any choice of local coordinates will do, so long as the metric reduces to Minkowski on $\gamma$; that is, $g_{\mu\nu}=\eta_{\mu\nu}+O(r)$, where $r\equiv\sqrt{\delta_{ij}x^ix^j}$ is the spatial distance from $\gamma$. For the explicit results in Sec.~\ref{explicit}, I adopt Fermi-Walker coordinates, in which $t$ is proper time on $\gamma$ and $r$ is the proper distance along a geodesic perpendicular to $\gamma$. But prior to that, excepting a brief detour in Sec.~\ref{nth_motion}, I leave the coordinates unspecified.

I write $\phi^{\mu\nu}_{(n)}$ as $\phi^{\mu\nu}_{(n)}(t,r,\vec n)$, where the unit vector $n^i\equiv x^i/r$. Favoring $n^a$ over the usual angles $(\theta,\phi)$ is useful because the smooth background metric $g_{\mu\nu}$ will naturally be a power series in $x^a=r n^a$. I assume that $\phi^{\mu\nu}_{(n)}(t,r,\vec n)$ possesses an expansion of the form
\begin{equation}
\phi^{\mu\nu}_{(n)}(t,r,n^a)=\sum_{p\geq-n}\sum_{\ell=0}^\lmax \sum_{q=0}^{\qmax}r^p(\ln r)^q\phi^{\mu\nu L}_{(npq\ell)}(t)\nhat_L, \label{hn_form}
\end{equation}
where the coefficients $\phi^{\mu\nu L}_{(npq\ell)}$ are smooth functions of time, and $\nhat_L\equiv n_{\langle i_1}\cdots n_{i_\ell\rangle}$ is STF with respect to $\delta_{ab}$. The decomposition in terms of $\nhat_L$ is equivalent to an expansion in spherical harmonics~\cite{Damour-Blanchet:86}; standard useful identities involving this decomposition are provided in Appendix~\ref{STF_decomposition}. The lower limit on the sum over $p$, which makes $r^{-n}$ the most singular power of $r$, comes from the matching condition, as noted in Sec.~\ref{intro}. The logarithmic terms arise from the perturbation of light cones. One can expect the solution to the exact EFE to propagate on (and within) null cones of the exact spacetime, and given that the mass of the body induces a logarithmic correction to the retarded time, logarithmic corrections then naturally appear in $\phi^{\mu\nu}_{(n)}$. This effect is well known from solutions to the EFE in harmonic coordinates (see, e.g., Refs.~\cite{Damour-Blanchet:86,Blanchet:87}). Other causes may also lead to logarithmic terms. For generality, I allow logarithms at any value of $n$, but I assume that for each finite $n$, $p$, and $\ell$, the highest power of $\ln r$ is a finite number $\qmax(n,p,\ell)$. It will be seen that this maximum is required to obtain a solution. For simplicity, to make sure that term-by-term differentiation is valid without worrying about issues of convergence, I also assume for a given, finite $n$ and $p$, $\ell$ has a maximum $\lmax(n,p)$.

I proceed by substituting this expansion into Eq.~\eqref{nth_wave} and solving for the coefficients $\phi^{\mu\nu L}_{(npq\ell)}(t)$. To facilitate that procedure, I split the wave operator into two pieces, $E^{\mu\nu}{}_{\rho\sigma}=g^\mu_\rho g^\nu_\sigma\partial^i\partial_i+\Delta E^{\mu\nu}{}_{\rho\sigma}$. The first operator is a flat-space Laplacian, which reduces a term of order $r^p$ to a term $\sim r^{p-2}$ (treating $r^p(\ln r)^q\sim r^p$). The second operator,
\begin{align}
\Delta E^{\mu\nu}{}_{\rho\sigma}\equiv E^{\mu\nu}{}_{\rho\sigma}-g^\mu_\rho g^\nu_\sigma\partial^i\partial_i,
\end{align}
takes a term of order $r^p$ and returns terms of order $r^{p-1}$ and higher. I break this down further into
\begin{equation}
\Delta E^{\mu\nu}{}_{\rho\sigma} = \sum_{p'\geq-1}{}^{(p')\!}\Delta E^{\mu\nu}{}_{\rho\sigma},
\end{equation}
where ${}^{(p')\!}\Delta E^{\mu\nu}{}_{\rho\sigma}$ takes a term of order $r^p$ and returns a term of order $r^{p+p'}$. The wave equation~\eqref{nth_wave} can then be written in the form of a Poisson equation,
\begin{align}
\partial^i\partial_i \phi^{\mu\nu}_{(n)} &= -\Delta E^{\mu\nu}{}_{\rho\sigma}\phi^{\rho\sigma}_{(n)}+C^{\mu\nu}_{(n)}-16\pi\tau^{\mu\nu}_{(n)}\\
&\equiv P^{\mu\nu}_{(n)}. \label{nth_Poisson}
\end{align}
Because $\tau^{\mu\nu}_{(n)}$ is made up of nonlinear combinations of $\phi^{\mu\nu}_{(n')}$ (where $n'<n$), it necessarily has an expansion of the form
\begin{align}
\tau^{\mu\nu}_{(n)} &= \sum_{p\geq-n}\sum_{q,\ell}r^{p-2}(\ln r)^q \tau^{\mu\nu L}_{(n,p,q,\ell)}\nhat_L.
\end{align}
Similarly, because $\Delta E^{\mu\nu}{}_{\rho\sigma}\phi^{\rho\sigma}_{(n)}$ is simply a combination of unit vectors and derivatives acting upon $\phi^{\rho\sigma}_{(n)}$, it has an analogous expansion. Assuming $C^{\mu\nu}_{(n)}$ can likewise be expanded as
\begin{align}
C^{\mu\nu}_{(n)} &= \sum_{p\geq-n}\sum_{q,\ell}r^{p-2}(\ln r)^q C^{\mu\nu L}_{(n,p,q,\ell)}\nhat_L,
\end{align}
the source in the Poisson equation then also has an expansion
\begin{align}
P^{\mu\nu}_{(n)} &= \sum_{p\geq-n}\sum_{\ell=0}^\lSmax\sum_{q=0}^\qSmax r^{p-2}(\ln r)^q P^{\mu\nu L}_{(n,p,q,\ell)}\nhat_L,
\end{align}
where $\lSmax=\lSmax(n,p)$ and $\qSmax=\qSmax(n,p,\ell)$ are finite for finite $n$, and the coefficients are given by
\begin{align}
P^{\mu\nu L}_{(n,p,q,\ell)} &= -r^{2-p}(\ln r)^{-q}\sum_{p'=-n}^{p-1}\sum_{\ell',q'}\nonumber\\
&\quad \times\left[{}^{(p-p'-2)\!}\Delta E^{\mu\nu}{}_{\rho\sigma}r^{p'}(\ln r)^{q'}\phi^{\rho\sigma L'}_{(n,p',q',\ell')}\nhat_{L'}\right]^L_{(q)}\nonumber\\
&\quad +C^{\mu\nu L}_{(n,p,q,\ell)}-16\pi \tau^{\mu\nu}_{(n,p,q,\ell)}.\label{nth_Poisson_Source}
\end{align}
Here $[f]^L_{(q)}$ denotes the coefficient of $(\ln r)^q\nhat_L$ in the series $f=\sum_{\ell',q'}(\ln r)^q [f]^{L'} \nhat_{L'}$, where the coefficient $[f]^{L'}$ is given by Eq.~\eqref{lth_coeff}.

Because $\nhat_L$ is an eigenvector of the flat-space Laplacian, satisfying $r^2\partial^i\partial_i \nhat_L = -\ell(\ell+1)\nhat_L$ (see Appendix~\ref{STF_decomposition}), one can easily evaluate the left-hand side of Eq.~\eqref{nth_Poisson}. Doing so, and substituting  the expansion of $P^{\mu\nu}_{(n)}$ into the right-hand side, leads to 
\begin{align}
&\sum_{p\geq-n}\sum_{q=0}^{\qmax}\sum_{\ell=0}^\lmax r^{p-2}(\ln r)^q\nhat_L\nonumber\\
&\quad \times \left\lbrace \left[p(p+1)-\ell(\ell+1)\right]\phi^{\mu\nu L}_{(npq\ell)}\right.\nonumber\\
&\quad +(q+1)(2p+1)\phi^{\mu\nu L}_{(n,p,q+1,\ell)}\nonumber\\
&\quad \left.+(q+1)(q+2)\phi^{\mu\nu L}_{(n,p,q+2,\ell)}\right\rbrace\nonumber\\
&\quad = \sum_{p\geq-n}\sum_{\ell=0}^{\lSmax}\sum_{q=0}^{\qSmax}r^{p-2}(\ln r)^q\nhat_LP^{\mu\nu L}_{(npq\ell)}\label{nth_Poisson_expanded},
\end{align}
where I have defined $\phi^{\mu\nu L}_{(npq\ell)}\equiv0$ for $q>\qmax$. This can now be solved by equating coefficients. Since the coefficient of $r^{p-2}$ in $P^{\mu\nu}_{(n)}$ is constructed from $\phi_{(n')}^{\mu\nu}$ for $n'<n$ and $\phi^{\mu\nu L}_{(n,p',q,\ell)}$ for $p'<p$, a term $\phi^{\mu\nu}_{(n,p,q,\ell)}$ on the left-hand side is sourced by lower-order terms on the right-hand side, allowing one to solve order by order.

The form of the solution obviously differs depending on whether $p(p+1)-\ell(\ell+1)$ vanishes. I define 
\begin{align}\label{lp}
\ell_p = \left\{
	\begin{array}{ll}
		-p-1  & \mbox{if } p < 0, \\
		p & \mbox{if } p \geq 0,
	\end{array}
\right.
\end{align}
which is the unique non-negative solution to $p(p+1)-\ell_p(\ell_p+1)=0$. I call $\phi^{\mu\nu L_p}_{(np0\ell_p)}$ a \emph{homogeneous mode}, inspired by the fact that $r^p\nhat_{L_p}\phi^{\mu\nu L_p}_{(np0\ell_p)}$ satisfies the homogeneous Poisson equation. This can  easily be seen by ignoring the logarithm terms, in which case the homogeneous Poisson equation reads $\sum_{p,\ell}r^{p-2}\nhat_L\left[p(p+1)-\ell(\ell+1)\right]\phi^{\mu\nu L}_{(np0\ell)} = 0$.

Now, for a given $p$, first consider the modes for which $\ell\neq\ell_p$. Each mode $\phi^{\mu\nu L}_{(npq\ell)}$ is coupled to the two higher-$q$ modes $\phi^{\mu\nu L}_{(n,p,q+1,\ell)}$ and $\phi^{\mu\nu L}_{(n,p,q+2,\ell)}$. Start with $q=\qmax$, in which case the two higher-$q$ modes vanish by definition; this is the reason why one must assume finite $\qmax$ in order to obtain a solution. If $\qmax>\qSmax$, the Poisson equation reads $\left[p(p+1)-\ell(\ell+1)\right]\phi^{\mu\nu L}_{(npq\ell)}=0$, implying $\phi^{\mu\nu L}_{(np\qmax\ell)}=0$. It follows that for $\ell\neq\ell_p$, $\qmax=\qSmax$, and the $q=\qmax$ term in the Poisson equation reads $\left[p(p+1)-\ell(\ell+1)\right]\phi^{\mu\nu L}_{(np\qSmax\ell)}=P^{\mu\nu L}_{(np\qSmax\ell)}$, or
\begin{align}
\phi^{\mu\nu L}_{(n,p,\qSmax,\ell)} & = \frac{P^{\mu\nu L}_{(np\qSmax\ell)}}{p(p+1)-\ell(\ell+1)}.
\end{align}
For $q=\qSmax-1$, one then immediately finds
\begin{align}
\phi^{\mu\nu L}_{(n,p,\qSmax-1,\ell)} & = \frac{P^{\mu\nu L}_{(n,p,\qSmax-1,\ell)}}{p(p+1)-\ell(\ell+1)}\nonumber\\
							&\quad -\frac{\qSmax(2p+1)P^{\mu\nu L}_{(np\qSmax\ell)}}{\left[p(p+1)-\ell(\ell+1)\right]^2}.
\end{align}
These two results are sufficient to find $\phi^{\mu\nu L}_{(npq\ell)}$ for any $0\leq q<\qSmax-2$, since
\begin{align}
\phi^{\mu\nu L}_{(npq\ell)} & = \frac{1}{p(p+1)-\ell(\ell+1)}\Big[P^{\mu\nu L}_{(npq\ell)}\nonumber\\
						&\quad-(q+1)(2p+1)\phi^{\mu\nu L}_{(n,p,q+1,\ell)}\nonumber\\
						&\quad-(q+1)(q+2)\phi^{\mu\nu L}_{(n,p,q+2,\ell)}\Big].
\end{align}

Now consider modes for which $\ell=\ell_p$. In this case each mode $\phi^{\mu\nu L}_{(npq\ell)}$ is coupled to one higher-$q$ mode, $\phi^{\mu\nu L}_{(n,p,q+1,\ell)}$. Following the same procedure as for $\ell\neq\ell_p$, one finds that for $\ell=\ell_p$, $\qmax$ is at most $\qSmax+1$. For $q=\qSmax+1$,
\begin{align}
\phi^{\mu\nu L_p}_{(n,p,\qSmax+1,\ell_p)} & = \frac{P^{\mu\nu L_p}_{(np\qSmax\ell_p)}}{(\qSmax+1)(2p+1)}.
\end{align}
Note that this raises the power of the highest-order logarithm to one higher than that in the source. If the source contains no logarithms, logarithms arise in the solution only when $P^{\mu\nu L_p}_{(np0\ell_p)}\neq0$. Next, for any $q$ in the range $1\leq q\leq\qSmax$,
\begin{align}
\phi^{\mu\nu L_p}_{(npq\ell_p)} & = \frac{P^{\mu\nu L_p}_{(n,p,q-1,\ell_p)}}{q(2p+1)}-\frac{q+1}{2p+1}\phi^{\mu\nu L_p}_{(n,p,q+1,\ell_p)}.
\end{align}
Last, for $q=0$,
\begin{align}
\phi^{\mu\nu L_p}_{(np0\ell_p)} &\quad \mbox{is undetermined}.
\end{align}

When the general solution is constructed in this way, each coefficient $\phi^{\mu\nu L}_{(npq\ell)}$ in it is either a homogeneous mode or directly proportional to a mode $P^{\mu\nu L}_{(npq\ell)}$ in the source $P^{\mu\nu}_{(n)}$. Since $P^{\mu\nu L}_{(npq\ell)}$ is constructed from the lower-order modes $\phi^{\mu\nu L'}_{(n'p'q'\ell')}$ with either $n'<n$ or $p'<p$, it follows that each $\phi^{\mu\nu L}_{(npq\ell)}$ is either a homogeneous mode or constructed from combinations of homogeneous modes (or constructed from a mode of $C^{\mu\nu}$). Therefore, the set of homogeneous modes $\phi^{\mu\nu L_p}_{(np0\ell_p)}$ comprise all the freedom in the general solution.
 
The homogeneous modes can be divided into two categories: those for $p<0$ and those for $p\geq0$. For $p\geq0$, they behave as $r^\ell$, and they (combined with the higher-order-in-$r$ terms constructed directly from them) correspond to free wave solutions to the homogeneous wave equation, not directly influenced by the presence of the body. If $C^{\mu\nu}$ vanished and the buffer region did not surround a body, these modes would fully characterize the perturbation.

For $p<0$, the homogeneous modes are the coefficients in a standard large-$r$ multipole expansion far from a source, as arise in electromagnetism and post-Minkowski theory~\cite{Damour-Blanchet:86, Damour-Iyer:91}. A $1/r$ homogeneous mode is a monopole; a $1/r^2$ homogeneous mode, a dipole; and so on. I therefore define the (unnormalized) multipole moments
\begin{equation}
I^{\mu\nu L}_{(n)}\equiv \phi^{\mu\nu L}_{(n,-\ell-1,0,\ell)},
\end{equation}
which are symmetric in $\mu$ and $\nu$ and STF in $i_1\ldots i_\ell$. In the solution \eqref{hn_form}, these moments appear in terms $I^{\mu\nu L}_{(n)}\nhat_L/r^{\ell+1}$. Each moment (except the monopoles), can be split into a mass and current moment,
\begin{equation}
I^{\mu\nu L}_{(n)} = M^{\mu\nu L}_{(n)}+S^{\mu\nu L}_{(n)}
\end{equation}
where the mass moment $M^{\mu\nu L}_{(n)}$ is the even-parity part of $I^{\mu\nu L}_{(n)}$, satisfying $M^{\mu j L}_{(n)}=M^{\mu(j i_1)i_2\cdots i_\ell}_{(n)}$, and the current moment $S^{\mu\nu L}_{(n)}$ is the odd-parity part, satisfying $S^{\mu jL}_{(n)}=S^{\mu[ji_1]i_2\cdots i_\ell}_{(n)}$. 

Since the most singular term in $h^{(n)}_{\mu\nu}$, $h^{(n,-n)}_{\mu\nu}/r^n$, is equal to the term $g^{(n)}_{{\rm I}\mu\nu}/r^n$ in the large-$r$ expansion of the body's unperturbed metric, the multipole moments $I^{\mu\nu N-1}_{(n)}$ that appear in $h^{(n,-n)}_{\mu\nu}$ are related to the multipole moments of the inner background spacetime (defined by, e.g., Geroch~\cite{Geroch:70} and Hansen~\cite{Hansen:74}).\footnote{The Geroch-Hansen moments are defined only for stationary spacetimes, while $g_{{\rm I}\mu\nu}$ may depend on time for a physically realistic system. However, the time dependence is slow compared to the body's lengthscales. So the definition may still be used.} Therefore, specifying the moments $I^{\mu\nu N-1}_{(n)}$ is equivalent to specifying those of the body. The remaining moments, $I^{\mu\nu L}_{(n)}$ for $\ell<N-1$, are gravitationally induced corrections to those of the unperturbed body. They will be discussed further in the following section.

This means that the homogeneous modes $\phi^{\mu\nu L_p}_{(np0\ell_p)}$ for $p<0$ are determined by the presence of the body. More precisely, the moments $I^{\mu\nu N-1}_{(n)}$ are fully determined by the matching condition, which plays the role of a boundary condition. The other moments, $I^{\mu\nu L}_{(n)}$ for $\ell<n-1$, will be found to involve these fully determined moments. Contrariwise, the homogeneous modes for $p\geq0$ are entirely undetermined by the matching condition; they are determined only once further boundary conditions are imposed.

We now have the following conclusions: I. In the buffer region, a solution of the form~\eqref{hn_form} to the $n$th-order wave equation~\eqref{nth_wave} can be obtained to any order in $\e$ and $r$. II. The solution may be constructed by a simple algorithm, solving order by order in $r$ and $\ln r$ in sequence (starting at the most singular order in each case). III. The highest power of $\ln r$ in the solution is increased when the source in the Poisson equation \eqref{nth_Poisson} contains a mode $P^{\mu\nu L_p}_{(npq\ell_p)}$, where $\ell_p$ is given by Eq.~\eqref{lp}. IV. Each term in the general solution is determined either by the source $C^{\mu\nu}_{(n)}-16\pi\tau^{\mu\nu}_{(n)}$, by the body's multipole moments or gravitational corrections to them, or by a homogeneous mode $\phi^{\mu\nu L_p}_{(np0\ell_p)}$, where $\ell_p=p>0$.

\subsection{Minimal solutions, singular/self-fields, and regular/effective fields}\label{nth_singular-regular}
Given that the homogeneous modes $\phi^{\mu\nu L_p}_{(n,p,0,\ell_p)}$ contain all the freedom in the solution to the wave equation, the solution can be naturally split into several pieces.

Consider a wave equation $E^{\mu\nu}{}_{\rho\sigma}\phi^{\rho\sigma}_{(n)}=F^{\mu\nu}$, with some source $F^{\mu\nu}$. I define the \emph{minimal inhomogeneous solution} $\phi^{{\rm IH}\rho\sigma}_{(n)}$ to this equation to be the particular solution in which all homogeneous modes vanish, $\phi^{\mu\nu L_p}_{(n,p,0,\ell_p)}=0$; it contains the minimal number of homogeneous modes of any particular solution.

Now take some other particular solution $\phi^{{\rm P}\rho\sigma}_{(n)}$ to the same equation. It differs from $\phi^{{\rm IH}\rho\sigma}_{(n)}$ in that it contains nonvanishing homogeneous modes, as determined by some choice of boundary conditions. Consider the homogeneous equation $E^{\mu\nu}{}_{\rho\sigma}\phi^{\rho\sigma}_{(n)}=0$. I define the \emph{minimal $p$th solution} ${}_{(p)}{\phi^{\mu\nu}_{(n)}}$ to be the solution containing $\phi^{\mu\nu L_p}_{(n,p,0,\ell_p)}$ as its only nonvanishing homogeneous mode, with $\phi^{\mu\nu L_{p'}}_{(n,p',0,\ell_{p'})}=0$ for all $p'\neq p$. We then have 
\begin{equation}
\phi^{{\rm P}\rho\sigma}_{(n)}=\phi^{{\rm IH}\rho\sigma}_{(n)}+\sum_{p}{}_{(p)}\phi^{\mu\nu}_{(n)},
\end{equation}
where in each ${}_{(p)}{\phi^{\mu\nu}_{(n)}}$, $\phi^{\mu\nu L_p}_{(n,p,0,\ell_p)}$ takes the value it has in $\phi^{{\rm P}\rho\sigma}_{(n)}$.
In the simple case where no logarithms arise,\footnote{I have proved by brute-force analysis of $E^{\mu\nu}{}_{\rho\sigma}$ that any minimal $p$th solution contains no logarithms if $p\geq-1$, though I omit the proof due to its tediousness. I hypothesize that the same holds true if $p<-1$, but I have not proved that to be the case.} one has ${}_{(p)}\phi^{\mu\nu}_{(n)}(t,r,\vec n)=\sum_{p'\geq p}r^{p'}{}_{(p)}\phi^{\mu\nu L}_{(n,p')}(t,\vec n)$, where the lowest-order term is the homogeneous mode itself, 
\begin{align}
{}_{(p)}\phi^{\mu\nu}_{(n,p)} &= \phi^{\mu\nu L_p}_{(n,p,0,\ell_p)}\nhat_{L_p},
\end{align}
and the higher-order terms are
\begin{align}
r^{p'}&{}_{(p)}\phi^{\mu\nu}_{(n,p')} \nonumber\\
	& = -\sum_\ell\sum_{p''=p}^{p'-1}
					\frac{\left[{}^{(p'-p'')\!}\Delta E^{\mu\nu}{}_{\rho\sigma}r^{p''}{}_{(p)}\phi^{\rho\sigma}_{(n,p'')}\right]^L\nhat_{L}}{p(p+1)-\ell(\ell+1)},
\end{align}
where $[f]^L$ is given by Eq.~\eqref{lth_coeff}.

With these preliminary definitions, I now define an $n$th-order singular/self-field $\phi^{{\rm S}\mu\nu}_{(n)}$ and regular/effective field $\phi^{{\rm R}\mu\nu}_{(n)}$, along with their sums
\begin{align}
\phi^{{\rm S}\mu\nu} &\equiv \sum_n\e^n\phi^{{\rm S}\mu\nu}_{(n)},\\
\phi^{{\rm R}\mu\nu} &\equiv \sum_n\e^n\phi^{{\rm R}\mu\nu}_{(n)},
\end{align}
as certain pieces of the physical solution to Eq.~\eqref{nth_wave}. By `physical solution' I simply mean \eqref{hn_form} with all the coefficients determined. In other words, given a metric in the buffer region around the body, extract the field $\phi^{\mu\nu}$ from it and decompose that field into the form \eqref{hn_form}; I define $\phi^{{\rm S}\mu\nu}$ and $\phi^{{\rm R}\mu\nu}$ as certain pieces of the result.

Every term in the $n$th-order solution to Eq.~\eqref{nth_wave} is a combination of homogeneous modes (and modes of $C^{\mu\nu}_{(n)}$). I wish the regular field to contain everything in the solution (constructed in the manner of Sec.~\ref{nth_buffer_solution}) not directly involving any of the body's multipole moments. At first order, where $C^{\mu\nu}_{(1)}$ is the only source in Eq.~\eqref{nth_wave}, this leads to the simple definition that $\phi^{{\rm R}\mu\nu}_{(1)}$ is the sum of minimal $p$th solutions for $p\geq0$, with the homogeneous modes therein taking the same values as in the physical field $\phi^{\mu\nu}_{(1)}$. Note that this $\phi^{{\rm R}\mu\nu}_{(1)}$ is a smooth field at $r=0$. At the first two orders in $r$, it reads $\phi^{{\rm R}\mu\nu}_{(1)}=\phi^{\mu\nu}_{(1,0,0,0)}+r\phi^{\mu\nu i}_{(1,1,0,1)}n_i+O(r^2)$. Beyond first order in $\e$, one could define the regular field in the same way, as the sum of $n$th-order minimal $p$th solutions for $p\geq0$. Again, this would be a smooth field at $r=0$. However, in the spirit of Detweiler and Whiting~\cite{Detweiler-Whiting:02}, I wish the total regular field $\phi^{{\rm R}\mu\nu}$ to be a solution to the homogeneous EFE. To that end, I define $\phi^{{\rm R}\mu\nu}_{(n)}$ to be the sum of (i) all $n$th-order minimal $p$th solutions with $p\geq0$ (again, with the homogeneous modes taking the same value as in the full field $\phi^{\mu\nu}_{(n)}$) and (ii) the minimal inhomogeneous solution to 
\begin{equation}
E^{\mu\nu}{}_{\rho\sigma}\phi^{{\rm R}\rho\sigma}_{(n)} = -16\pi\tau^{\mu\nu}_{(n)}[\phi^{\rm R}],
\end{equation}
where $\tau^{\mu\nu}_{(n)}[\phi]$ is as defined in Sec.~\ref{wave_gauges}. It follows that $\phi^{{\rm R}\mu\nu}$ satisfies
\begin{equation}
E^{\mu\nu}{}_{\rho\sigma}\phi^{{\rm R}\rho\sigma} = -16\pi\tau^{\mu\nu}[\phi^{\rm R}].\label{nth_regular_wave}
\end{equation}
Since every function involved is smooth at $r=0$, $\phi^{{\rm R}\mu\nu}$ is a smooth solution to the homogeneous wave equation not only in the buffer region but also in the region $\mathcal{B}\subset\mathcal{M}_{\rm E}$ that is surrounded by the buffer. Note, however, that my definition of $\phi^{{\rm R}\mu\nu}$ is local, in terms of a certain series solution in a certain coordinate system.

I next define the $n$th-order singular/self-field to comprise everything else in $\phi^{\mu\nu}_{(n)}$; this means that in the solution to the $n$th-order wave equation constructed in the manner of Sec~\ref{nth_buffer_solution}, the singular field contains every term directly involving at least one of the body's multipole moments. More precisely, $\phi^{{\rm S}\mu\nu}_{(n)}$ consists of the sum of all $n$th-order minimal $p$th solutions with $p<0$, together with the minimal inhomogeneous solution to 
\begin{equation}
E^{\mu\nu}{}_{\rho\sigma}\phi^{{\rm S}\rho\sigma}_{(n)} = C^{\mu\nu}_{(n)} -16\pi(\tau^{\mu\nu}_{(n)}-\tau^{\mu\nu}_{(n)}[\phi^{\rm R}]).\label{nth_singular_wave}
\end{equation}
Again, the homogeneous modes defining the minimal $p$th solutions are given the values they have in the full field $\phi^{\mu\nu}_{(n)}$. From that full solution, the singular/self-field contains all the terms that diverge at $r=0$ (along with the higher-order terms that follow from its definition). At first order, where $\tau^{\mu\nu}_{(1)}=0=\tau^{\mu\nu}_{(1)}[\phi^{\rm R}]$, $\phi^{{\rm S}\rho\sigma}_{(1)}$ is a solution to the governing wave equation~\eqref{nth_wave} in $\Omega$. At higher orders, it ceases to be, and it begins to directly involve the lower-order regular fields. 

With these definitions, I have split the $n$th-order field in the buffer region into a singular and a regular piece,
\begin{align}
\phi^{\mu\nu}_{(n)}=\phi^{{\rm S}\mu\nu}_{(n)}+\phi^{{\rm R}\mu\nu}_{(n)},
\end{align}
and, likewise, the full solution has been split as
\begin{align}
\phi^{\mu\nu}=\phi^{{\rm S}\mu\nu}+\phi^{{\rm R}\mu\nu}.
\end{align}
As desired, $\phi^{{\rm R}\mu\nu}$ shares the properties of the Detweiler-Whiting regular field. It is a smooth solution to the homogeneous wave equation even at $r=0$. As a homogeneous solution, it propagates independently of the body and is determined by global boundary conditions that specify, for example, the incoming wave content in the full solution. The singular/self-field is the rest of the full field; that is, it is the part that does not propagate independently, instead being bound to the body.

All of this puts aside enforcing the gauge condition. Ideally, one can impose separate conditions on the singular and regular fields:
\begin{align}
\nabla_{\!\nu}\phi^{{\rm S}\mu\nu} & = Z^\mu{}_{\rho\sigma}\phi^{{\rm S}\rho\sigma}+D^\mu,\label{singular_gauge}\\
\nabla_{\!\nu}\phi^{{\rm R}\mu\nu} & = Z^\mu{}_{\rho\sigma}\phi^{{\rm R}\rho\sigma}\label{regular_gauge}.
\end{align} 
This ensures the total field $\phi^{\mu\nu}$ satisfies $\nabla_{\!\nu}\phi^{\mu\nu} = Z^\mu{}_{\rho\sigma}\phi^{\rho\sigma}+D^\mu$, but it puts all the dependence on $D^\mu$ into the singular field, in accordance with how $C^{\mu\nu}$ appears only in the wave equation for the singular field. In Eqs.~\eqref{nth_regular_wave} and \eqref{nth_singular_wave}, $\tau^{\mu\nu}_{(n)}[\phi^{\rm R}]$ is written with the gauge condition on $\phi^{{\rm R}\mu\nu}$ already imposed, which ensures that $\phi^{{\rm R}\mu\nu}$ is a solution to the full vacuum EFE, not merely its relaxed version.

However, there is a potential difficulty in imposing gauge conditions separately on the singular and regular fields: since the singular field contains none of the minimal $p$th solutions with $p\geq0$, it may not have sufficient freedom to satisfy its gauge condition. If this issue arises, one may abandon the goal of imposing the separate gauge conditions \eqref{singular_gauge} and \eqref{regular_gauge} on each of the fields, thereby abandoning the goal of having $\phi^{{\rm R}\mu\nu}$ satisfy the full vacuum EFE. Alternatively, one may incorporate some particular minimal $p$th solution (or however many are necessary) into $\phi^{{\rm S}\mu\nu}$ that ensures it satisfies its gauge condition. $\phi^{{\rm R}\mu\nu}_{(n)}$ will then contain the minimal $p$th solutions with $\phi^{\mu\nu L_p}_{(n,p,0,p)}$ replaced by $\phi^{\mu\nu L_p}_{(n,p,0,p)}-\phi^{{\rm S}\mu\nu L_p}_{(n,p,0,p)}$, where $\phi^{{\rm S}\mu\nu L_p}_{(n,p,0,p)}$ is the homogeneous mode introduced into $\phi^{{\rm S}\mu\nu}_{(n)}$ in order to satisfy the gauge condition. $\phi^{{\rm S}\mu\nu L_p}_{(n,p,0,p)}$ will not be uniquely specified in that case, since, among other reasons, one could always add to it another minimal $p$th solution of the same $p$ but contributing nothing to Eq.~\eqref{singular_gauge}. Therefore, this will make the split into singular and regular fields non-unique. But any convenient choice may be made, and $\phi^{{\rm R}\mu\nu}$ will remain a smooth solution to the vacuum EFE in $\B$ and in the buffer region surrounding it. In any case, this issue does not arise at the orders explicitly considered in Sec.~\ref{explicit}.

\subsection{Gauge condition, corrected multipole moments, and equation of motion}\label{nth_motion}
This consideration of the gauge condition leads well into a more detailed discussion of satisfying that condition. As shown above, the homogeneous modes contain all the freedom in the solution of the form \eqref{hn_form}. And from the perspective of the wave equation \eqref{nth_wave}, these homogenous modes can be arbitrary (smooth) functions of time. However, they \emph{are} constrained by the gauge condition.

We can analyze this condition in a manner similar to the analysis of the wave equation. First, note that 
\begin{align}
\nabla_{\!\nu}\phi^{\mu\nu}_{(n)} = \partial_b\phi^{\mu b}_{(n)}+\partial_t\phi^{\mu t}_{(n)}
									+\left(\Gamma^\mu_{\nu\rho}+\Gamma^{\sigma}_{\sigma\nu}g^{\mu}_{\rho}\right)\phi^{\nu\rho}_{(n)}.
\end{align}
After substituting the ansatz \eqref{hn_form} and making use of the identities \eqref{nnhat} and \eqref{PDnhat}, the first term becomes
\begin{align}
\partial_b\phi^{\mu b}_{(n)} &=\sum_{\substack{p\geq-n\\q,\ell}}r^{p-1}(\ln r)^q
								\bigg[(1-\delta_{\ell0})(p-\ell+1)\phi^{\mu\langle L\rangle}_{(n,p,q,\ell-1)}\nonumber\\
							&\quad	+(1-\delta_{\ell0})(q+1)\phi^{\mu\langle L\rangle}_{(n,p,q+1,\ell-1)}\nonumber\\
							&\quad	+\frac{(\ell+1)(\ell+p+2)}{2\ell+3}\delta_{ij}\phi^{\mu ijL}_{(n,p,q,\ell+1)}\nonumber\\
							&\quad	+\frac{(\ell+1)(q+1)}{2\ell+3}\delta_{ij}\phi^{\mu ijL}_{(n,p,q+1,\ell+1)}\bigg]\nhat_L.
\end{align}
Since both $Z^\mu{}_{\nu\rho}$ and the Christoffel symbols of the background are smooth, they can be expanded in Taylor series $Z^\mu{}_{\nu\rho}=\sum_{s\geq0}\frac{r^sn^S}{s!}Z^\mu{}_{\nu\rho,S}\big|_{r=0}$ and $\Gamma^\mu_{\nu\rho}=\sum_{s\geq0}\frac{r^sn^S}{s!}\Gamma^\mu_{\nu\rho,S}\big|_{r=0}$. Putting these results together, one finds that the $n$th-order gauge condition \eqref{nth_gauge} becomes
\begin{align}
0 &= \sum_{s=0}^{n-1}\bar\delta^s\Bigg\lbrace (1-\delta_{\ell0})(p-\ell+1)\phi^{\mu\langle L\rangle}_{(n-s,p,q,\ell-1)}\nonumber\\
							&\quad +(1-\delta_{\ell0})(q+1)\phi^{\mu\langle L\rangle}_{(n-s,p,q+1,\ell-1)}\nonumber\\
							&\quad +\frac{(\ell+1)(\ell+p+2)}{2\ell+3}\delta_{ij}\phi^{\mu ijL}_{(n-s,p,q,\ell+1)}\nonumber\\
							&\quad +\frac{(\ell+1)(q+1)}{2\ell+3}\delta_{ij}\phi^{\mu ijL}_{(n-s,p,q+1,\ell+1)}\nonumber\\
							&\quad +\sum^{p-1}_{p'=-n}\sum_{\ell'}
							\frac{\left(\Gamma^\mu_{\nu\rho}+\Gamma^{\sigma}_{\sigma\nu}g^{\mu}_{\rho}-Z^\mu{}_{\nu\rho}\right)_{,P-P'-1}}{(p-p'-1)!}\nonumber\\
							&\quad\times\left[n_{L'}{}^{P-P'-1}\phi^{\nu\rho L'}_{(n-s,p',q,\ell')}\right]^L\nonumber\\
							&\quad +\partial_t\phi^{\mu tL}_{(n-s,p-1,q,\ell)}-D^{\mu L}_{(n-s,p,q,\ell)}\Bigg\rbrace,\label{nth_gauge_expanded}
\end{align}
where $[f]^L$ indicates the coefficient of $\nhat_L$ in the expansion $f=\sum [f]^L\nhat_L$, given by Eq.~\eqref{lth_coeff}. Here we have an equation for $\phi^{\mu\nu L}_{(n,p,q,\ell)}$ and $\phi^{\mu\nu L+2}_{(n,p,q,\ell+2)}$ in terms of $\phi^{\mu\nu L}_{(n,p,q+1,\ell)}$, $\phi^{\mu\nu L+2}_{(n,p,q+1,\ell+2)}$, and terms of lower $n$ and/or $p$.

Equation \eqref{nth_gauge_expanded} is solved in the same manner as was the Poisson equation \eqref{nth_Poisson_expanded}, proceeding order by order in $r$, beginning with the most singular terms, $\phi^{\mu\nu L}_{(n,-n,q,\ell)}$. Since the gauge condition couples modes of degree $\ell$ to those of degree $\ell\pm2$, doing so is not quite as straightforward as in the case of the wave equation. However, it may still be readily accomplished.

Because the homogeneous modes contain all the freedom in the solution, I am interested only in how Eq.~\eqref{nth_gauge_expanded} constrains those modes. It could be the case that the homogeneous modes provide insufficient freedom to satisfy the gauge condition. However, one may always impose any desired generalized wave gauge, since the transformation from any other gauge can be constructed by solving a sequence of wave equations for the gauge vectors.\footnote{In Ref.~\cite{Pound:10b}, I showed that the Lorenz gauge condition could be imposed at all orders by solving a sequence of wave equations. There, I showed that if the condition could be imposed in the buffer region, then it could be imposed everywhere in the external universe outside the buffer region. It follows from the results of Sec.~\ref{nth_buffer_solution} that the condition can be imposed to arbitrarily high order in the buffer region by solving the relevant wave equations, which completes the proof.} Since Eq.~\eqref{hn_form} is a very general form, it is very unlikely that it would be unable to satisfy the gauge condition, most of which is satisfied automatically by virtue of the Bianchi identity.

So let us examine how Eq.~\eqref{nth_gauge_expanded} constrains a given homogeneous mode. The equation yields at least three constraints on each homogeneous mode: For $q=0$ and $\ell-1=\ell_p$, the first term in Eq.~\eqref{nth_gauge_expanded} involves $\phi^{\mu\langle iL_p\rangle}_{(n,p,0,\ell_p)}$, a certain part of the homogeneous mode $\phi^{\mu\nu L_p}_{(n,p,0,\ell_p)}$. The equation is then one for this quantity in terms of lower-order modes and three modes of the same $n$ and $p$, $\phi^{\mu\nu L_p+2}_{(n,p,0,\ell_p+2)}$, $\phi^{\mu\nu L_p}_{(n,p,1,\ell_p)}$, and $\phi^{\mu\nu L_p+2}_{(n,p,1,\ell_p+2)}$. For $q=0$ and $\ell+1=\ell_p$, the third term is a different piece of the same homogeneous mode, $\phi^{\mu\nu L_p}_{(n,p,0,\ell_p)}$, and we arrive at an equation for that quantity. For $q=0$ and $\ell=\ell_{p-1}$, we have an equation for the time derivative of the homogeneous mode. Therefore, the tensorial form of the modes will be constrained by the first two equations, and its time dependence by the third.

The constraints on the homogeneous modes appearing in the regular/effective field are not of particular interest, since the regular field will be calculated numerically from a puncture scheme. The constraints on the homogeneous modes appearing in the singular/self-field, however, are essential, because the singular field will be used to define the puncture, the form of which must be determined analytically. Furthermore, the homogeneous modes in the singular field are the multipole moments $I^{\mu\nu L}_{(n)}$. They contain information about the body and its interaction with the geometry, and in the case of a material body, the constraints on them should correspond to the conservation of the body's stress-energy tensor.

The multipole moments come in two types: the body's unperturbed moments, which behave as if the body were isolated, and corrections to them. As mentioned in Sec.~\ref{nth_buffer_solution}, the moment occurring in the most divergent piece of $\phi^{\mu\nu}_{(n)}$ at a given $n$ is an unperturbed moment, corresponding to the $\ell=n-1$th moment of the inner background spacetime $g_{{\rm I}\mu\nu}$. This moment may be freely specified modulo the constraints placed on it by the gauge condition. For example, in the case $Z^\mu{}_{\nu\rho}=D^\mu=0$, Eq.~\eqref{nth_gauge_expanded} with $q=0$, $p=-1$, and $\ell=1$ constrains the first-order monopole moment to have the form~\cite{Pound:10a} 
\begin{equation}
I^{\mu\nu}_{(1)}=4mu^\mu u^\nu\label{mass_monopole},
\end{equation}
where the matching condition implies $m$ is the ADM mass of $g_{{\rm I}\mu\nu}$; and Eq.~\eqref{nth_gauge_expanded} with $q=0$, $p=0$, and $\ell=0$ constrains $m$ to be constant, but its value is freely specifiable.

For $\ell<n-1$, $I^{\mu\nu L}_{(n)}$ is a correction to the body's $\ell$th moment. From the perspective of the wave equation, this could be freely set to zero, effectively incorporating the correction into the body's unperturbed moment. Therefore, the finite corrections come entirely from the gauge condition. A consequence of this is that the corrections, as obtained entirely within the buffer region, without specifying global boundary conditions, will always directly involve the unperturbed moments, as mentioned in Sec.~\ref{nth_buffer_solution}; terms in the gauge condition involving only the homogeneous modes with $p\geq0$ cannot have negative powers of $r$, and therefore cannot constrain the modes with $p<0$. The corrections to the multipole moments may arise from the nonlinear effect of couplings between multipole moments or between a multipole moment and the regular field. They may also arise as the linear effect of coupling between a multipole moment and the external curvature. However, these corrections will not be uniquely determined, because one can always add a minimal $p$th solution satisfying the homogeneous equation $\bar\delta^0(\nabla_{\!\nu}\phi^{\mu\nu}_{(n)}-Z^\mu{}_{\nu\rho}\phi^{\nu\rho}_{(n)})=0$. That is, one can always add a field indistinguishable from that due to the body's unperturbed moments. Such additional terms may be freely specified.

In addition to these constraints on the multipole moments, the gauge condition also determines an equation of motion for the body, in the sense of providing a relationship between the acceleration of $\gamma$ and the field in the buffer region. For simplicity, consider the case $Z^\mu{}_{\nu\rho}=D^\mu=0$. Also, take the background metric to be $g_{\mu\nu}=\eta_{\mu\nu}-2a_ix^iu_\mu u_\nu+O(r^2)$, which is the form it takes in Fermi-Walker coordinates. Now, with these simplifications, consider the contribution of $\phi^{\mu\nu}_{(1)}$ to the gauge condition. Ignoring the $\bar\delta$, the $\ell=0$, $p=-1$ piece of Eq.~\eqref{nth_gauge_expanded} yields 
\begin{align}
\frac{1}{4\pi}\int\nabla_{\!\nu}\phi^{\mu\nu}_{(1)}d\Omega &= 4\partial_t mu^\alpha + 4ma^\alpha\\
															&= 4u^\beta\nabla_{\!\beta}(mu^\alpha),\label{ma}
\end{align}
where the integral over a sphere picks off the $\ell=0$ piece, and I have made use of Eq.~\eqref{mass_monopole}. Equation~\eqref{ma} tells us that to the $p=-1$, $\ell=0$ part of the gauge condition, the first-order field $\phi^{\mu\nu}_{(1)}$ contributes what looks like the left-hand side of an equation of motion. The right-hand side of that equation of motion will be contributed by higher-order fields. Write the $n$th-order gauge condition \eqref{nth_gauge} as 
\begin{equation}
\bar\delta^n\nabla_{\!\nu}\phi^{\mu\nu}_{(1)}=-\sum_{n'=1}^n\bar\delta^{(n-n')}\nabla_{\!\nu}\phi^{\mu\nu}_{(n'+1)}.
\end{equation}
Picking out the $p=-1$ term using Eq.~\eqref{nth_gauge_expanded}, integrating over a sphere to pick off the $\ell=0$ piece, and making use of Eqs.~\eqref{ma} and \eqref{identity}, one finds
\begin{align}
4ma^\mu_{(n)} &= -\sum_{n'=1}^n\bar\delta^{n-n'}\Bigg\lbrace \frac{1}{3}\delta_{ij}\big(2\phi^{\mu ij}_{(n'+1,0,0,1)}+\phi^{\mu ij}_{(n'+1,0,1,1)}\big)\nonumber\\
			&\quad +\partial_t I^{\mu t}_{(n'+1)}+a^\mu I^{tt}_{(n'+1)}+a_iI^{\mu i}_{(n'+1)}\nonumber\\
			&\quad +\sum_{p'=-(n'+1)}^{-2}\sum_{p''=0}^{\lfloor -\frac{p'+1}{2}\rfloor}\frac{(-p'-1)!}{(2p'')!(-2p'-2p''-1)!!}\nonumber\\
			&\quad \times\partial_{i_1}{}^{i_1}{}_{\ldots i_{p''}}{}^{i_{p''}}{}_{(-P'-1-2P'')}
					\left(\Gamma^\mu_{\nu\rho}+\Gamma^{\sigma}_{\sigma\nu}g^{\mu}_{\rho}\right)\nonumber\\
			&\quad \times \phi^{\nu\rho(-P'-1-2P'')}_{(n'+1,p',0,-p'-1-2p'')}\Bigg\rbrace,\label{nth_acceleration}
\end{align}
where I have used the first-order gauge condition to set $\partial_t m=0$. The full equation of motion then reads $ma^\mu=m(a^\mu_{(0)}+\e a^\mu_{(1)}+\e^2 a^\mu_{(2)}+\ldots)$.

To turn this relationship into a useful equation of motion, one must actually solve the wave equation and gauge condition to sufficiently high order to determine the content of the coefficients $\phi^{\nu\rho L}_{(n,p,q,\ell)}$ appearing in Eq.~\eqref{nth_acceleration}. Furthermore, one must impose some condition on the worldline in order to ensure that it accurately represents the body's motion. This will take the form of a condition on the mass dipole moments $M^{\mu\nu i}_{(n)}$.

The first mass dipole moment that appears in $\phi^{\mu\nu}$, $M^{\mu\nu i}_{(2)}$, corresponds to the mass dipole of the inner background spacetime. It is a measure of the `position' of the unperturbed mass $m$ relative to the origin $r=0$; a translation $x^a\to x^a+\delta x^a(t)$ applied to the $m/r$ monopole induces a $M^{\mu\nu i}n_i/r^2$ dipole. (Less intuitively, a transformation $x^a\to x^a+\delta x^b(t)\nhat^a{}_b$ also induces a mass dipole.) Equivalently, the mass dipole may be shifted by adopting a slightly different worldline about which to center the coordinates.  Hence, this mass dipole is set to zero to ensure that $\gamma$ represents the body's bulk motion. Analogously, the corrections to the mass dipole, $M^{\mu\nu i}_{(n)}$ for $n>2$, are measures of the `position' of the \emph{corrections} to the mass $m$. In other words, in the local coordinates $(t,x^a)$ centered on $\gamma$, the mass dipole terms are a measure of the position of the center of energy, including `gravitational energy', relative to $\gamma$. The lowest-order mass dipole seems to be on firmer ground as a measure of center of mass than are its corrections as measures of center of energy, since it can be related to a well-defined notion of the mass dipole of the inner background spacetime. But for an observer sitting in the buffer region, the corrections to the mass dipole are indistinguishable from the unperturbed mass dipole, since they take the identical form in the metric. More strictly speaking, a certain piece of the corrections is indistinguishable from the unperturbed mass dipole: the piece satisfying the same constraint imposed by the gauge condition, $\partial_j(M^{\mu j i}_{(n)}n_i/r^2)=0$. A good center-of-mass condition might be to set this piece to zero at all orders. One can always do so, since one can always add a term to $M^{\mu\nu i}_{(n)}$ that satisfies the same constraint as the unperturbed mass dipole and has arbitrary time dependence. 

Further investigation will be necessary to verify that this is a good condition to impose. It would not set the whole of the corrected mass dipole moment to zero, and one should examine the physical content of the nonzero portion. To that end, one should examine the physical content of the corrections to the monopole moment, $I^{\mu\nu}_{(n)}$, which are symmetric tensors of rank two rather than scalars as the mass is, and one should verify how the corrections to the mass dipole are related to a small translation applied $I^{\mu\nu}_{(n)}/r$. Alternative conditions to impose will certainly be possible, and at some order, there may not be a `best' choice: any meaningful measure of the `center of energy' will likely cease to coincide with the `center of mass', in which case it would not be possible to set both to zero simultaneously.

\subsection{Effective stress-energy tensor}\label{nth_stress-energy}
Thus far, everything has, for the most part, been confined to the buffer region. In the region surrounded by the buffer, where $r\sim m$, the equations are not valid for multiple reasons: the expansions used are not an accurate approximation, since the body's field dominates over the external background in that region; the region $\B$ in the external background spacetime may not be diffeomorphic to the region surrounded by the buffer in the full spacetime; and unless the body is a black hole, the stress-energy tensor of the body will be finite somewhere in the region. However, we may ask the question `\emph{if} the solutions are taken to be valid for all $r>0$, what stress-energy sources them?' This may or may not correspond to a small-$\e$ expansion of the body's actual stress-energy tensor, but it will take a physically meaningful form. It will also be relevant in Sec.~\ref{nth_puncture} when defining an effective-source equation that may be solved numerically for the regular field.

Now, by construction, the minimal inhomogeneous solution in the solution to Eq.~\eqref{nth_wave} is sourced by $C^{\mu\nu}_{(n)}-16\pi\tau^{\mu\nu}_{(n)}$, and the minimal $p$th solutions for $p\geq0$ are homogeneous solutions even at $r=0$. This leaves only the minimal $p$th solutions for $p<0$---that is, the parts of the singular field that form homogeneous solutions for $r>0$. For each such ${}_{(p)}\phi^{\rho\sigma}_{(n)}$, I follow the approach taken by Gralla and Wald at first order~\cite{Gralla-Wald:08}, defining the distributional effective stress-energy tensor 
\begin{equation}
{}_{(p)}T^{\mu\nu}_{(n)} \equiv -\frac{1}{16\pi}E^{\mu\nu}{}_{\rho\sigma}{}_{(p)}\phi^{\rho\sigma}_{(n)}.\label{Tp}
\end{equation}
The right-hand side can be split as 
\begin{align}\label{split}
E^{\mu\nu}{}_{\rho\sigma}{}_{(p)}\phi^{\rho\sigma}_{(n)} = \partial^i\partial_i \frac{I^{\mu\nu L_p}_{(n)}\nhat_{L_p}}{r^p}
															+\Delta\left(E^{\mu\nu}{}_{\rho\sigma}{}_{(p)}\phi^{\rho\sigma}_{(n)}\right),
\end{align}
where the first term is the most singular and $\Delta\left(E^{\mu\nu}{}_{\rho\sigma}{}_{(p)}\phi^{\rho\sigma}_{(n)}\right)$ carries all the less singular terms. 

First examine the most singular term. Using the identities $\partial_L r^{-1}=(-1)^\ell(2\ell-1)!!\frac{\nhat_L}{r^{\ell+1}}$ and $\partial^i\partial_i r^{-1}=-4\pi\delta^3(\vec x)$, we have 
\begin{align}
\partial^i\partial_i\frac{I^{\mu\nu L}_{(n)}\nhat_L}{r^{\ell+1}} &= \frac{4\pi(-1)^{\ell+1}I^{\mu\nu L}_{(n)}}{(2\ell-1)!!}\partial_L\delta^3(\vec x).\label{ddI}
\end{align}
One can find the covariant form of the right-hand side by integrating against a test function $\psi_{\mu\nu}$. Doing so, we have
\begin{align}
\int \psi_{\mu\nu}&I^{\mu\nu L}_{(n)}\partial_L\delta^3(\vec x)\sqrt{-g}d^3xdt \nonumber\\
				&= (-1)^\ell\int_\gamma\partial_L\left(\sqrt{-g}\psi_{\mu\nu} I^{\mu\nu L}_{(n)}\right)dt\\
				&= (-1)^\ell\int_\gamma\left(\psi_{\mu\nu} I^{\mu\nu\alpha_1\cdots\alpha_\ell}_{(n)}\right)_{;\alpha_1\cdots\alpha_\ell}dt,\label{delta}
\end{align}
where $g={\rm det}(g_{\mu\nu})$, and in going from the second line to the third I have utilized the identity $\partial_i\sqrt{-g}=\Gamma^\beta_{\beta i}\sqrt{-g}$. Here I have defined $I^{\mu\nu \alpha_1\cdots\alpha_\ell}_{(n)}$ to be the tensor that agrees with $I^{\mu\nu L}_{(n)}$ if all $\alpha_i$ are spatial indices and zero otherwise, meaning that $I^{\mu\nu \alpha_1\cdots\alpha_\ell}_{(n)}$ is STF with respect to $g_{\mu\nu}$ and that $I^{\mu\nu \alpha_1\cdots\alpha_i\cdots\alpha_\ell}_{(n)} u_{\alpha_i}=0$ for all $1\leq i\leq\ell$. Equation~\eqref{delta} shows that
\begin{equation}
I^{\mu\nu L}_{(n)}\partial_L\delta^3(\vec x) = \int_\gamma I^{\mu\nu\alpha_1\cdots\alpha_\ell}_{(n)}\delta(x,z)_{;\alpha_1\cdots\alpha_\ell}d\tau,
\end{equation}
where
\begin{equation}
\delta(x,z)=\frac{\delta^4\left(x^\mu-z^\mu(\tau)\right)}{\sqrt{-g(x)}}=\frac{\delta^4\left(x^\mu-z^\mu(\tau)\right)}{\sqrt{-g(z)}}
\end{equation}
is the covariant Dirac delta function, and I have made use of the fact that $t=\tau$ on the worldline.

Equation~\eqref{split} now reads
\begin{align}
E^{\mu\nu}{}_{\rho\sigma}{}_{(p)}\phi^{\rho\sigma}_{(n)}	&= 
		\frac{4\pi(-1)^{\ell+1}}{(2\ell-1)!!}\int_\gamma I^{\mu\nu \alpha_1\cdots\alpha_\ell}_{(n)}\delta(x,z)_{;\alpha_1\cdots\alpha_\ell}d\tau\nonumber\\
 		&\quad -\Delta\left(E^{\mu\nu}{}_{\rho\sigma}{}_{(p)}\phi^{\rho\sigma}_{(n)}\right).
\end{align}
Now note that, by construction, $\Delta\left(E^{\mu\nu}{}_{\rho\sigma}{}_{(p)}\phi^{\rho\sigma}_{(n)}\right)$ vanishes pointwise for $r>0$. If it is nonvanishing as a distribution, it must have support only on $\gamma$, in which case it must be proportional to $\delta^3(x)$ or a derivative thereof; but from the calculation just performed, that would lead to a homogeneous function in the solution. By definition, no such functions do appear in the minimal $p$th solution. So $\Delta\left(E^{\mu\nu}{}_{\rho\sigma}{}_{(p)}\phi^{\rho\sigma}_{(n)}\right)$ vanishes as a distribution. This should also be intuitively obvious, since each term in the minimal $p$th solution is constructed directly from its most singular term, $I^{\mu\nu L}_{(n)}\nhat_L/r^{\ell+1}$; only that most singular term needs a source.

Therefore, Eq.~\eqref{Tp} becomes 
\begin{equation}
{}_{(p)}T^{\mu\nu}_{(n)}=\frac{(-1)^\ell}{4(2\ell-1)!!}\int_\gamma I^{\mu\nu \alpha_1\cdots\alpha_\ell}_{(n)}\delta(x,z)_{;\alpha_1\cdots\alpha_\ell}d\tau.
\end{equation}
Since $\phi^{\mu\nu}_{(n)}$ contains minimal $p$th solutions for $p=-n,\ldots,-1$, the total $n$th-order stress-energy tensor is given by
\begin{equation}
T^{\mu\nu}_{(n)}=\sum_{\ell=0}^{n-1}\frac{(-1)^\ell}{4(2\ell-1)!!}\int_\gamma I^{\mu\nu \alpha_1\cdots\alpha_\ell}_{(n)}(\tau)\delta(x,z)_{;\alpha_1\cdots\alpha_\ell}d\tau.\label{nth_stress-energy-tensor}
\end{equation}
This includes both the moments of the unperturbed body and the corrections to them due to `gravitational energy'. If the corrections are added to the moments of the unperturbed body, then one has the total stress-energy tensor
\begin{align}
T^{\mu\nu} &= \sum_n T^{\mu\nu}_{(n)} = \sum_\ell\int_\gamma I^{\mu\nu\alpha_1\cdots\alpha_\ell}\delta(x,z)_{;\alpha_1\cdots\alpha_\ell}d\tau,\label{total_T}
\end{align}
where I have defined the normalized corrected moments
\begin{equation}
I^{\mu\nu \alpha_1\cdots\alpha_\ell} \equiv \sum_n\frac{(-1)^\ell}{4(2\ell-1)!!}I^{\mu\nu \alpha_1\cdots\alpha_\ell}_{(n)}.
\end{equation}
It is noteworthy that Eq.~\eqref{total_T} agrees with the traditional form for the multipolar expansion of a material body's stress-energy tensor~\cite{Mathisson:37, Tulczyjew:59, Steinhoff-Puetzfeld:10}.

Equations~\eqref{nth_stress-energy-tensor} and \eqref{total_T} have a physically intuitive form: in the limit of small size, a body appears as its skeleton of multipole moments (including gravitational corrections to them). However, the fields sourced by this stress-energy tensor are only meaningful approximations in the buffer region and beyond, and it sources only the homogeneous (for $r>0$) solutions to the wave equation. So we can say more precisely that at distances $r\gg m$, the body affects the gravitational field only through its multipole structure, and in that sense, it appears as a point particle equipped with multipole moments.

\subsection{Puncture scheme}\label{nth_puncture}
In practice, of course, one is interested in obtaining a global solution, not merely a local solution in the buffer region. Furthermore, solving the field equations in the buffer region will not determine the regular field $\phi^{{\rm R}\mu\nu}$; to do so, one must impose boundary conditions, over and above the matching condition. To obtain such a global solution numerically, and at the same time determine the regular field, one can use a puncture scheme. Here, I follow the description of Dolan and Barack~\cite{Dolan-Barack:11}, but once a local, analytical expression for the singular field is constructed, any puncture scheme may be used to numerically find a global solution for the physical field.

First, define a $k$th-order \emph{puncture function} $\phi^{\mathcal{P}\mu\nu}_{[k](n)}$ to be $\phi^{{\rm S}\mu\nu}_{(n)}$ truncated at order $r^k$:
\begin{equation}
\phi^{\mathcal{P}\mu\nu}_{[k](n)}=\sum_{p=-n}^{k}r^p\phi^{{\rm S}\mu\nu}_{(n,p)}.
\end{equation}
This function may be obtained analytically by transforming the buffer-region expression for $\phi^{{\rm S}\mu\nu}_{(n)}$ into convenient coordinates. Similarly, define a \emph{residual function}
\begin{equation}
\phi^{\mathcal{R}\mu\nu}_{[k](n)}\equiv\phi^{\mu\nu}_{(n)}-\phi^{\mathcal{P}\mu\nu}_{[k](n)}.
\end{equation}
With these definitions, $\phi^{\mathcal{R}\mu\nu}_{[k](n)}$ is a $C^{k}$ function, and it agrees with $\phi^{{\rm R}\mu\nu}_{(n)}$ through order $r^k$.\footnote{My notation differs slightly from that of Ref.~\cite{Dolan-Barack:11}, where the label $[k]$ corresponded to a puncture containing $k$ orders, which would agree with $\phi^{{\rm S}\mu\nu}_{(n)}$ through order $r^{k-n-1}$.} 

Next, define a hollow timelike tube $\Gamma$ of some numerically reasonable radius around the body. Inside $\Gamma$, one can replace the physical problem, in which one would encounter the material body or black hole, with the effective problem
\begin{align}
E^{\mu\nu}{}_{\rho\sigma}\phi^{\mathcal{R}\rho\sigma}_{[k](n)} &= C^{\mu\nu}_{(n)} -16\pi (T^{\mu\nu}_{(n)}+\tau^{\mu\nu}_{(n)})\nonumber\\
&\quad-E^{\mu\nu}{}_{\rho\sigma}\phi^{\mathcal{P}\rho\sigma}_{[k](n)}\label{effective_field_equation}\\
&\equiv -16\pi \mathcal{S}^{\mu\nu}_{[k](n)}.
\end{align}
This problem can be solved for $\phi^{\mathcal{R}\rho\sigma}_{[k](n)}$. By construction, the singular terms in $E^{\mu\nu}{}_{\rho\sigma}\phi^{\mathcal{P}\rho\sigma}_{[k](n)}$ cancel those in $C^{\mu\nu}_{(n)} -16\pi (T^{\mu\nu}_{(n)}+\tau^{\mu\nu}_{(n)})$. Since the source terms contain two spatial derivatives of the field, the effective source $\mathcal{S}^{\mu\nu}_{[k](n)}$ is a $C^{k-2}$ function. So we require $k\geq2$ to obtain a $C^0$ source for the residual field. However, a source that is finite but discontinuous will suffice in practice, meaning a $k=1$ puncture will suffice.

Outside $\Gamma$, one may solve the physical problem
\begin{equation}
E^{\mu\nu}{}_{\rho\sigma}\phi^{\rho\sigma}_{(n)} = C^{\mu\nu}_{(n)}-16\pi\tau^{\mu\nu}_{(n)} 
\end{equation}
for the physical field $\phi^{\rho\sigma}_{(n)}$. When crossing $\Gamma$, one may change variables from the residual field to the full field via 
\begin{equation}
\phi^{\mu\nu}_{(n)}=\phi^{\mathcal{R}\rho\sigma}_{[k](n)}+\phi^{\mathcal{P}\rho\sigma}_{[k](n)};
\end{equation}
that is, across $\Gamma$ one goes from the effective problem to the physical problem by, roughly speaking, `re-inserting' the body, adding the analytical expression for $\phi^{\mathcal{P}\rho\sigma}_{[k](n)}$ to the numerical result for $\phi^{\mathcal{R}\rho\sigma}_{[k](n)}$. Likewise, when crossing in the other direction, into the region bounded by $\Gamma$, one may change variables from the full field to the residual field by subtracting the puncture.

Recall that $\phi^{{\rm S}\rho\sigma}_{(n)}$ will depend on the lower-order regular fields $\phi^{{\rm R}\rho\sigma}_{(n'<n)}$. So at each timestep in a numerical evolution, one must first calculate the first-order residual field from the first-order puncture, then use that residual field to calculate the second-order effective source, and so forth. 

There is obviously great freedom in formulating and implementing a puncture scheme. For example, rather than taking the truncated expression for  $\phi^{{\rm S}\mu\nu}_{(n)}$ as a puncture, one could use any other function that agrees with $\phi^{{\rm S}\mu\nu}_{(n)}$ through order $r^k$ for small $r$. Additionally, the puncture function could be chosen  to smoothly go to zero at large distances~\cite{Barack-Golbourn:07} or at some finite distance~\cite{Vega-Detweiler:07} from $\gamma$. In that case, at sufficient distance the residual function would become the physical field, meaning the physical field could be obtained by solving Eq.~\eqref{effective_field_equation} globally rather than only inside $\Gamma$.


\section{Explicit outer solution in buffer region through second order}\label{explicit}


I now present the explicit results sufficient to implement a second-order puncture scheme. I choose
\begin{align}
\phi^{\mu\nu} & \equiv \bar h^{\mu\nu},\\
Z^\mu{}_{\rho\sigma} & \equiv 0,\\
D^\mu & \equiv 0.
\end{align}
The first- and second-order wave equations then read
\begin{align}
E^{\mu\nu}{}_{\rho\sigma}\bar h^{\rho\sigma}_{(1)} &= 0,\label{1st_wave}\\
E^{\mu\nu}{}_{\rho\sigma}\bar h^{\rho\sigma}_{(2)} &= -2\delta^2 G^{\mu\nu}[\bar h^{(1)}].\label{2nd_wave}
\end{align}
The second-order Einstein tensor, $\delta^2 G^{\mu\nu}$, can be written with the gauge condition on $\bar h^{\mu\nu}$ already imposed or not. Here, I use the full expression, \eqref{second-order_Einstein}, without simplifying it via the gauge condition. The solutions to these equations are constrained by the first- and second-order gauge conditions
\begin{align}
\bar\delta^0\nabla_{\!\nu}\bar h^{\mu\nu}_{(1)} &= 0,\label{1st_gauge}\\
\bar\delta^0\nabla_{\!\nu}\bar h^{\mu\nu}_{(2)} &= - \bar\delta^1\nabla_{\!\nu}\bar h^{\mu\nu}_{(1)};\label{2nd_gauge}
\end{align}
that is, in coordinates centered on $\gamma$ the divergence of $\bar h^{\mu\nu}_{(1)}$ at time $t$ must vanish if $a^\mu(t)$ is set to $a^\mu_{(0)}(t)$, and the divergence of $\e \bar h^{\mu\nu}_{(1)}+\e^2\bar h^{\mu\nu}_{(2)}$ at time $t$ must vanish if $a^\mu(t)$ is set to $\e a^\mu_{(1)}(t)$ in the first expression and $a^\mu_{(0)}$ in the second.

In the buffer region, I use Fermi-Walker coordinates, in which the spatial coordinates $x^a$ span a three-surface that intersects $\gamma$ orthogonally, and the time coordinate is equal to proper time on $\gamma$ at the point of intersection. More explicitly, at a point $x\in\mathcal{M}_{\rm E}$ near $\gamma$, the spatial coordinates are given by $x^a(x)=-e^a_{\alpha'}(x')\sigma^{;\alpha'}(x,x')$, where $x'$ is a point on $\gamma$ connected to $x$ by a unique spatial geodesic $\beta$ that intersects $\gamma$ orthogonally, $e^\alpha_a$ is a spatial triad  orthogonal to $u^\alpha$ on $\gamma$,  and $\sigma(x,x')$ is one half the squared geodesic distance from $x'$ to $x$. The time coordinate is given by $t(x)=\tau(x')$, the radial distance is given by $r(x)=\sqrt{\delta_{ab}x^a(x)x^b(x)}=\sqrt{2\sigma(x,x')}$, and the radial unit vectors are $n^a(x)=x^a(x)/r(x)$. Further details about the construction of the coordinates can be found in Ref.~\cite{Poisson-Pound-Vega:11}.

Through order $r^3$, the metric in Fermi-Walker coordinates is given by
\begin{align}
g_{tt} &= -1-2a_ix^i-\left(R_{0i0j}+a_ia_j\right)x^ix^j\nonumber\\
&\quad-\frac{1}{3}\left(4R_{0i0j}a_k+R_{0i0j|k}\right)x^ix^jx^k+O(r^4),\\
g_{ta} &= -\frac{2}{3}R_{0iaj}x^ix^j-\frac{1}{3}R_{0iaj}a_kx^ix^jx^k\nonumber\\
&\quad -\frac{1}{4}R_{0iaj|k}x^ix^jx^k+O(r^4),\\
g_{ab} &= \delta_{ab}-\frac{1}{3}R_{aibj}x^ix^j-\frac{1}{6}R_{aibj|k}x^ix^jx^k+O(r^4),
\end{align}
where the pieces of the Riemann tensor are evaluated on the worldline and contracted with members of the tetrad $(u^\alpha,e^\alpha_a)$. For example, $R_{0iaj|k}\equiv R_{\alpha\mu\beta\nu;\rho}\big|_\gamma u^\alpha e^\mu_ie^\beta_ae^\nu_je^\rho_k$. An overdot will indicate a covariant derivative along the worldline, $\dot R_{0iaj}\equiv R_{\alpha\mu\beta\nu;\rho}\big|_\gamma u^\alpha e^\mu_ie^\beta_ae^\nu_ju^\rho$. Because the background is Ricci-flat, the components of the Riemann tensor and its first derivatives can be written in terms of Cartesian STF tensors $\E_{ab}$, $\B_{ab}$, $\E_{abc}$, and $\B_{abc}$:
\begin{align}
\E_{ab} &\equiv R_{0a0b},\\
\B_{ab} &\equiv \frac{1}{2}\epsilon^{pq}{}_{(a}R_{b)0pq}, \\
\E_{abc} &\equiv \mathop{\rm STF}_{abc}R_{0a0b|c},\\
\E_{abc} &\equiv \frac{3}{8}\mathop{\rm STF}_{abc}\epsilon^{pq}{}_{a}R_{b0pq|c},
\end{align}
where `STF' denotes the STF combination of the indicated indices. $\E_{ab}$ and $\B_{ab}$ are the even- and odd-parity tidal quadrupole moments of the background spacetime in the neighbourhood of $\gamma$, and $\E_{abc}$ and $\B_{abc}$ are the even- and odd-parity tidal octupole moments. Identities for decomposing each component of the Riemann tensor and its derivatives in terms of these tidal moments can be found in Appendix D3 of Ref.~\cite{Poisson-Vlasov:09}.

Following the procedure of the previous sections, in terms of these coordinates I assume the trace-reversed fields have the expansion
\begin{align}
\bar h^{\mu\nu}_{(n)}&= \sum_{p\geq-n,q,\ell}r^p(\ln r)^q \bar h^{\mu\nu L}_{(n,p,q,\ell)}(t)\nhat_L.
\end{align}
An advantage of Fermi-Walker coordinates is that their simple geometric properties afford an easy transformation to arbitrary coordinate systems. Such transformations are available in the literature~\cite{Klein-Collas:07}. Alternatively, one may use the geometric definitions of the Fermi-Walker coordinates to express the above expansion in terms of the covariant quantity $\sigma(x,x')$; from there, one can express the result in any desired coordinates.
Therefore, once the singular field is known in Fermi coordinates, it may be easily transformed into any desired coordinates to form a puncture function useful for numerical implementation.

In order to obtain a solution for $\bar h^{\mu\nu}_{(2)}$ through order $r$, as is necessary for a puncture scheme, I require four orders in $r$ in all my expansions, since $\bar h^{\mu\nu}_{(2)}$ begins at order $1/r^2$. Because the general procedure was shown in Sec.~\ref{nth_outer_expansion}, and the first three orders in $r$ were derived step by step in Ref.~\cite{Pound:10a},\footnote{Reference \cite{Pound:10a} obtained results for $h^{(1)}_{\mu\nu}$ and $h^{(2)}_{\mu\nu}$ rather than for $\bar h_{(1)}^{\mu\nu}$ and $\bar h_{(2)}^{\mu\nu}$, and in the case of $h^{(2)}_{\mu\nu}$ it did so only after substituting the expansion for $a^\mu$. However, the steps involved in the calculation are essentially identical. The results presented here were actually obtained by trace-reversing the results reported in Ref.~\cite{Pound:12}; I present the trace-reversed fields to mesh with the foregoing discussion of the $n$th-order problem.} here I shall simply state the results for the most part.

\subsection{First order}\label{1st_explicit}
Following the steps outlined in Sec.~\ref{nth_buffer_solution}, one finds the general solution to Eq.~\eqref{1st_wave} in the buffer region has the form
\begin{align}
\bar h^{\mu\nu}_{(1)} &= \frac{1}{r}\bar h^{\mu\nu}_{(1,-1,0,0)}+\sum_{\ell=0}^1\bar h^{\mu\nu L}_{(1,0,0,\ell)}\nhat_L+r\sum_{\ell=0}^2\bar h^{\mu\nu L}_{(1,1,0,\ell)}\nhat_L\nonumber\\
&\quad +r^2\sum_{\ell=0}^3\bar h^{\mu\nu L}_{(1,2,0,\ell)}\nhat_L+O(r^3),\label{1st_form}
\end{align}
where $\bar h^{\mu\nu}_{(1,-1,0,0)}$, $\bar h^{\mu\nu}_{(1,0,0,0)}$, $\bar h^{\mu\nu i}_{(1,1,0,1)}$, and $\bar h^{\mu\nu ij}_{(1,2,0,2)}$ are the homogeneous modes, each of which defines a minimal $p$th solution to Eq.~\eqref{1st_wave}, and of which the other functions $\bar h^{\mu\nu L}_{(1,p,0,\ell)}$ are linear combinations.

\subsubsection{Singular and regular field}
The general solution can be conveniently split into two:
\begin{equation}
\bar h_{(1)}^{\mu\nu}=\bar h_{(1)}^{{\rm S} \mu\nu}+\bar h_{(1)}^{{\rm R}\mu\nu}.
\end{equation} 
Following the definitions introduced in Sec.~\ref{nth_singular-regular}, I define the singular field to be the piece of $\bar h_{(1)}^{\mu\nu}$ made from a minimal $p$th solution for $p=-1$---that is, the solution fully determined by $\bar h^{\mu\nu}_{(1,-1,0,0)}$ alone. It contains the following pieces of the full first-order field \eqref{1st_form}:
\begin{align}
\bar h^{{\rm S}\mu\nu}_{(1)} &= \frac{1}{r}\bar h^{\mu\nu}_{(1,-1,0,0)}+\bar h^{\mu\nu i}_{(1,0,0,1)}n_i+r\sum_{\ell=0,2}\bar h^{\mu\nu L}_{(1,1,0,\ell)}\nhat_L\nonumber\\
&\quad +r^2\sum_{\ell=1,3}\bar h^{\mu\nu L}_{(1,1,0,\ell)}\nhat_L+O(r^3).\label{hS1_form}
\end{align}
The monopole moment $I^{\mu\nu}_{(1)}\equiv\bar h^{\mu\nu}_{(1,-1,0,0)}$ is constrained by the gauge condition \eqref{nth_gauge_expanded} to be of the form $I_{(1)}u^\mu u^\nu$, where $I_{(1)}$ is a constant. The matching condition $h^{(1,-1)}_{\mu\nu}=g^{(1)}_{{\rm I}\mu\nu}$ then determines $I_{(1)}=4m$, where $m$ is the ADM mass of the inner background spacetime $g_{{\rm I}\mu\nu}$. Given these constraints, Eq.~\eqref{hS1_form} explicitly reads
\begin{align}
\bar h^{{\rm S}tt}_{(1)} &= \frac{4m}{r}-10ma^in_i\nonumber\\
&\quad +mr\left[\frac{28}{3}a_ia^i + \frac{7}{6}(15 a^ia^j-4\E^{ij})\nhat_{ij}\right]\nonumber\\
&\quad +mr^2\left[\frac{1}{2}\left(19a_j\E^{ij}-\partial_t^2a^i-56 a_ja^ja^i\right)n_i\right.\nonumber\\
&\quad +\left.\left(15a^i\E^{jk}-\frac{3}{2}\E^{ijk}-\frac{105}{4}a^ia^ja^k\right)\nhat_{ijk}\right],\label{hS1tt}
\end{align}
\begin{align}
\bar h^{{\rm S}ta}_{(1)} &= 2mr\left(\partial_t a^a+\B^{ic}\epsilon^a{}_c{}^j\nhat_{ij}\right) \nonumber\\
&\quad +mr^2\left[\left(\frac{73}{30}\epsilon^a{}_{bd}a^b\B^{id}-\frac{7}{30}\epsilon^{ai}{}_da_b\B^{bd}\right.\right.\nonumber\\
&\quad \left.-\frac{7}{6}\epsilon^{bi}{}_d a_b\B^{ad}+\frac{8}{15}\dot\E^{ai}-5a^i\partial_ta^a-2a^a \partial_ta^i\right)n_i\nonumber\\
&\quad \left.+\frac{1}{9}\!\left(\!2\delta^{ai}\dot\E^{jk}-7\epsilon^a{}_c{}^ia^j\B^{ck}+8\epsilon^a{}_b{}^i\B^{bjk}\right)\!\nhat_{ijk}\right],
\end{align}
\begin{align}
\bar h^{{\rm S}ab}_{(1)} &= 4mr\left(a^{a}a^{b}-\E^{ab}\right)+ 2mr^2\bigg[(\E^{ab}-3a^{a}a^{b})a^i \nonumber\\
&\quad -\mathcal{E}^{abi}+2a^{(a}\E^{b)i} -\frac{2}{3}\dot\B^{d(a}\epsilon^{b)i}{}_{d}\bigg] \hat{n}^{c},\label{hS1ab}
\end{align}
where $\partial_t m=0$.

The regular field is the sum of minimal $p$th solutions for $p\geq0$ appearing in $\bar h_{(1)}^{\mu\nu}$---that is, the part of $\bar h_{(1)}^{\mu\nu}$ that is determined by global boundary conditions rather than the matching condition. It contains the following pieces of the full field: 
\begin{align}
\bar h^{{\rm R}\mu\nu}_{(1)} &= \bar h_{(1,0,0,0)}^{\mu\nu} + r\bar h_{(1,1,0,1)}^{\mu\nu i}n_i\nonumber\\
&\quad +r^2\left(\bar h_{(1,2,0,2)}^{\mu\nu ij}\nhat_{ij} + \bar h_{(1,2,0,0)}^{\mu\nu}\right) + O(r^3).\label{h1R}
\end{align}
We can write this as a Taylor series
\begin{align}
\bar h^{{\rm R}\mu\nu}_{(1)} &= \left.\bar h^{{\rm R}\mu\nu}_{(1)}\right|_{r=0} + \left.\bar h^{{\rm R}\mu\nu}_{(1)}{}_{,i}\right|_{r=0}x^i\nonumber\\
&\quad +\frac{1}{2}\left.\bar h^{{\rm R}\mu\nu}_{(1)}{}_{,ij}\right|_{r=0}x^ix^j + O(r^3),
\end{align}
where the coefficients are related to the pieces of the full metric as
\begin{align}
\left.\bar h^{{\rm R}\mu\nu}_{(1)}\right|_{r=0} &= \bar h_{(1,0,0,0)}^{\mu\nu},\\
\left.\bar h^{{\rm R}\mu\nu,i}_{(1)}\right|_{r=0} &= \bar h_{(1,1,0,1)}^{\mu\nu i},\\
\left.\bar h^{{\rm R}\mu\nu,\langle ij\rangle}_{(1)}\right|_{r=0} &= 2\bar h_{(1,2,0,2)}^{\mu\nu ij},\\
\left.\bar h^{{\rm R}\mu\nu}_{(1)}{}_{,i}{}^i\right|_{r=0} &= 6\bar h_{(1,2,0,0)}^{\mu\nu}.
\end{align}
As noted above, all of the functions on the right-hand sides must be independently determined by boundary conditions, except $\bar h_{(1,2,0,0)}^{\mu\nu}$, which is given in terms of the others by 
\begin{align}
6\bar h_{(1,2,0,0)}^{tt} &= -\frac{5}{2} \bar h^{{\rm R}tt,a}_{(1)} a{}_{a} -  \mathcal{E}{}^{ab} \bar h^{{\rm R}}_{(1)ab} 
						+ a{}^{a} a{}^{b} \bar h^{{\rm R}}_{(1)ab} \nonumber\\
						&\quad - a{}_{a} a{}^{a} \bar h^{{\rm R}tt}_{(1)} + \bar h^{{\rm R}ta}_{(1)} a{}_{a}{}_{,t} 
						+ 2 a{}^{a} \bar h^{{\rm R}t}_{(1)a,t} \nonumber\\
						&\quad + \frac{1}{2} \bar h^{{\rm R}tt}_{(1)}{}_{,tt},
\end{align}
\begin{align}
6\bar h_{(1,2,0,0)}^{ta} &= -\frac{3}{2} \bar h^{{\rm R}ta,b}_{(1)}a{}_{b} + \mathcal{B}{}^{bc} \epsilon{}^{a}{}_{cd} \bar h^{{\rm R}d}_{(1)b} 
						+ \frac{3}{2} a{}^{a} a{}^{b} \bar h^{{\rm R}t}_{(1)b} \nonumber\\
						&\quad - \frac{5}{6} \mathcal{E}{}^{a}{}_{b} \bar h^{{\rm R}tb}_{(1)} + \frac{1}{2} \bar h^{{\rm R}tt}_{(1)} a{}^{a}{}_{,t} 
						+ \frac{1}{2} \bar h^{{\rm R}ab}_{(1)} a{}_{b}{}_{,t}\nonumber\\
						&\quad + a{}^{b} \bar h^{{\rm R}a}_{(1)b,t} + a{}^{a} \bar h^{{\rm R}tt}_{(1)}{}_{,t} 
						+ \frac{1}{2} \bar h^{{\rm R}ta}_{(1)}{}_{,tt},
\end{align}
\begin{align}
6\bar h_{(1,2,0,0)}^{ab} &= -\frac{1}{2} \bar h^{{\rm R}ab}_{(1)}{}_{,c} a{}^{c} + \frac{7}{3} \mathcal{E}{}^{c(a} \bar h^{{\rm R}b)}_{(1)c} 
						+ a^{c}a^{(a}\bar h^{{\rm R}b)}_{(1)c} \nonumber\\
						&\quad - \mathcal{E}{}^{cd} \delta{}^{ab} \bar h^{{\rm R}}_{(1)cd}  - \mathcal{E}{}^{ab} \bar h^{{\rm R}c}_{(1)c} 
						- 2\mathcal{B}{}^{d(a} \epsilon{}^{b)}{}_{cd} \bar h^{{\rm R}tc}_{(1)} \nonumber\\
						&\quad - \mathcal{E}{}^{ab} \bar h^{{\rm R}tt}_{(1)} + a{}^{a} a{}^{b} \bar h^{{\rm R}tt}_{(1)} 
						+ \bar h^{{\rm R}t(a}_{(1)}a^{b)}{}_{,t} \nonumber\\
						&\quad + 2a^{(a} \bar h^{{\rm R}tb)}_{(1)}{}_{,t} + \frac{1}{2} \bar h^{{\rm R}ab}_{(1)}{}_{,tt}.
\end{align}
Here $\bar h^{{\rm}\mu\nu}_{(1)}$ and its derivatives are evaluated at $r=0$.

\subsubsection{Stress-energy tensor and puncture scheme}
From Eq.~\eqref{nth_stress-energy-tensor}, the body's effective stress-energy tensor is determined to be
\begin{align}
T^{\mu\nu}_{(1)} &= \int_{\gamma}mu^{\mu}u^\nu\delta(x,z)d\tau,\label{1st_stress-energy}
\end{align}
which is the stress-energy of a point mass moving on $\gamma$ in the external background spacetime. Therefore, at first order and at distances in the buffer region or greater, the body appears as a point mass.

Following Sec.~\ref{nth_puncture}, one may use a puncture scheme to obtain the unknown pieces of the field (i.e., $\bar h^{{\rm R}\mu\nu}_{(1)}$) in the buffer region, along with the global solution $\bar h^{\mu\nu}_{(1)}$ everywhere outside the buffer region. After transforming it into a numerically useful coordinate system and truncating it at order $r^2$, the singular field given in Eqs.~\eqref{hS1tt}--\eqref{hS1ab} defines a puncture $\bar h^{\mathcal{P}\rho\sigma}_{[2](1)}$. Inside a tube $\Gamma$ about the body, the residual field $\bar h^{\mathcal{R}\rho\sigma}_{[2](1)}\equiv \bar h^{\rho\sigma}_{(1)}-\bar h^{\mathcal{P}\rho\sigma}_{[2](1)}$ can be found by solving the effective-source equation
\begin{align}
E^{\mu\nu}{}_{\rho\sigma}\bar h^{\mathcal{R}\rho\sigma}_{[2](1)} &= -16\pi T^{\mu\nu}_{(1)}-E^{\mu\nu}{}_{\rho\sigma}\bar h^{\mathcal{P}\rho\sigma}_{[2](1)}\label{1st_effective_wave}\\
&\equiv \mathcal{S}^{\mu\nu}_{[2](1)},
\end{align}
where $\mathcal{S}^{\mu\nu}_{[2](1)}$ is a $C^0$ source that can be calculated analytically. Outside $\Gamma$, the physical field $\bar h^{\rho\sigma}_{(1)}$ can be found by solving \eqref{1st_wave}. When crossing from one side of $\Gamma$ to the other, one changes variables between $\bar h^{\rho\sigma}_{(1)}$ and $\bar h^{\mathcal{R}\rho\sigma}_{(1)}$ using 
\begin{equation}
\bar h^{\mu\nu}_{(1)}=\bar h^{\mathcal{R}\rho\sigma}_{[2](1)}+\bar h^{\mathcal{P}\rho\sigma}_{[2](1)}.
\end{equation}
Given reasonable initial conditions and a forward evolution in time, the field $\bar h^{\mu\nu}_{(1)}$ obtained with this procedure will describe the retarded solution to Eq.~\eqref{1st_wave} with the appropriate singularity structure, and on $\gamma$ the value of the residual field $\bar h^{\mathcal{R}\mu\nu}_{[2](1)}$ and its first and second derivatives will agree with those of the regular field $\bar h^{{\rm R}\mu\nu}_{(1)}$. Details of the puncture scheme can be found in Refs.~\cite{Barack-Golbourn:07,Dolan-Barack:11,Dolan-Barack-Wardell:11}. Alternative, equivalent first-order puncture schemes are described in Refs.~\cite{Vega-Detweiler:07,Vega-Wardell-Diener:11,Wardell-etal:11}.

Note that because the point-particle stress-energy tensor~\eqref{1st_stress-energy} is a well-defined distribution, the global solution for the physical field may be obtained directly from 
\begin{align}
E^{\mu\nu}{}_{\rho\sigma}\bar h^{\rho\sigma}_{(1)} &= -16\pi T^{\mu\nu}_{(1)}.
\end{align}
This field equation was traditionally the starting point for approaching the first-order problem. A variety of numerical methods exist for solving it~\cite{Poisson-Pound-Vega:11}, the above puncture scheme being just one among them. Analytically, with arbitrary initial conditions on the Cauchy surface $\Sigma$ intersecting $\gamma$, the retarded solution\footnote{By `retarded solution', I mean the causal solution obtained with fields propagating forward in time from arbitrary initial data; I do not mean the particular causal solution corresponding to zero incoming radiation at the infinite past. Given that the influence of initial data decays with time, any causal solution should approach the latter.} reads
\begin{align}
\bar h^{\mu\nu}_{(1)} &= 4m\int_\gamma G^{\mu\nu}{}_{\mu'\nu'}u^{\mu'}u^{\nu'}d\tau +\bar h^{\mu\nu}_{(1)\Sigma},\label{1st_analytic}
\end{align}
where $G^{\mu\nu}{}_{\mu'\nu'}$ is the retarded Green's function for the wave operator $E^{\mu\nu}{}_{\rho\sigma}$ (with the normalization used in Ref.~\cite{Poisson-Pound-Vega:11}), the integral over $\gamma$ is truncated at $\Sigma$, and $\bar h^{\mu\nu}_{(1)\Sigma}$ is the contribution from initial data. As was shown in Ref.~\cite{Pound:10a},\footnote{In Ref.~\cite{Pound:10a}, the equality was shown only for $a^\mu=0$; it is shown for arbitrary acceleration in Ref.~\cite{Poisson-Pound-Vega:11}.} when one writes this solution in Fermi-Walker coordinates and decomposes it into the multipolar form~\eqref{1st_form}, $\bar h_{(1)}^{{\rm R}\mu\nu}$ consists of tail terms (integrals over the past history of the worldline and contributions from initial data) plus local terms involving the acceleration and tidal moments. Furthermore, one finds that at least through order $r$, and modulo any incoming waves contained in the initial data, $\bar h_{(1)}^{{\rm R}\mu\nu}$ is equal to the Detweiler-Whiting regular field $4m\int_\gamma G^{{\rm R}\mu\nu}{}_{\mu'\nu'}u^{\mu'}u^{\nu'}d\tau$, where $G^{{\rm R}\mu\nu}{}_{\mu'\nu'}$ is the Detweiler-Whiting regular Green's function~\cite{Detweiler-Whiting:02}. Correspondingly, through order $r$, $\bar h_{(1)}^{{\rm S}\mu\nu}$ is equal to the Detweiler-Whiting singular field.

\subsubsection{Equation of motion}
Evaluating Eq.~\eqref{nth_acceleration}, one finds that the zeroth-order term in the acceleration is
\begin{equation}
a^\mu_{(0)} = 0.
\end{equation}
Therefore, $a^\mu=O(\e)$, and at leading order the body behaves as a test mass, satisfying the geodesic equation in the external background spacetime. Corrections to the geodesic equation appear only at subleading order.

\subsection{Second order}\label{2nd_explicit}
Again following the steps outlined in Sec.~\ref{nth_buffer_solution}, one finds the general solution to Eq.~\eqref{2nd_wave} in the buffer region has the form
\begin{align}
\bar h^{\mu\nu}_{(2)} &= \frac{1}{r^2}\sum_{\ell=0}^2\bar h_{(2,-2,0,\ell)}^{\mu\nu L}\nhat^L
+\frac{1}{r}\sum_{\ell=0}^3 \bar h_{(2,-1,0,\ell)}^{\mu\nu L}\nhat^L\nonumber\\
&\quad +\sum_{\ell=0}^4\bar h_{(2,0,0,\ell)}^{\mu\nu L}\nhat^L+r\sum_{\ell=0}^5 \bar h_{(2,1,0,\ell)}^{\mu\nu L}\nhat^L\nonumber\\
&\quad +\ln r \big[\bar h_{(2,0,1,0)}^{\mu\nu}+r\bar h_{(2,1,1,1)}^{\mu\nu i}n_i\big]+O(r^2),\label{h2}
\end{align}
where $\bar h_{(2,-2,0,1)}^{\mu\nu i}=I^{\mu\nu i}_{(2)}$, $\bar h_{(2,-1,0,0)}^{\mu\nu}=I^{\mu\nu}_{(2)}$, $\bar h_{(2,0,0,0)}^{\mu\nu}$, and $\bar h_{(2,1,0,1)}^{\mu\nu i}$ are the homogeneous modes. The other functions $\bar h^{\mu\nu L}_{(2,p,0,\ell)}$ are linear combinations of those homogeneous modes plus quadratic combinations of the homogeneous modes appearing in $\bar h^{\mu\nu}_{(1)}$. 

\subsubsection{Singular and regular field}
Following Sec.~\ref{nth_singular-regular}, I split the second-order general solution as
\begin{equation}
\bar h_{(2)}^{\mu\nu}=\bar h_{(2)}^{{\rm S} \mu\nu}+\bar h_{(2)}^{{\rm R}\mu\nu}.
\end{equation} 
The singular field  $\bar h_{(2)}^{{\rm S} \mu\nu}$ is the sum of three terms: 
\begin{equation}
\bar h^{{\rm S}\mu\nu}_{(2)}={}_{(-2)}\bar h^{\mu\nu}_{(2)} +{}_{(-1)}\bar h^{\mu\nu}_{(2)}+\bar h^{{\rm IH}\mu\nu}_{(2)},
\end{equation}
where each of the terms is given through order $r$ in Appendix~\ref{singular_field}. ${}_{(-2)}\bar h^{\mu\nu}_{(2)}$ and ${}_{(-1)}\bar h^{\mu\nu}_{(2)}$ are the minimal $p$th solutions for $p=-2$ and $-1$, and $\bar h^{{\rm IH}\mu\nu}_{(2)}$ is the minimal inhomogeneous solution to
\begin{equation}
E^{\mu\nu}{}_{\rho\sigma}\bar h_{(2)}^{{\rm S} \rho\sigma} = -2\left(\delta^2G^{\mu\nu}[\bar h_{(1)}]
																-\delta^2G^{\mu\nu}[\bar h^{\rm R}_{(1)}]\right)\label{hS2_wave}.\\ 
\end{equation}
In other words,  $\bar h_{(2)}^{{\rm S} \mu\nu}$ is the sum of all terms in $\bar h^{\mu\nu}_{(2)}$ that directly involve the moments $I^{\mu\nu}_{(1)}$, $I^{\mu\nu i}_{(2)}$, and $I^{\mu\nu}_{(2)}$. The regular field $\bar h_{(2)}^{{\rm R} \mu\nu}$ is the sum of all minimal $p$th solutions for $p\geq0$, plus the minimal inhomogeneous solution to
\begin{align}
E^{\mu\nu}{}_{\rho\sigma}\bar h_{(2)}^{{\rm R} \rho\sigma} &= -2\delta^2G^{\mu\nu}[\bar h^{\rm R}_{(1)}].\label{hR2_wave}
\end{align}
That is, $\bar h_{(2)}^{{\rm R} \mu\nu}$ contains all terms in $\bar h^{\mu\nu}_{(2)}$ that are not determined without imposing boundary conditions beyond the matching condition. The regular field $\bar h^{{\rm R} \mu\nu}=\e\bar h_{(1)}^{{\rm R} \mu\nu}+\e^2\bar h_{(2)}^{{\rm R} \mu\nu}$ is a free gravitational field, a smooth solution to the homogeneous EFE through second order in $\e$.

With these definitions, the singular field $\bar h^{{\rm S}\mu\nu}_{(2)}$ contains the following pieces of the general solution:
\begin{align}
\bar h^{{\rm S}\mu\nu}_{(2)} &= \frac{1}{r^2}\sum_{\ell=0}^2\bar h_{(2,-2,0,\ell)}^{\mu\nu L}\nhat^L
+\frac{1}{r}\sum_{\ell=0}^3 \bar h_{(2,-1,0,\ell)}^{\mu\nu L}\nhat^L\nonumber\\
&\quad +\sum_{\ell=1}^4\bar h_{(2,0,0,\ell)}^{\mu\nu L}\nhat^L+r\!\!\!\!\!\!\!\sum_{\ell=0,2,3,4,5}\!\!\!\!\! \bar h_{(2,1,0,\ell)}^{\mu\nu L}\nhat^L\nonumber\\
&\quad +\ln r \big[\bar h_{(2,0,1,0)}^{\mu\nu}+r\bar h_{(2,1,1,1)}^{\mu\nu i}n_i\big]+O(r^2).\label{h2S}
\end{align}
First examine ${}_{(-2)}\bar h^{\mu\nu}_{(2)}$, which is the homogeneous solution of the form $I^{\mu\nu i}_{(2)}n_i/r^2+O(r^{-1})$, defined to contain only those terms in $\bar h^{\mu\nu}_{(2)}$ that are directly proportional to the dipole moment $I^{\mu\nu i}_{(2)}$. After splitting the dipole moment into its mass and current pieces, $I^{\mu\nu i}_{(2)}=M^{\mu\nu i}_{(2)}+S^{\mu\nu i}_{(2)}$, the gauge condition \eqref{nth_gauge_expanded} yields $M^{\mu bi}_{(2)}\nhat_{bi}=0$, which constrains the mass dipole to have the form $M^{\mu\nu i}_{(2)}=u^\mu u^\nu M^{tti}_{(2)} + \frac{2}{3}u^{(\mu}e^{\nu)i} \delta_{ij}M^{tij}_{(2)}$. A gauge refinement can be used to eliminate the $ta$ component of $M^{\mu\nu i}_{(2)}$ and the $ab$ component of $S^{\mu\nu i}_{(2)}$~\cite{Pound:10a}, leaving the dipole moment in the canonical form 
\begin{equation}
I^{\mu\nu i}_{(2)} = 4u^\mu u^\nu M^i+4u^{(\mu}e^{\nu)}_a\epsilon^{aji}S_j,
\end{equation}
where I have defined $M^i\equiv M^{tti}_{(2)}/4$ and $S_j\equiv\epsilon_{jib}S^{tbi}_{(2)}/4$. The factors of 4 are chosen based on the matching condition $h^{(2,-2)}_{\mu\nu}=g^{(2)}_{{\rm I}\mu\nu}$, which determines $S^i$ to be the ADM angular momentum, and $M^i$ the standard mass dipole, of the inner background spacetime $g_{{\rm I}\mu\nu}$. I set $M^i\equiv0$ to enforce the condition that the body is effectively centered on $\gamma$, which leads to 
\begin{equation}
{}_{(-2)}\bar h^{\mu\nu}_{(2)} = \frac{4u^{(\mu}e^{\nu)}_a\epsilon^{aji}S_jn_i}{r^2}+O(r^{-1}).
\end{equation}
The gauge condition \eqref{nth_gauge_expanded} further determines that the spin is constant; $\partial_tS^i=0$. 

Next, the monopole field ${}_{(-1)}\bar h^{\mu\nu}_{(2)}$ is the homogeneous solution containing only those terms in $\bar h^{\mu\nu}_{(2)}$ that are directly proportional to the monopole moment $I^{\mu\nu}_{(2)}$, which I now denote $\delta m^{\mu\nu}$. It has the form
\begin{equation}
{}_{(-1)}\bar h^{\mu\nu}_{(2)} = \frac{\delta m^{\mu\nu}}{r}+O(1),
\end{equation}
where the gauge condition constrains $\delta m^{\mu\nu}$ to be
\begin{align}
\delta m^{tt} &= -\frac{1}{3} m \bar h^{{\rm R}a}_{(1)}{}_{a} + m \bar h^{{\rm R}tt}_{(1)},\label{dmtt}\\
\delta m^{ta} &= -\frac{4}{3} m \bar h^{{\rm R}ta}_{(1)},\label{dmta}\\
\delta m^{ab} &= \frac{2}{3} m \bar h^{{\rm R}ab}_{(1)} - \frac{2}{3} m \delta{}^{ab} \bar h^{{\rm R}c}_{(1)}{}_{c} 
				- \frac{4}{3} m \delta{}^{ab} \bar h^{{\rm R}tt}_{(1)}.\label{dmab}
\end{align}
This is a gravitationally induced monopole; it can be thought of as the perturbation of the smooth field $\bar h^{{\rm R}\mu\nu}_{(1)}$ by the body's local field $\bar h^{{\rm S}\mu\nu}_{(1)}$, or as a correction to the body's local field $\bar h^{{\rm S}\mu\nu}_{(1)}$ by the smooth field $\bar h^{{\rm R}\mu\nu}_{(1)}$. Mathematically, it arises because $\bar h^{{\rm IH}\mu\nu}_{(2)}$ requires the addition of a $1/r$ monopole in order to satisfy the gauge condition. One could add an arbitrary, time-independent monopole of the form $u^\mu u^\nu\delta m $---that is, one that behaves exactly like the first-order mass---but I choose to incorporate that term into the definition of $m$.

The regular field contains the remaining pieces of the general solution:
\begin{equation}
\bar h_{(2)}^{{\rm R}\mu\nu} = \bar h_{(2,0,0,0)}^{\mu\nu}+r\bar h_{(2,1,0,1)}^{\mu\nu i}n_i+O(r^2);\label{h2R}
\end{equation}
terms quadratic in $\bar h_{(1)}^{{\rm R}\mu\nu}$, arising from the source in Eq.~\eqref{hR2_wave}, would appear at order $r^2$. The functions in Eq.~\eqref{h2R} are constrained by the gauge condition \eqref{2nd_gauge}, which relate them to each other and to the first-order regular field, but they are otherwise undetermined until further boundary conditions are imposed. As at first order, we can write this as the Taylor series
\begin{align}
\bar h^{{\rm R}\mu\nu}_{(2)} &= \left.\bar h^{{\rm R}\mu\nu}_{(2)}\right|_{r=0} + \left.\bar h^{{\rm R}\mu\nu}_{(1)}{}_{,i}\right|_{r=0}x^i + O(r^2),
\end{align}
where the coefficients are related to the pieces of the full metric  perturbation as
\begin{align}
\left.\bar h^{{\rm R}\mu\nu}_{(2)}\right|_{r=0} &= \bar h_{(2,0,0,0)}^{\mu\nu},\\
\left.\bar h^{{\rm R}\mu\nu,i}_{(2)}\right|_{r=0} &= \bar h_{(2,1,0,1)}^{\mu\nu i}.
\end{align}

\subsubsection{Stress-energy tensor and puncture scheme}\label{2nd_puncture}
From Eq.~\eqref{nth_stress-energy-tensor} the body's second-order effective stress-energy tensor is 
\begin{align}\label{T2}
T_{(2)}^{\mu\nu} &= \int_\gamma u^{(\mu}S^{\nu) \alpha}\delta(x,z)_{;\alpha}d\tau +\int_\gamma \frac{1}{4}{\delta m}^{\mu\nu}\delta(x,z)d\tau,
\end{align}
where I have defined $S^{\alpha\beta}\equiv e^\alpha_ae^\beta_b\epsilon^{abi}S_i$ (not to be confused with $S^{\mu\nu L}_{(n)}$). The first term is the effective stress-energy due to the body's spin; the second, that due to the induced monopole $\delta m^{\mu\nu}$. If the spin vanishes, then at distances in the buffer region or greater, the body still appears as a monopolar point mass, but with a nonlinear gravitational correction to the monopole field.

Again, a puncture scheme can be used to obtain the unknown pieces of the field in the buffer region, $\bar h_{(2,0,0,0)}^{\mu\nu}$ and $\bar h_{(2,1,0,1)}^{\mu\nu i}$, along with the physical field globally. The singular field given in Appendix~\eqref{singular_field}, truncated at order $r$, defines a puncture $\bar h^{\mathcal{P}\rho\sigma}_{[1](2)}$. Inside the tube $\Gamma$, the residual field $\bar h^{\mathcal{R}\rho\sigma}_{[1](2)}\equiv \bar h^{\rho\sigma}_{(2)}-\bar h^{\mathcal{P}\rho\sigma}_{[1](2)}$ can be found by solving the effective-source equation
\begin{align}
E^{\mu\nu}{}_{\rho\sigma}\bar h^{\mathcal{R}\rho\sigma}_{[1](2)} &= -16\pi T^{\mu\nu}_{(2)}-2\delta^2G^{\mu\nu}[\bar h_{(1)}]\nonumber\\
			&\quad-E^{\mu\nu}{}_{\rho\sigma}\bar h^{\mathcal{P}\rho\sigma}_{[1](2)}\label{2nd_effective_wave}\\
			&\equiv \mathcal{S}^{\mu\nu}_{[1](2)},
\end{align}
where $\mathcal{S}^{\mu\nu}_{[1](2)}$ is a discontinuous but finite source that can be calculated analytically. The analytical expression for the source will involve the first-order residual field, which is to be determined numerically at each time step along with the second-order residual field. Outside $\Gamma$, the physical field $\bar h^{\rho\sigma}_{(2)}$ can be found by solving \eqref{2nd_wave}. When crossing $\Gamma$, one may change variables between $\bar h^{\rho\sigma}_{(2)}$ and $\bar h^{\mathcal{R}\rho\sigma}_{(2)}$ using 
\begin{equation}
\bar h^{\mu\nu}_{(2)}=\bar h^{\mathcal{R}\rho\sigma}_{[1](2)}+\bar h^{\mathcal{P}\rho\sigma}_{[1](2)}.
\end{equation}

This scheme is essentially identical to that at first order. In principle, it can be implemented almost immediately in existing codes. One need only transform the puncture function into a convenient coordinate system and then calculate an analytical expression for the effective source in those coordinates.

Unlike at first order, where one could write down an analytical solution in terms of a retarded Green's function, at second order no analytical solution presents itself. The part of the retarded solution that is sourced by $\delta^2G^{\mu\nu}$ cannot be written as an integral over all space, of the form $\int G^{\mu\nu}{}_{\rho'\sigma'}\delta^2 G^{\rho'\sigma'}dV$, since the $1/r^4$ singularity in $\delta^2 G^{\mu\nu}$ would make that integral diverge at every point in space. The actual solution consists of a volume integral over $\Omega$ plus a surface integral over $\partial\Omega$~\cite{Pound:10a}:
\begin{align}\label{formal_solution}
\bar h^{\alpha\beta}_{(2)} &= \frac{1}{4\pi}\!\oint\limits_{\partial\Omega}\!\!\Big(G^{\alpha\beta}{}_{\gamma'\delta'} \nabla_{\!\mu'}\bar h^{\gamma'\delta'}_{(2)}-\bar h^{\gamma'\delta'}_{(2)}\nabla_{\!\mu'} G^{\alpha\beta}{}_{\gamma'\delta'}\Big) dS^{\mu'}\nonumber\\
&\quad -\frac{1}{2\pi}\int_\Omega G^{\alpha\beta}{}_{\gamma'\delta'}\delta^2G^{\gamma'\delta'}[\bar h_{(1)}]dV',
\end{align}
which requires one to know the field on the timelike part of $\partial\Omega$ in the buffer region. Therefore this formal solution is not especially useful. However, one could use to it to find analytical expressions for the homogeneous modes that appear in $\bar h^{{\rm R}\mu\nu}_{(2)}$. Using the analytically determined solution in the buffer region as data on the part of $\partial\Omega$ embedded therein, the integrals could be expanded for small $r$ with methods from Ref.~\cite{Pound:10a}; decomposing the result into STF pieces, one could then read off the homogeneous modes. This small-$r$ expansion is complicated by the fact that an expansion of the Green's function would rely on the Hadamard decomposition~\cite{Poisson-Pound-Vega:11}, which is valid only in a convex normal neighbourhood, while the volume integral in Eq.~\eqref{formal_solution} generically extends over a much larger region. So, to evaluate the volume integral at a point $x$, one must restrict one's attention to points $x$ very near the initial time-slice $\Sigma$, such that the volume in which $G^{\alpha\beta}{}_{\gamma'\delta'}$ has support inside the integral lies entirely within the convex normal neighbourhood of $x$. I leave this calculation to future work.

However, the part of the retarded field that is sourced by $T^{\mu\nu}_{(2)}$ can be immediately written down. It reads
\begin{equation}
\int_\gamma\left[4\nabla_{\!\alpha'}(G^{\mu\nu}{}_{\rho'\sigma'}u^{\rho'}S^{\sigma'\alpha'})+G^{\mu\nu}{}_{\rho'\sigma'}\delta m^{\rho'\sigma'}\right]d\tau. 
\end{equation}
This can be expanded for small $r$, decomposed into multipoles, and its contribution to $\bar h^{{\rm R}\mu\nu}_{(2)}$ determined. (Its contribution to $\bar h^{{\rm S}\mu\nu}_{(2)}$ is simply the minimal $p$th solutions for $p=-2$ and $p=-1$.) At least through order $r$, the $\delta m^{\mu\nu}$ term in $\bar h^{{\rm R}\mu\nu}_{(2)}$ is found to agree with $\int_\gamma G^{{\rm R}\mu\nu}{}_{\rho'\sigma'}\delta m^{\rho'\sigma'}d\tau$. Explicitly, this consists of (i) tails of tails, since $\delta m^{\rho\sigma}$ is made up of tails, and (ii) tails multiplied by local factors of the tidal moments and acceleration. The spin term in $\bar h^{{\rm R}\mu\nu}_{(2)}$ can be expected to agree with $4\int_\gamma\nabla_{\!\alpha'}(G^{{\rm R}\mu\nu}{}_{\rho'\sigma'}u^{\rho'}S^{\sigma'\alpha'})d\tau$ at least through order $r$, but I have not performed the explicit expansion.

\subsubsection{Equation of motion}
Evaluating Eq.~\eqref{nth_acceleration} and then making use of the gauge condition $\bar\delta^0(\partial_t\bar h^{at}_{(1)}+\partial_b\bar h^{ab}_{(1)})\big|_{r=0}=0$, one finds the first-order term in the acceleration is 
\begin{equation}
ma_{(1)}^i=\frac{1}{4} m\bar\delta^0(\delta_{ab}\bar h^{abi}_{(1,1,0,1)} + \bar h^{tti}_{(1)}) + m\bar\delta^0\partial_t\bar h^{ti}_{(1)}-\B^{ij}S_j.\label{a1_full}
\end{equation}
In terms of the regular field $\bar h^{{\rm R}\mu\nu}_{(1)}$, the result reads
\begin{equation}
ma_{(1)}^i = \frac{1}{4} m\bar\delta^0 \partial^i(\bar h^{{\rm R}a}_{(1)}{}_{a} + \bar h^{{\rm R}tt}_{(1)}) + m \bar\delta^0\partial_t\bar h^{{\rm R}ti}_{(1)}
			-\B^{ij}S_j.\label{a1_R}
\end{equation}
This is the usual first-order result. The first three terms are the first-order gravitational self-force, and the last term is the Mathisson-Papapetrou spin force. 

Summing the terms in the acceleration as $a^\mu=a^\mu_{(0)}+\e a^\mu_{(1)}+O(\e^2)$ and converting from $\bar h^{{\rm R}\mu\nu}$ to $h^{{\rm R}\mu\nu}$, one arrives at
\begin{align}
a^\mu &= -\frac{1}{2}(g^{\mu\nu}+u^\mu u^\nu)(2h^{{\rm R}}_{\rho\nu;\sigma}-h^{{\rm R}}_{\rho\sigma;\nu})u^\rho u^\sigma \nonumber\\
       &\quad + \frac{1}{2m}R^\mu{}_{\nu\rho\sigma}u^\nu S^{\rho\sigma}+O(\e^2).\label{a}
\end{align}
Through order $\e$ and excluding the spin, Eq.~\eqref{a} is the geodesic equation in the effective metric $g_{\mu\nu}+h^{\rm R}_{\mu\nu}$~\cite{Poisson-Pound-Vega:11}. Through that order and with the spin, it is the equation of motion of a test body in $g_{\mu\nu}+h^{\rm R}_{\mu\nu}$. In the puncture scheme described in the preceding section, it should be used (after dropping the ``$O(\e^2)$") to self-consistently evolve the position of the body.\footnote{In fact, to be consistent with my equations for the second-order singular field, one \emph{must} use an equation of motion at least as accurate as Eq.~\eqref{a}. If instead the less accurate equation $a^\mu=0$ is used, then the mass dipole terms in the singular field cannot be set to zero, since if a worldline $\gamma$ of vanishing acceleration is chosen, then the mass dipole evolves according to~\cite{Pound:10a}
\begin{align} \label{M_evolution}
\frac{D^2}{dt^2}M^\mu &= R^\mu{}_{\nu\rho\sigma}u^\nu u^\rho M^\sigma + \frac{1}{2m}R^\mu{}_{\nu\rho\sigma}u^\nu S^{\rho\sigma}\nonumber\\
       &\quad -\frac{1}{2}(g^{\mu\nu}+u^\mu u^\nu)(2h^{{\rm R}(1)}_{\rho\nu;\sigma}-h^{{\rm R}(1)}_{\rho\sigma;\nu})u^\rho u^\sigma.
\end{align}
If one is interested in finding a solution to the second-order EFE on a very short timescale, one can use $a^\mu=0$, include the $M^i$ terms in the singular field (presented through order $r^0$ in Ref.~\cite{Pound:10a}), and evolve $M^i$ via Eq.~\eqref{M_evolution}.}

In going from Eq.~\eqref{a1_R} to Eq.~\eqref{a}, I have replaced $\e h^{{\rm R}(1)}_{\mu\nu}$ with $h^{\rm R}_{\mu\nu}$ and dropped the $\bar \delta^0$. Doing so does not alter the acceleration at order $\e$. However, recall that $\bar\delta^0$ indicates evaluation at $a^\mu=0$. This evaluation is to be performed only at time $t$, leaving the past history of $\gamma$ unchanged. If $\bar h^{{\rm R}\mu\nu}_{(1)}$ is evaluated analytically, in terms of tails, then $-\frac{1}{2}(g^{\mu\nu}+u^\mu u^\nu)(2h^{{\rm R}}_{\rho\nu;\sigma}-h^{{\rm R}}_{\rho\sigma;\nu})u^\rho u^\sigma$ includes tail terms, which depend on the past history, plus local terms involving the acceleration. In particular, there is a term $-\frac{11}{3}m \dot a^\mu$,\footnote{This is easily found by using the explicit results for $h^{{\rm R}}_{\rho\sigma}(r=0)$ and $\partial_\nu h^{{\rm R}}_{\rho\sigma}(r=0)$ in Fermi coordinates in Ref.~\cite{Pound:10b} (also reproduced in Ref.~\cite{Poisson-Pound-Vega:11}). But note that Table I in Ref.~\cite{Pound:10b} and Table II in Ref.~\cite{Poisson-Pound-Vega:11} are missing a factor of 4 from the $ma_a$ term in the quantity $\hat C_a^{(1,0)}=h^{{\rm R}}_{ta}(r=0)$. The missing 4 appears, correctly situated, in Eq.~(E.9) from the former reference and (23.10) from the latter.} which is the traditional gravitational anti-damping term found by Havas~\cite{Havas:57} (as corrected in Ref~\cite{Havas-Goldberg:62}). It might seem that going from Eq.~\eqref{a1_R} to Eq.~\eqref{a} has thus transformed a second-order ODE into a third-order, non-physical one. However, in practice no third-order ODE will arise; the regular field, not the tail field, will be found from a puncture scheme (or the various other numerical regularization schemes used at first order), and Eq.~\eqref{a} will be evolved directly, as a second-order ODE. Evidence from the scalar field case suggests that no problems will arise when performing a self-consistent evolution in this manner: when written in terms of tails, the scalar self-force involves $\dot a^\mu$ terms that would lead to runaway evolutions~\cite{Poisson-Pound-Vega:11}, but self-consistent numerical evolutions, using an equation of the form~\eqref{a}, show no evidence of runaway behavior~\cite{Diener-etal:12}. Alternatively, it should be possible to numerically evolve $a^{\mu}=\e a^\mu_{(1)}$ directly, with $a^\mu_{(1)}$ given by Eq.~\eqref{a1_full} or \eqref{a1_R}. Finally, it is important to note that the acceleration in the form~\eqref{a} would have followed directly from the gauge condition~\eqref{2nd_gauge_approx}. Therefore, assuming Eq.~\eqref{a} is well behaved, in the sense that the $\dot a^\mu$ term that is hidden within the right-hand side gives rise to no non-physical behavior, then the assumed expansion \eqref{a_expansion} of the acceleration on $\gamma$ is unnecessary: at least through the orders explicitly examined here, the assumptions on the field alone suffice to arrive at a well-behaved equation of motion. These issues will be discussed further in the sequel~\cite{Pound:12c}.


\section{Discussion}\label{discussion}

The analytical calculation of the second-order gravitational field outside a small, compact body is now complete, in the sense that sufficient information has been found to numerically obtain a global solution outside the body. I have defined a split of the second-order field into singular and regular pieces, and I have found an explicit expression for the singular field through order $r$ in a local expansion. In an earlier work~\cite{Pound:10a}, to derive the first-order equation of motion I found the second-order singular field through order $r^0$, but that is insufficient to find a global solution, and I also dropped terms with explicit acceleration dependence. The full results through order $r$, first reported in Ref.~\cite{Pound:12} and made explicit here, do suffice to obtain a global solution via a puncture scheme. Reference~\cite{Pound:12} also showed that the regular field, as I have defined it, is entirely responsible for the second-order force; further details of that derivation will be presented in the sequel~\cite{Pound:12c}. 

Note that I could have bypassed the split into singular and regular fields entirely, simply defining a puncture and a residual field. From the perspective of the equation of motion, the only matter of practical importance is that through order $r$, the residual field contains every term in the field that is smooth at $r=0$; that is, given the field in the form \eqref{hn_form}, one can simply take the $n$th-order puncture to be everything but $\phi^{\mu\nu}_{(2,0,0,0)}+r\phi^{\mu\nu i}_{(2,1,0,1)}n_i+O(r^2)$, where the choice of the $O(r^2)$ terms has no practical impact. However, the definition of the singular and regular fields through all orders in $r$ affords a prettier picture, in that the regular field, as I have defined it, is a smooth solution to the vacuum Einstein equation, and the effective metric $g_{\mu\nu}+h^{\rm R}_{\mu\nu}$ in which the motion is geodesic is therefore a physically meaningful spacetime.

More generally, I have shown how the same procedure works at arbitrary order in a large class of gauges. I defined an $n$th-order split of the field around a small body into a singular/self-field and a regular/effective field, where the singular field involves the body's multipole moments and corrections thereto, and the regular field is a smooth solution to the vacuum Einstein equation, which is determined by global boundary conditions. I also described how to use the former to find the latter (and the global solution) with a puncture scheme. In one sense, this split into self-field and effective field marks an improvement over the non-perturbative split defined by Harte~\cite{Harte:12}, since my regular field satisfies the vacuum EFE at all orders in $\e$, while his does not. However, it lacks the elegance and covariant utility of his definition, since it is defined in terms of a series solution in a particular coordinate system.

I also found the form of the body's $n$th-order stress-energy tensor as it appears (in a sense) from distances in the buffer region and beyond. And I have shown how an $n$th-order equation of motion follows from the $n$th-order gauge condition, though an appropriate constraint on the `corrected dipole moment' must be imposed to make this a meaningful and useful representation of the body's motion. 

From these considerations, one sees that a specification of the body's multipole moments (including the aforementioned dipole moments and together with initial data) is sufficient to find the physical metric everywhere outside the body and to self-consistently evolve the body's position.


\subsection{Comparison with other work}
Besides Refs.~\cite{Pound:10a,Pound:12} and the present paper, the main work done on the second-order self-force problem has been performed by Rosenthal~\cite{Rosenthal:06a,Rosenthal:06b}, Detweiler~\cite{Detweiler:12}, and Gralla~\cite{Gralla:12}. The most immediate, though relatively minor, difference between those works and the present paper is that they assume the small body is a monopole through second order, while the present work is valid for a spinning body. However, all of these studies share common aspects: gathering information about the singular behavior of the metric perturbation by matching to an inner expansion; and expressing the force in terms of some specified, regular part of the field. Here I will discuss only the methods of finding the second-order perturbation, leaving discussion of the force itself to the sequel~\cite{Pound:12c}.

Rosenthal found an analytic, global (excluding $\gamma$ itself) solution to the second-order Einstein  equation~\cite{Rosenthal:06a}, along with the general form of its divergent pieces near $\gamma$~\cite{Rosenthal:06b}. Since he focused on finding an expression for the self-force, he did not derive an explicit local expansion for a singular field that could be used within a numerical puncture scheme. However, the larger disadvantage of his approach is that his solution, and his method of solution, relies on a peculiar choice of gauge in which $\bar h^{{\rm R}\mu\nu}_{(1)}$ and $\partial_\rho\bar h^{{\rm R}\mu\nu}_{(1)}$ both vanish on $\gamma$. This means, most strikingly, that the first-order self-force vanishes. (It also means that $\delta m^{\mu\nu}$ vanishes.) In such a gauge, there is no obvious way to calculate the metric perturbation in practice. Also, the perturbation will grow large with time.

Recently, Detweiler has taken the opposite focus, not deriving an expression for the force but instead discussing general features of a second-order split into singular and regular fields~\cite{Detweiler:12}. Like the present paper, he advocates using a locally obtained singular field together with a puncture scheme.\footnote{The equation he presents for the second-order retarded field is purely formal, since it includes $\delta^2 G^{\mu\nu}[h^{(1)}]$ at all points down to $r=0$; because $\delta^2 G^{\mu\nu}[h^{(1)}]\sim1/r^4$, this has no clear distributional meaning. But the essential aspect is his effective-source equation for the regular field, which is well defined and takes the same general form as that presented here.} His discussion of the second-order problem has the advantage of being gauge-independent, though it does not provide an expression for a second-order singular field or an algorithm for constructing one. If his singular field is to be taken from the metric of a tidally perturbed black hole, as indicated in Appendix A of Ref.~\cite{Detweiler:12}, then the metric perturbation will be in an inconvenient gauge. When transforming to another gauge, one must then decide how to split the transformed perturbation into singular and regular pieces. Depending on how that is done, the motion may or may not be geodesic in $g_{\mu\nu}+h^{\rm R}_{\mu\nu}$; for example, with Gralla's choice of $h^{\rm S}_{\mu\nu}$ and $h^{\rm R}_{\mu\nu}$, the motion is not quite geodesic in $g_{\mu\nu}+h^{\rm R}_{\mu\nu}$~\cite{Gralla:12}. Of course, any choice of singular and regular fields will do, so long as one has a convenient equation of motion in terms of that particular choice. 

Besides these differences in scope, one detail of Detweiler's approach appears to be at odds with the results presented here. Based on the fact that the particle obeys the geodesic equation in $g_{\mu\nu}+\e h^{\rm R(1)}_{\mu\nu}$, he assumes that through second order, the stress-energy tensor of the body is that of a point particle in $g_{\mu\nu}+\e h^{{\rm R}(1)}_{\mu\nu}$:
\begin{equation}\label{Detweiler_Tab}
T^{\mu\nu}[h^{{\rm R}(1)}_{\rho\sigma}] = \int_\gamma mu^\mu u^\nu \frac{\delta^4(x^\alpha-z^\alpha)}{\sqrt{{\rm det}\left(g_{\rho\sigma}+\e h^{{\rm R}(1)}_{\rho\sigma}\right)}}d\tau.
\end{equation}
This expression for the stress-energy does not agree with my result for $\e T^{\mu\nu}_{(1)}+\e^2 T^{\mu\nu}_{(2)}$. One can see that the results differ simply by noting that Eq.~\eqref{Detweiler_Tab} contains no components orthogonal to the worldline,\footnote{Note that Detweiler advocates using an ordinary power series expansion rather than a self-consistent one, meaning Eq.~\eqref{Detweiler_Tab} is to be expanded not only for small $h^{{\rm R}(1)}_{\mu\nu}$, but also for $z^\mu\to z^\mu_{(0)}+\e z^\mu_{(1)}$, where $z^\mu_{(0)}$ is the geodesic initially tangential to $z^\mu$. To compare my stress-energy tensor to that expanded form, one would similarly expand $\e T^{\mu\nu}_{(1)}+\e^2T^{\mu\nu}_{(2)}$. Section~IIB of Ref.~\cite{Pound:10a} contains details of such an expansion, the resulting second-order EFE, and its formal solution, in the case where $h^{{\rm R}(1)}_{\rho\sigma}$ is replaced by $h^{(1)}_{\rho\sigma}$ in Eq.~\eqref{Detweiler_Tab}.} while Eqs.~\eqref{T2}, \eqref{dmta}, and \eqref{dmab} show that $T^{\mu\nu}_{(2)}$ does contain such components. The source of the difference is not clear. It might be a quirk in my choice of gauge; the components of the regular field, which determine the form of $\delta m^{\mu\nu}$ and therefore of $T^{\mu\nu}_{(2)}$, can be adjusted via a smooth gauge transformation. However, it is not obvious that Eq.~\eqref{Detweiler_Tab} should be the correct choice. If a metric perturbation $h_{\mu\nu}$ is taken to naively satisfy the ill-defined second-order EFE $\delta G^{\mu\nu}[h]=8\pi T^{\mu\nu}[h]-\delta^2 G^{\mu\nu}[h]$, then geodesic motion in $g_{\mu\nu}+h_{\mu\nu}$ at linear order in $h_{\mu\nu}$ formally follows from the Bianchi identity $0=\nabla_{\!\nu}(8\pi T^{\mu\nu}[h]-\delta^2 G^{\mu\nu}[h])$, as shown in, e.g, Sec.~IIB of Ref.~\cite{Pound:10a}; but if the metric perturbation is taken to naively satisfy the (also ill-defined) EFE $\delta G^{\mu\nu}[h]=8\pi T^{\mu\nu}[h^{\rm R}]-\delta^2 G^{\mu\nu}[h]$, then geodesic motion in $g_{\mu\nu}+h^{{\rm R}}_{\mu\nu}$ does not so follow from the Bianchi identity $0=\nabla_{\!\nu}(8\pi T^{\mu\nu}[h^{{\rm R}}]-\delta^2 G^{\mu\nu}[h])$. Of course, neither of these Bianchi identities are well defined on $\gamma$. So I leave the issue open.

Most recently, within the context of a power series expansion rather than a self-consistent one, Gralla~\cite{Gralla:12} has obtained results very similar to my own: an expression for a singular field accurate through order $r$ in a local coordinate system (centered on a geodesic $\gamma_0$, rather than the $\gamma$ of the present paper) a prescription for obtaining the regular remainder of the field via a puncture scheme, and a second-order equation of motion written in terms of that regular field. His method of finding the second-order field (as well as the equation of motion) begins by finding an inner expansion in a gauge that is mass-centered on the geodesic $\gamma_0$. That is sensible within the context of the short timescales of Gralla's approach, because on those timescales, the body's deviation from $\gamma_0$ is of order $\e$ or smaller, meaning the body can be translated to sit atop $\gamma_0$ via a gauge transformation. The solution in that mass-centered gauge, though it applies to any spherical body, is equivalent to typical expressions for a tidally perturbed Schwarzschild black hole, such as those in Ref.~\cite{Poisson:05}; it is also similar to the gauge used by Rosenthal. After converting this inner expansion into an outer expansion, the solution in any arbitrary, smoothly related gauge is written as the sum of the original solution plus terms involving the arbitrary gauge vector. That gauge vector is implicitly determined, up to a free choice of some initial values, by one's choice of gauge for the regular field, and it is found by solving a set of simple transport equations on $\gamma_0$. As such, Gralla's second-order field is more general than that presented here, since it is expressed in a class of gauges rather than a single gauge.

The more obvious difference is that his approximation is inherently limited to short times. To determine long-term effects it will need to be combined with a scheme such as the two-timescale expansion of Hinderer and Flanagan~\cite{Hinderer-Flanagan:08}. In Gralla's second-order field this limitation manifests as secularly growing mass dipole terms representing the body's movement away from $\gamma_0$. These are the terms involving $A^i$ (corresponding to $M^i$ in my notation) in his Eqs.~(B4) and (B6); in my approach, $\gamma$ is chosen to ensure that these terms vanish. Less notably, as mentioned above, his choice of singular and regular fields differs from my own and leads to an equation of motion that is not quite the expanded geodesic equation in $g_{\mu\nu}+h^{{\rm R}}_{\mu\nu}$. To determine whether his results agree with mine, one would have to find a gauge transformation from the Lorenz gauge to his class of gauges and perform an expansion of the worldline $\gamma$ about a geodesic $\gamma_0$. Alternatively, one could avoid the expansion of the worldline by setting $a^\mu$ to zero in all my expressions and reinserting the mass dipole terms that were set to zero (through order $r^0$, these terms can be found in Ref.~\cite{Pound:10a}). Doing so would immediately yield the metric centered on $\gamma_0$. In either case, given the very similar underlying methods, disagreement is unlikely.

\subsection{Future work}
At second order, in the case of a compact body with slow internal dynamics, the present paper (together with its sequel~\cite{Pound:12c}) and that of Gralla essentially represent a complete solution to the analytical portion of the problem. The only obvious analytical work remaining in my treatment is to find a closed form expression for the singular and regular fields (or for two fields that agree with them through order $r$)---in other words, to find second-order equivalents to the Detweiler-Whiting fields. One might do this by finding suitable quadratic combinations of first-order fields, as in Ref.~\cite{Rosenthal:06a}, to make up a singular field. As mentioned in Sec.~\ref{2nd_puncture}, to find a closed form for the regular field, one could write the retarded solution in integral form outside a worldtube embedded in the buffer region and then devise a decomposition of the original integrals into two pieces, such that one of them, when expanded for small $r$, contributes to the $O(r^0)$ and $O(r)$ pieces of $\bar h^{{\rm R}\mu\nu}_{(2)}$ and the other does not. Details of such integral representations and their expansion are contained in Ref.~\cite{Pound:10a}.

Closed-form expressions might allow numerical alternatives to the puncture scheme. And they would aid in comparing my results with other analytical methods, such as those of Harte~\cite{Harte:12} and Galley~\cite{Galley:12}, once those other methods are applied at second order. However, this is not strictly necessary. So at second order there is only one major remaining goal: to numerically implement a puncture scheme. Using the results of this paper alone, one can implement such a scheme by simultaneously solving Eqs.~\eqref{1st_effective_wave} and \eqref{2nd_effective_wave} for the regular field in a region effectively covering the body (in the sense described in the introduction), Eqs.~\eqref{1st_wave} and \eqref{2nd_wave} for the full field everywhere else, and Eq.~\eqref{a} for the position of the body. Doing so would yield a solution to the EFE through second order, since the relaxed EFE \eqref{wave_equation} and the gauge condition \eqref{generalized_gauge} would be satisfied up to third-order errors. However, to obtain a solution accurate on longer timescales, a preferable prescription would be to replace \eqref{a} with a second-order-accurate equation of motion. For the case of a non-spinning body, that equation is given by Eq.~(17) in Ref.~\cite{Pound:12}. Further details of its derivation will be presented in the sequel~\cite{Pound:12c}.

At $n$th order, much work remains. Alternative gauges could be explored. A different choice of field variable $\phi^{\mu\nu}$ and gauge terms $Z^{\mu}{}_{\nu\rho}$ and $D^\mu$ might simplify the equations. In particular, one might find a combination of $\phi^{\mu\nu}$, $Z^\mu{}_{\nu\rho}$, and $D^\mu$ to eliminate logarithmic terms from the solution, as in post-Newtonian theory~\cite{Blanchet:87}. Among other benefits, this would allow one to construct a strict power series expansion in a generalized wave gauge.

One should be able to explicitly relate the multipole moments $I^{\mu\nu L}$ (including the corrections to the body's unperturbed moments) in the outer expansion to properties of the matter and gravity inside the body, analogous to how radiative multipole moments are related to the source moments in post-Newtonian theory~\cite{Blanchet:06}. This might be most easily done in the case of a material body, where the moments could be expressed as integrals over the body's interior. In particular, one would like to ensure that the corrected mass dipole provides a meaningful measure of the center of energy relative to the worldline $\gamma$. It may be helpful to relate the corrected mass dipole to a derivative of some linear momentum, defined as an integral over a surface around the body~\cite{Thorne-Hartle:85}. 

Mass dipole aside, concrete relationships between the multipole moments and properties of the body would render their meaning more transparent and aid in formulating physically realistic models for their magnitude and evolution. Models for both the spin and quadrupole moments will be of use in modeling binaries, since those moments appear in the equations of motion at the same order as the first- and second-order self-force. See Ref.~\cite{Steinhoff-Puetzfeld:12} for a recent discussion of modeling the quadrupole moment of the small body in an EMRI.

One could also consider a wider class of small bodies. The assumed inner expansion, in which all lengths are scaled by $\e$ and time is unscaled relative to the external scales, restricts the approximation scheme to compact bodies with slow internal dynamics. The single scaling of lengths means that the body's linear size $d$ is of the same order as its mass. This makes its multipole moments scale as $m^{\ell+1}$, as opposed to the more natural scaling $m d^\ell$. From the perspective of the outer expansion, one could simply let the multipole moments scale with independent powers of $\e$ to describe a more generic body. However, there would still be an underlying limitation in the approximation scheme, since it assumes the buffer region is outside the body; if the body is very diffuse, such that $d\gtrsim\mathcal{R}$, then the scheme cannot be expected to be accurate. Not scaling time in the inner expansion means that no functions in it depend on the fast timescale $t/\e$ in the way they depend on the short lengthscale $r/\e$. Allowing fast internal dynamics in the body would result in rapid oscillations in the metric. Removing this restriction would likely require a more intensive and delicate approach to the outer expansion.  

Finally, the outer expansion cannot be expected to be valid on an infinite domain (though it may turn out to be). Since in the end we seek the waveforms produced by the body, we may need to match the outer expansion to an outgoing-wave solution at infinity. 

\begin{acknowledgments}
I wish to thank Leor Barack for helpful discussions. This work was supported by the Natural Sciences and Engineering Research Council of Canada.
\end{acknowledgments}

\appendix

\section{Second-order curvature tensors}\label{second-order_tensors}
The second-order Ricci tensor $\delta^2R_{\alpha\beta}$ corresponding to a metric perturbation $h_{\mu\nu} $is given by
\begin{align}
\delta^2R_{\alpha\beta} &=-\tfrac{1}{2}\bar h^{\mu\nu}{}_{;\nu}\left(2h_{\mu(\alpha;\beta)}-h_{\alpha\beta;\mu}\right) 
					+\tfrac{1}{4}h^{\mu\nu}{}_{;\alpha}h_{\mu\nu;\beta}\nonumber\\
					&\quad +\tfrac{1}{2}h^{\mu}{}_{\beta}{}^{;\nu}\left(h_{\mu\alpha;\nu} -h_{\nu\alpha;\mu}\right)\nonumber\\
					&\quad-\tfrac{1}{2}h^{\mu\nu}\left(2h_{\mu(\alpha;\beta)\nu}-h_{\alpha\beta;\mu\nu}-h_{\mu\nu;\alpha\beta}\right).\label{second-order_Ricci}
\end{align}
The second-order Einstein tensor with indices up can be obtained using
\begin{align}
\delta^2G^{\alpha\beta}&=\delta\left\lbrace\delta\left[\left(g^{\alpha\mu}g^{\beta\nu}-\frac{1}{2}g^{\alpha\beta}g^{\mu\nu}\right)R_{\mu\nu}\right]\right\rbrace\\
					&= \left(h^{\alpha\beta}g^{\mu\nu}+g^{\alpha\beta}h^{\mu\nu}
					-2h^{\alpha\mu}g^{\beta\nu}-2g^{\alpha\mu}h^{\beta\nu}\right)\delta R_{\mu\nu}\nonumber\\
					&\quad +\left(g^{\alpha\mu}g^{\beta\nu}-\frac{1}{2}g^{\alpha\beta}g^{\mu\nu}\right)\delta^2 R_{\mu\nu},\label{second-order_Einstein}
\end{align}
where I have specialized to a Ricci-flat background and utilized the identity $\delta (g^{\mu\nu}) = -h^{\mu\nu}$. $\delta^2G^{\alpha\beta}$ can be written as a functional of two arguments, $\delta^2G^{\alpha\beta}[h,h]$, such that, for example,
\begin{align}
\delta^2G^{\alpha\beta}&[h^{\rm S}+h^{\rm R},h^{\rm S}+h^{\rm R}] \nonumber\\
		&= \delta^2G^{\alpha\beta}[h^{\rm S},h^{\rm S}]
		+\delta^2G^{\alpha\beta}[h^{\rm S},h^{\rm R}]\nonumber\\
		&\quad +\delta^2G^{\alpha\beta}[h^{\rm R},h^{\rm S}]+\delta^2G^{\alpha\beta}[h^{\rm R},h^{\rm R}].
\end{align}
With an abuse of notation, I write $\delta^2G^{\alpha\beta}[h]=\delta^2G^{\alpha\beta}[h,h]$ when both arguments are the same. With a greater abuse of notation, I write $\delta^2G^{\alpha\beta}[\phi]$ to denote $\delta^2G^{\alpha\beta}[h[\phi]]$, meaning $h_{\mu\nu}$ is replaced by its expression in terms of $\phi_{\mu\nu}$. For example, $\delta^2G^{\alpha\beta}[\bar h]$ denotes the right-hand side of Eq.~\eqref{second-order_Einstein} with $h_{\mu\nu}$ replaced by $h_{\mu\nu}[\bar h]=\bar h_{\mu\nu}-\frac{1}{2}g_{\mu\nu}g^{\rho\sigma}\bar h_{\rho\sigma}$.

Analogously, $\delta^n G^{\alpha\beta}$ denotes the piece of $G^{\alpha\beta}[g_{\mu\nu}+h_{\mu\nu}]$ that contains $n$ factors of $h_{\mu\nu}$ and its derivatives, and I write it as $\delta^n G^{\alpha\beta}[h,...,h]$ or $\delta^n G^{\alpha\beta}[h]$.

\section{STF decompositions}\label{STF_decomposition}
I here reproduce standard formulas from Refs.~\cite{Damour-Blanchet:86, Damour-Iyer:91}; Eq.~\eqref{identity} is the only identity not contained therein.

Any Cartesian tensor $T^S(\theta,\phi)$ on a sphere can be expanded along STF combinations of unit vectors as
\begin{equation}
T^S(\theta,\phi)=\sum_{\ell\geq0}T^{S\langle L\rangle}\nhat_L,\label{nhat_decomposition}
\end{equation}
where the coefficients are given by 
\begin{equation}
T^{S\langle L\rangle} = \frac{(2\ell+1)!!}{4\pi\ell!}\int T^S(\theta,\phi)\nhat^Ld\Omega,\label{lth_coeff}
\end{equation}
where $x!!=x(x-2)\cdots1$.

These coefficients can then be put into an irreducible form. For example, for $s=1$, we have
\begin{equation}\label{s=1_decomposition}
T^{a\langle L\rangle} = \hat T_{(+)}^{aL}+\epsilon^{ja\langle i_\ell}\hat T_{(0)}^{L-1\rangle}{}_j+\delta^{a\langle i_\ell}\hat T_{(-)}^{L-1\rangle},
\end{equation}
where the $\hat T^{(n)}$'s are STF tensors given by
\begin{align}
\hat T_{(+)}^{L+1} & \equiv T^{\langle L+1\rangle}, \\
\hat T_{(0)}^{L} & \equiv \frac{\ell}{\ell+1}T^{pq\langle L-1}\epsilon^{i_\ell\rangle}{}_{pq}, \\
\hat T_{(-)}^{L-1} & \equiv \frac{2\ell-1}{2\ell+1}T_j{}^{jL-1}.
\end{align} 
Similarly, for a symmetric tensor $T_S$ with $s=2$, we have
\begin{align}\label{s=2_decomposition}
T_{ab\langle L\rangle} & = \delta_{ab}\hat K_L+\hat T^{(+2)}_{abL}\nonumber\\
						&\quad +\mathop{\STF}_L\mathop{\STF}_{ab}\Big(\epsilon^p{}_{ai_\ell}\hat T^{(+1)}_{bpL-1}+\delta_{ai_\ell}\hat T^{(0)}_{b L-1}\nonumber\\
						&\quad +\delta_{a i_\ell}\epsilon^p{}_{bi_{\ell-1}}\hat T^{(-1)}_{pL-2} +\delta_{ai_\ell}\delta_{bi_{\ell-1}}\hat T^{(-2)}_{L-2}\Big),
\end{align}
where{\allowdisplaybreaks
\begin{align}
\hat T^{(+2)}_{L+2} & \equiv T_{\langle L+2\rangle}, \\
\hat T^{(+1)}_{L+1} & \equiv \frac{2\ell}{\ell+2}\mathop{\STF}_{L+1}(T_{\langle pi_\ell\rangle qL-1}\epsilon_{i_{\ell+1}}{}^{pq}), \\
\hat T^{(0)}_L & \equiv \frac{6\ell(2\ell-1)}{(\ell+1)(2\ell+3)}\mathop{\STF}_L(T_{\langle ji_\ell\rangle}{}^j{}_{L-1}), \\
\hat T^{(-1)}_{L-1} & \equiv \frac{2(\ell-1)(2\ell-1)}{(\ell+1)(2\ell+1)}\mathop{\STF}_{L-1}(T_{\langle jp\rangle q}{}^j{}_{L-2}\epsilon_{i_{\ell-1}}{}^{pq}), \\
\hat T^{(-2)}_{L-2} & \equiv \frac{2\ell-3}{2\ell+1}T_{\langle jk\rangle}{}^{jk}{}_{L-2} \\
\hat K_L & \equiv \tfrac{1}{3}T^j{}_{jL}.
\end{align}}
These decompositions are equivalent to the formulas for addition of angular momenta, $J=S+L$, which results in terms with angular momentum $\ell-s\leq j\leq \ell+s$; the superscript labels $(\pm n)$ in these formulas indicate by how much each term's angular momentum differs from $\ell$.

When manipulating quantities of the form \eqref{nhat_decomposition}, the following identities are useful:
\begin{align}
n_c\nhat_L &= \nhat_{cL}+\frac{\ell}{2\ell+1}\delta_{c\langle i_\ell}\nhat_{L-1\rangle}, \label{nnhat}\\
r\partial_c\nhat_L &= -\ell\nhat_{cL} + \frac{\ell(\ell+1)}{2\ell+1}\delta_{c\langle i_\ell}\nhat_{L-1\rangle}. \label{PDnhat}
\end{align}
From the latter, one finds that $\nhat_L$ is an eigenvector of the flat-space Laplacian, satisfying
\begin{equation}
r^2\partial^i\partial_i \nhat_L = -\ell(\ell+1)\nhat_L.
\end{equation}

Equation \eqref{lth_coeff} is related to the integral identity 
\begin{equation}
\frac{1}{4\pi}\int n^S\nhat^L d\Omega = \mathop{\STF}_L\frac{\delta_{\lbrace i_1i_2}\cdots\delta_{i_{s+\ell-1}i_{s+\ell}\rbrace}}{(s+\ell+1)!!} 
\end{equation}
if $s+\ell$ is even and zero otherwise. Here the curly braces indicate the smallest symmetric (unnormalized) combination of the enclosed indices. With a bit of combinatorics one can use this identity to show
\begin{equation}
A_S\hat B_L\frac{1}{4\pi}\int n^S\nhat^L d\Omega = \frac{s!A_{i_1}{}^{i_1}{}_{\cdots i_p}{}^{i_p}{}_L\hat B^L}{(s-\ell)!(s+\ell+1)!!},\label{identity}
\end{equation}
for a symmetric tensor $A_S$ and an STF tensor $\hat B_L$ satisfying $\ell=s-2p$, where $p$ is a nonnegative integer. If $\ell\neq s-2p$, then the left-hand side vanishes.

\section{Normalization of logarithms}\label{logarithms}
In my ansatz \eqref{hn_form} for the field, I write the logarithms with no normalization in their argument. In actuality, they must read $\ln\frac{r}{r_0}$, where $r_0$ must be determined by boundary conditions. At least at low orders, the normalization can be expected to be $r_0=2\e m$, which is what would arise from the effect of the mass on the shape of the light cones (i.e., the shift from $u=t-r$ to $u=t-r^*$ as a lightcone coordinate). However, within the present context, the actual choice is arbitrary. Consider the case when $C^{\mu\nu}_{(n)}$ contains no logarithms. Then a logarithm first arises in the solution when the source for the Poisson equation~\eqref{nth_Poisson} contains a mode with $\ell=\ell_p$; all other logarithms will either arise in the same way, or be sourced by logarithms that arose in that way. So we may consider those logarithms alone. Suppose we begin with a normalization $r_0$ but then change to $r'_0$. Then $r^p\nhat_{L_p}\ln(r/r_0)\phi^{\mu\nu L_p}_{np1\ell_p}$ is replaced with $r^p\nhat_{L_p}\ln(r/r'_0)\phi^{\mu\nu L_p}_{np1\ell_p}+r^p\nhat_{L_p}\ln(r'_0/r_0)\phi^{\mu\nu L_p}_{np1\ell_p}$. The term $\ln(r'_0/r_0)\phi^{\mu\nu L_p}_{np1\ell_p}$ is then simply absorbed into the homogeneous mode $\phi^{\mu\nu L_p}_{np0\ell_p}$. Therefore the normalization $r_0$ does not introduce additional freedom into the solution; the homogeneous modes carry all the freedom. The value of $r_0$ will  effectively be determined by determining the homogeneous modes, either by specifying the material composition of the body to find its multipole moments or by specifying global boundary conditions to find the pieces of the regular field.

For the purposes of numerical calculations, one would likely use a normalization $r_0\sim\mathcal{R}$ (say, $r_0=M$ in the case of an EMRI). Since the `natural' normalization can be expected to be $2\e m$, terms proportional to $\ln\e$ will likely appear in the homogeneous modes $\bar h^{\mu\nu}_{(2,0,0,0)}$ and $\bar h^{\mu\nu}_{(2,1,0,1)}$, corresponding to the logarithmic terms $\ln r\bar h^{\mu\nu}_{(2,0,1,0)}$ and $r\ln r\bar h^{\mu\nu i}_{(2,1,1,1)}n_i$ in the second-order field. This would mean $\ln\e$ terms would appear in the regular field $\bar h^{{\rm R}\mu\nu}_{(2)}$. Whether or not that is the case can be determined analytically by evaluating the integrals in Eq.~\eqref{formal_solution}. But note that these potential $\ln\e$ terms, and any that appear at higher order in $\e$, would not disrupt the ordering of the perturbations: $\e^n(\ln\e)^q\gg\e^{n'}(\ln\e)^{q'}$ for any $n'>n$, regardless of the values of $q$ and $q'$.

Note that Ref.~\cite{Gralla:12} disallows $\ln r$ terms in $\bar h^{\mu\nu}_{(n)}$ because they will correspond to $\ln\e$ terms in the inner expansion, where $\ln r$ would become $\ln(\e\tilde r)$. If the logarithms occur due to the standard retardation effects just described, then the $\ln\e$ terms appear instead in the outer expansion. Here, since I assume only the existence of a certain asymptotic series, rather than smoothness in $\e$ at $\e=0$, and since the logarithms are generic, I do not disallow them. A wave gauge in which they do not appear would certainly be advantageous, however, particularly at orders where logarithmic terms might lead to $\ln\e$ terms in a multipole moment $I^{\mu\nu L}_{(n)}$, potentially obscuring the moment's meaning.

\begin{widetext}
\section{Second-order singular field}\label{singular_field}
The second-order singular field can be written as the sum of three pieces, $\bar h^{{\rm S}\mu\nu}_{(2)}={}_{(-2)}\bar h^{\mu\nu}_{(2)} +{}_{(-1)}\bar h^{\mu\nu}_{(2)}+\bar h^{{\rm IH}\mu\nu}_{(2)}$. Those three pieces are given below through order $r$, in Fermi-Walker coordinates with the mass dipole set to zero. I have also made the second-order field available in a Mathematica notebook at http://www.personal.soton.ac.uk/ap8e11/second-order-fields.html. The notebook utilizes the tensor-manipulation package xAct.
\begin{align}
{}_{(-2)}\bar h^{tt}_{(2)} &= \left(2\epsilon_{ab}{}^cS^ba_{c,t}-\tfrac{12}{5} S^b \B_{ab}\right)n^a +\tfrac{2}{3}S^a\B^{bc}\nhat_{abc}
							+r\Big(\tfrac{4}{3}S^aa^b\B_{ba}-\tfrac{2}{3}S^aa^b\epsilon_{ba}{}^ca_{c,t}+7S^aa^b\epsilon_{ac}{}^da_{d,t}\nhat_b{}^c \nonumber\\
							&\quad + 2S^aa^b\epsilon_{bac}a_{d,t}\nhat^{cd}+\tfrac{19}{21}S^aa^b\B_b{}^c\nhat_{ac}+\tfrac{152}{21}S^aa^b\B_a{}^c\nhat_{bc}
							-\tfrac{22}{21}a^bS_b\B^{ac}\nhat_{ac}-\tfrac{8}{7}S^b\B_{bac}\nhat^{ac} 
							+\tfrac{1}{3}S^b\dot\E^{ac}\epsilon_{bcd}\nhat_a{}^d\nonumber\\
							&\quad +\tfrac{4}{9}S^b\B^{acd}\nhat_{bacd}+\tfrac{25}{9}S^aa^b\B^{cd}\nhat_{bacd}\Big)+O(r^2),
\end{align}
\begin{align}
{}_{(-2)}\bar h^{ta}_{(2)} &= - \frac{2\epsilon^{aij}S_jn_i}{r^2}+ \frac{3 \epsilon^{ajd}S_{d}a^i\hat{n}_{ij}}{r}
							+\left(\tfrac{1}{3} S^c\E_{b}{}^d\epsilon^{a}{}_{cd}-\tfrac{49}{15}S^c\E^{ad}\epsilon_{bcd}\right)n^b
							+3 a^aa^b S^d\epsilon_{bcd}n^c+\tfrac{1}{3}S^b\E^{cd}\epsilon_{bc}{}^i\nhat^a{}_{di}\nonumber\\
							&\quad -S^b\E^{cd}\epsilon^a{}_b{}^i\nhat_{cdi} +\tfrac{15}{4} a^ba^cS^d\epsilon^a{}_d{}^i\nhat_{bci}
							+r\Big(\tfrac{1}{4}S^b\epsilon^a{}_{bc}a_{d,tt}\nhat^{cd}+\tfrac{1}{3}S^b\epsilon^a{}_b{}^ca_{c,tt}
							-\tfrac{1}{18}S^b\dot\B^{cd}\nhat^a{}_{bcd}-\tfrac{11}{63}S^a\dot\B^{bc}\nhat_{bc}\nonumber\\
							&\quad +\tfrac{34}{63}S^b\dot\B_b{}^c\nhat^a{}_c+\tfrac{41}{63}S^b\dot\B^{ac}\nhat_{bc}+\tfrac{16}{15}S^b\dot\B^a{}_b
							-\tfrac{1}{2}a^bS^c\E^{di}\epsilon_{cd}{}^j\nhat^a{}_{bij}+\tfrac{5}{2}a^bS^c\E^{di}\epsilon^a{}_c{}^j\nhat_{bdij}
							+\tfrac{3}{7}a^aS^b\E^{cd}\epsilon_{bdi}\nhat_c{}^i\nonumber\\
							&\quad +\tfrac{2}{21}a^bS^c\E^{di}\epsilon_{bci}\nhat^a{}_d
							+\tfrac{3}{7}a^bS^c\E_b{}^d \epsilon_{cdi}\nhat^{ai}+\tfrac{13}{42}a^bS^c\E^{di}\epsilon^a{}_{ci}\nhat_{bd}
							+\tfrac{69}{14}a^bS^c\E^{ad}\epsilon_{cdi}\nhat_b{}^i+\tfrac{17}{21}a^bS^c\E^{di}\epsilon^a{}_{bc}\nhat_{di}\nonumber\\
							&\quad +\tfrac{25}{84}a^bS^c\E_b{}^d\epsilon^a{}_{ci}\nhat_d{}^i-\tfrac{7}{3}a^bS^c\E^{ad}\epsilon_{bci}\nhat_d{}^i
							+\tfrac{2}{15}a^bS^c\E_b{}^d\epsilon^a{}_{cd}+\tfrac{191}{45}a^bS^c\E^{ad}\epsilon_{bcd}
							+\tfrac{1}{6}S^b\epsilon_{bc}{}^j\E^{cdi}\nhat^a{}_{dij}\nonumber\\
							&\quad-\tfrac{1}{4}S^b\epsilon^a{}_b{}^j\E^{cdi}\nhat_{cdij}+\tfrac{1}{6}S^b\epsilon^a{}_{bi}\E_{cd}{}^i\nhat^{cd}
							+\tfrac{61}{42}S^b\E^a{}_c{}^i\epsilon_{bdi}\nhat^{cd}-\tfrac{35}{8}a^ba^ca^dS^i\epsilon^a{}_i{}^j\nhat_{bcdj}
							+\tfrac{15}{2}a^aa^ba^cS^d\epsilon_{cdi}\nhat_b{}^i\Big)\nonumber\\
							&\quad +O(r^2),
\end{align}
\begin{align}
{}_{(-2)}\bar h^{ab}_{(2)} &= \left(2S^d a^{(a}{}_{,t}\epsilon^{b)}{}_{cd}+\tfrac{4}{5}S^d \delta^{ab} \B_{cd}-\tfrac{24}{5}S^{(a}\B^{b)}{}_{c}
							+\tfrac{12}{5}S_c\B^{ab}\right)n^c+\tfrac{8}{5} S^c\B^{(a}{}_cn^{b)} 
							+2\left(\B^{cd}S^{(a}+S^c\B^{d(a}\right)\nhat^{b)}{}_{cd}\nonumber\\
							&\quad -2 S^c\B_c{}^d\nhat^{ab}{}_d  -2S^c\delta^{ab}\B^{di}\nhat_{cdi}+r\Big(\tfrac{16}{5}\B^{(a}{}_cS^{b)}a^c
							+\tfrac{58}{21}\B_c{}^da^cS^{(a}\nhat^{b)}{}_d+\tfrac{33}{7}\B^{cd}a^{(a}S^{b)}\nhat_{cd}
							-\tfrac{95}{21}\B^{d(a}S^{b)}a^c\nhat_{cd}\nonumber\\ 
							&\quad-\tfrac{64}{21}S^{(a}\B^{b)}{}_{cd}\nhat^{cd}+\tfrac{8}{9}\B^{cdi}S^{(a}\nhat^{b)}{}_{cdi}
							+\tfrac{59}{9}\B^{di}a^cS^{(a}\nhat^{b)}{}_{cdi}-\tfrac{8}{5}\B^{ab}a^cS_c+\tfrac{22}{21}a^cS_c\B^{d(a}\nhat^{b)}{}_d
							-\tfrac{41}{21}\B^{di}a^cS_c\delta^{ab}\nhat_{di}\nonumber\\
							&\quad-\tfrac{16}{9}\B^{ab}{}_cS^c+\tfrac{2}{5}\dot\E^{d(a}\epsilon^{b)}{}_{cd}S^c+\tfrac{4}{15}\B^{(a}{}_ca^{b)}S^c
							+\tfrac{2}{7}\dot\E^{di}S^c\epsilon^{(a}{}_{ci}\nhat^{b)}{}_d-\tfrac{31}{7}\B_c{}^dS^ca^{(a}\nhat^{b)}{}_d
							+\tfrac{32}{21}S^c\B^{(a}{}_{cd}\nhat^{b)d}\nonumber\\
							&\quad+\tfrac{4}{3}S^c\epsilon_{cdi}\dot\E^{d(a}\nhat^{b)i}+\tfrac{29}{21}S^c\B^{d(a}a^{b)}\nhat_{cd}
							+\tfrac{8}{7}\B^{ab}{}_dS^c\nhat_c{}^d-\tfrac{22}{21}\dot \E^{d(a}\epsilon^{b)}{}_{ci}S^c\nhat_d{}^i
							-\tfrac{2}{3}\dot\E^{di}\epsilon_{cij}\delta^{ab}S^c\nhat_d{}^j+\tfrac{8}{21}\B_{cdi}\delta^{ab}S^c\nhat^{di}\nonumber\\
							&\quad-\tfrac{8}{9}\B_c{}^{di}S^c\nhat^{ab}{}_{di}+\tfrac{8}{9}S^c\B^{di(a}\nhat^{b)}{}_{cdi}
							+\tfrac{2}{9}\dot\E^{di}S^c\epsilon^{(a}{}_{cj}\nhat^{b)j}{}_{di}-\tfrac{8}{9}\B^{dij}S^c\delta^{ab}\nhat_{cdij}
							-\tfrac{8}{5}\B_{cd}a^c\delta^{ab}S^d-\tfrac{22}{21}\B_{cd}a^cS^d\nhat^{ab}\nonumber\\
							&\quad+\tfrac{13}{21}a^cS^d\B^{(a}{}_d\nhat^{b)}{}_c
							+\tfrac{22}{21}S^da^c\B^{(a}{}_c\nhat^{b)}{}_d-\tfrac{55}{21}\B^{ab}a^cS^d\nhat_{cd}
							+\tfrac{19}{7}\B_d{}^iS^da^c\delta^{ab}\nhat_{ci}-\tfrac{26}{21}\B_c{}^ia^c\delta^{ab}S^d\nhat_{di}\nonumber\\
							&\quad-\tfrac{59}{9}\B_d{}^ia^cS^d\nhat^{ab}{}_{ci}+\tfrac{59}{9}a^cS^d\B^{i(a}\nhat^{b)}{}_{cdi}
							-\tfrac{11}{9}\B^{ij}a^c\delta^{ab}S^d\nhat_{cdij}+3a^cS^da^{(a}{}_{,t}\epsilon^{b)}{}_{di}\nhat_c{}^i
							+\tfrac{2}{3}a^{(a}\epsilon^{b)}{}_c{}^dS^c a_{d,t}\nonumber\\
							&\quad+\tfrac{1}{2}a^{(a}\epsilon^{b)}{}_{cd}a_{i,t}S^c\nhat^{di}\Big)+O(r^2),
\end{align}
\begin{align}
{}_{(-1)}\bar h^{tt}_{(2)} &= \frac{\delta m^{tt}}{r}-\tfrac{5}{2}\delta m^{tt}a^an_a +r\Big(\tfrac{1}{2}\delta m^{tt}{}_{,tt}+2 a^b\delta m^{t}{}_{b,t}
							+\delta m^{tb}a_{b,t}+\tfrac{7}{3} \delta m^{tt}a_ba^b-\delta m^{ba}\E_{ba}+a^aa^b \delta m_{ba}\nonumber\\
						&\quad +\tfrac{35}{8}\delta m^{tt}a^aa^b\nhat_{ba}-\tfrac{1}{3}\delta m^{tb}\B^{ac}\epsilon_{bcd} \nhat_{a}{}^d
							-\tfrac{7}{6}\delta m^{tt}\E^{ba}\nhat_{ba}\Big)+O(r^2),
\end{align}
\begin{align}
{}_{(-1)}\bar h^{ta}_{(2)} &= \frac{\delta m^{ta}}{r}-\tfrac{3}{2}\delta m^{ta}a^bn_b+r\Big(a^b\delta m^a{}_{b,t}+\tfrac{1}{2}\delta m^{ab}a_{b,t}
							+\tfrac{1}{2}\delta m^{ta}{}_{,tt}-\tfrac{17}{18}\delta m^{tb}\E^a{}_b+\B^{bc}\delta m_b{}^d\epsilon^a{}_{cd}
							+a^a\delta m^{tt}{}_{,t}\nonumber\\
						&\quad +\tfrac{1}{2}\delta m^{tt}a^a{}_{,t}+\tfrac{3}{2}a^a\delta m^{tb}a_b+\delta m^{ta}a_ba^b
							+\tfrac{1}{6}\delta m^{ad}\B^{bc}\epsilon_{cdi}\nhat_{b}{}^i+\tfrac{1}{2}\delta m^{tt}\B^{bc}\epsilon^a{}_{cd}\nhat_b{}^d
							-\tfrac{1}{6}\delta m^{tb}\E_b{}^c\nhat^a{}_c \nonumber\\
						&\quad-\tfrac{1}{6}\delta m^{tb}\E^{ac}\nhat_{bc}
							-\tfrac{1}{2}\delta m^{ta}\E^{bc}\nhat_{bc}+\tfrac{15}{8}\delta m^{ta}a^ba^c\nhat_{bc}\Big)+O(r^2),
\end{align}
\begin{align}
{}_{(-1)}\bar h^{ab}_{(2)} &= \frac{\delta m^{ab}}{r}-\tfrac{1}{2} a^c\delta m^{ab}n_c+r\Big(\tfrac{1}{2}\delta m^{ab}{}_{,tt}
							+\delta m^{t(a}\epsilon^{b)}{}_{di}\B^{cd}\nhat_c{}^i-2\delta m^{tc}\B^{d(a}\epsilon^{b)}{}_{cd}
							-\delta^{ab}\delta m^{cd}\E_{cd}-\tfrac{1}{3}\E_c{}^d\delta m^{c(a}\nhat^{b)}_d\nonumber\\
						&\quad  -\tfrac{1}{3}\E^{d(a}\delta m^{b)c}\nhat_{cd}+\tfrac{1}{6}\delta m^{ab}\E^{cd}\nhat_{cd}+\tfrac{3}{8}a^ca^d\delta m^{ab}\nhat_{cd}
							-\delta m^c{}_c\E^{ab}+\tfrac{19}{9}\E^{(a}{}_c\delta m^{b)c}+a^ca^{(a}\delta m^{b)}{}_c\nonumber\\
						&\quad -\delta m^{tt}\E^{ab}+2a^{(a}\delta m^{b)t}{}_{,t}+\delta m^{t(a}a^{b)}{}_{,t}+\delta m^{tt}a^aa^b\Big)+O(r^2),
\end{align}
\begin{align}
\bar h^{{\rm IH}tt}_{(2)} &= \frac{3 m^2}{r^2}-\frac{1}{r}\left(17m^2 a^in_i + m\bar{h}^{{\rm R}ij}_{(1)}\hat{n}_{ij}\right)
							-m\left(\tfrac{11}{5}\bar h^{\rm R}_{(1)a}{}^b{}_{,b} + \tfrac{1}{10}\bar h^{\rm R}_{(1)b}{}^b{}_{,a}
							+\bar h^{{\rm R}t}_{(1)a,t}\right)n^a + ma^a\bar h^{{\rm R}}_{(1)ab}n^b\nonumber\\
						&\quad  + m\left(\tfrac{3}{2}\bar h^{{\rm R}tt,a}_{(1)}+\bar h^{{\rm R}tt}_{(1)}a^a
							-\tfrac{1}{3}a^a\bar h^{{\rm R}b}_{(1)}{}_{b}\right)n_a+m^2\left(\tfrac{113}{3}a^aa^b
							-\tfrac{7}{3}\E^{ab}\right)\nhat_{ab}+m\left(\tfrac{5}{2}a^a\bar h^{{\rm R}bc}_{(1)}
							-\tfrac{1}{2}\bar h^{{\rm R}ab,c}_{(1)}\right)\nhat_{abc}\nonumber\\
						&\quad+\ln(r)\tfrac{20}{3} m^2 a{}_{a} a{}^{a} +r\bigg[\tfrac{1}{252}m \left(150\bar h^{{\rm R}}_{(1)}{}^{ac}{}_{,b} a{}^{b} 
							+ 90\bar h^{{\rm R}}_{(1)}{}_{b}{}^{a,c} a{}^{b} 
							+ 315 a{}_{b} a{}^{b} \bar h^{{\rm R}}_{(1)}{}^{ac} + 644\mathcal{E}{}^{ba} \bar h^{{\rm R}}_{(1)}{}_{b}{}^{c} 
							- 1344a{}^{a} a{}^{b} \bar h^{{\rm R}}_{(1)}{}_{b}{}^{c}\right) \hat{n}{}_{ac} \nonumber\\
						&\quad + \tfrac{1}{72}m\Bigl(108 \bar h^{{\rm R}tt}_{(1)}{}^{,ba} + \mathcal{E}{}^{ba} \bigl(-76 \bar h^{{\rm R}}_{(1)}{}^{c}{}_{c} 
							+ 96 \bar h^{{\rm R}tt}_{(1)}\bigr) + 3 a{}^{a} \bigl(-42 \bar h^{{\rm R}tt}_{(1)}{}^{,b} 
							+ a{}^{b} (44\bar h^{{\rm R}}_{(1)}{}^{c}{}_{c} - 68\bar h^{{\rm R}tt}_{(1)})\bigr)\Bigr) \hat{n}{}_{ba}\nonumber\\
						&\quad + \tfrac{5}{9} m \mathcal{B}{}^{ac} \epsilon{}_{bcd} \bar h^{{\rm R}t}_{(1)}{}^{b} \hat{n}{}_{a}{}^{d} 
							+ \tfrac{1}{84}m\Bigl(15 \bar h^{{\rm R}}_{(1)}{}^{a}{}_{a}{}^{,c}+ 450\bar h^{{\rm R}}_{(1)}{}^{ac}{}_{,a}\Bigr)a{}^{b}\hat{n}{}_{bc}
							+ \tfrac{1}{60}m^2\Bigl(-87\mathcal{E}{}^{bac}+4a{}^{b}\bigl(298 \mathcal{E}{}^{ac} 
							- 963 a{}^{a} a{}^{c}\bigr)\Bigr)\hat{n}{}_{bac}\nonumber\\
						&\quad + \tfrac{1}{72}m \Bigl(-24\bar h^{{\rm R}}_{(1)}{}^{ba,cd} + 90\bar h^{{\rm R}}_{(1)}{}^{ac,d} a{}^{b} 
							+ 84\mathcal{E}{}^{ba} \bar h^{{\rm R}}_{(1)}{}^{cd} - 315a{}^{a} a{}^{b} \bar h^{{\rm R}}_{(1)}{}^{cd}\Bigr)\hat{n}{}_{bacd}
							+ \tfrac{1}{6} m \hat{n}{}_{b}{}^{a} \Bigl(-5 \bar h^{{\rm R}t}_{(1)}{}^{b} a{}_{a}{}_{,t} 
							+ 11 a{}^{b} \bar h^{{\rm R}t}_{(1)}{}_{a}{}_{,t}\Bigr) \nonumber\\
						&\quad + \tfrac{1}{42} m \hat{n}{}^{ba} \Bigl(26 \bar h^{{\rm R}}_{(1)}{}_{ba}{}^{,c}{}_{c}- 78 \bar h^{{\rm R}}_{(1)}{}_{b}{}^{c}{}_{,ac}
							- 9 \bar h^{{\rm R}}_{(1)}{}^{c}{}_{c,ba} - 21 \bar h^{{\rm R}t}_{(2)}{}_{b,a}{}_{t} 
							- 7 \bar h^{{\rm R}}_{(1)}{}_{ba}{}_{,tt}\Bigr)\nonumber\\
						&\quad + \tfrac{1}{270}m \Bigl(-252 \bar h^{{\rm R}}_{(1)}{}^{ba}{}_{,ba} + 84 \bar h^{{\rm R}}_{(1)}{}^{b}{}_{b}{}^{,a}{}_{a} 
							- 765 \bar h^{{\rm R}tt}_{(1)}{}^{,b} a{}_{b} - 165 \bar h^{{\rm R}}_{(1)}{}^{a}{}_{a,b} a{}^{b} 
							+ 1170 \bar h^{{\rm R}}_{(1)}{}_{b}{}^{a}{}_{,a} a{}^{b} + 30 a{}_{b} a{}^{b} \bar h^{{\rm R}}_{(1)}{}^{a}{}_{a}\nonumber\\ 
						&\quad - 268 \mathcal{E}{}^{ba} \bar h^{{\rm R}}_{(1)}{}_{ba} - 210 a{}^{a} a{}^{b} \bar h^{{\rm R}}_{(1)}{}_{ba} 
							- 390 a{}_{b} a{}^{b} \bar h^{{\rm R}tt}_{(1)} + 630 \bar h^{{\rm R}t}_{(2)}{}^{b}{}_{,b}{}_{t} 
							+ 1410 \bar h^{{\rm R}t}_{(1)}{}^{b} a{}_{b}{}_{,t} + 1920 a{}^{b} \bar h^{{\rm R}t}_{(1)}{}_{b}{}_{,t}
							- 15 \bar h^{{\rm R}}_{(1)}{}^{b}{}_{b}{}_{,tt} \nonumber\\
						&\quad + 675 \bar h^{{\rm R}tt}_{(1)}{}_{,tt}\Bigr)\bigg]-r\ln(r)m^2\Big[\tfrac{344}{15} a{}_{a} a{}^{a} a{}^{b} \hat{n}{}_{b} 
							+ \tfrac{64}{15} \mathcal{E}{}_{ab} a{}^{a} \hat{n}{}^{b} + \tfrac{8}{3} \hat{n}{}^{a} a{}_{a}{}_{,tt}\Big]+O(r^2),
\end{align}
\begin{align}
\bar h^{{\rm IH}ta}_{(2)} &= -\frac{m\bar{h}^{R(1)ti}\nhat^{a}{}_i}{r}-m\bar h^{{\rm R}a}_{(1)}{}_{b,t}n^b
							+m\left(\tfrac{1}{2}\bar h^{{\rm R}b}_{(1)}{}_{b,t}-\tfrac{3}{5}\bar h^{{\rm R}tb}_{(1)}{}_{,b}
							-\tfrac{3}{2}\bar h^{{\rm R}tt}_{(1)}{}_{,t}-\tfrac{4}{5}a_b\bar h^{{\rm R}tb}_{(1)}\right)n^a
							-\tfrac{1}{2}m^2a_{b,t}\nhat^{ab}\nonumber\\
						&\quad+2m^2\B^{bc}\epsilon^a{}_{cd}\nhat_{b}{}^d +m\left(a^b\bar h^{{\rm R}tc}_{(1)}
							-\tfrac{1}{2}\bar h^{{\rm R}tb,c}_{(1)}\right)\nhat^a{}_{bc}+\tfrac{1}{5}m\left(2\bar h^{{\rm R}ta,b}_{(1)}
							-8\bar h^{{\rm R}tb,a}_{(1)}-9\bar h^{{\rm R}ta}_{(1)}a^b-4a^a\bar h^{{\rm R}tb}_{(1)}\right)n_b\nonumber\\
						&\quad +\ln(r)8 m^2 a{}^{a}{}_{,t} +r\bigg[\tfrac{1}{252}m \bigl(-108\bar h^{{\rm R}t}_{(1)}{}^{b}{}_{,b}{}^{c} 
							- 90\bar h^{{\rm R}t}_{(1)}{}^{b,c} a{}_{b}
							+ 162 \bar h^{{\rm R}t}_{(1)}{}^{b}{}_{,b} a{}^{c} - 18 a{}^{b} a{}^{c} \bar h^{{\rm R}t}_{(1)}{}_{b} 
							+ 260\mathcal{E}{}_{b}{}^{c}\bar h^{{\rm R}t}_{(1)}{}^{b} \nonumber\\
						&\quad + 243a{}_{b}a{}^{b}\bar h^{{\rm R}t}_{(1)}{}^{c}\bigr)\hat{n}{}^{a}{}_{c}	
							- \tfrac{40}{63} m \mathcal{B}{}^{bc} \epsilon{}_{cdi} \bar h^{{\rm R}}_{(1)}{}_{b}{}^{d} \hat{n}{}^{ai} 
							- \tfrac{32}{63} m \mathcal{B}{}^{bc} \epsilon{}^{a}{}_{ci} \bar h^{{\rm R}}_{(1)}{}^{di} \hat{n}{}_{bd} 
							- \tfrac{5}{9} m \mathcal{B}{}^{bc} \epsilon{}^{a}{}_{cd} \bar h^{{\rm R}tt}_{(1)} \hat{n}{}_{b}{}^{d} \nonumber\\
						&\quad + \tfrac{1}{42} m\mathcal{B}{}^{bc} \bigl(34 \epsilon{}_{cdi} \bar h^{{\rm R}}_{(1)}{}^{ad} 
							+ 40\epsilon{}^{a}{}_{ci} \bar h^{{\rm R}}_{(1)}{}^{d}{}_{d}\bigr) \hat{n}{}_{b}{}^{i} 
							+ \tfrac{1}{63}m \bigl(-21\mathcal{B}{}^{bc} \epsilon{}^{a}{}_{di} \bar h^{{\rm R}}_{(1)}{}_{b}{}^{d} 
							+ 16 \mathcal{B}{}^{ab} \epsilon{}_{bdi} \bar h^{{\rm R}}_{(1)}{}^{cd}\bigr) \hat{n}{}_{c}{}^{i} \nonumber\\
						&\quad - \tfrac{23}{21} m \mathcal{B}{}^{bc} \epsilon{}^{a}{}_{ci} \bar h^{{\rm R}}_{(1)}{}_{b}{}^{d} \hat{n}{}_{d}{}^{i} 
							+ \tfrac{47}{90} m^2 \dot{\mathcal{E}}{}^{bc} \hat{n}{}^{a}{}_{bc} 
							+ \tfrac{8}{15} m^2 \mathcal{B}{}^{cd} \epsilon{}_{bc}{}^{i} a{}^{b} \hat{n}{}^{a}{}_{di} 
							- \tfrac{128}{9} m^2 \mathcal{B}{}^{cd} \epsilon{}^{a}{}_{c}{}^{i} a{}^{b} \hat{n}{}_{bdi} 
							+ \tfrac{14}{9} m^2 \mathcal{B}{}^{bcd} \epsilon{}^{a}{}_{b}{}^{i} \hat{n}{}_{cdi} \nonumber\\
						&\quad + \tfrac{1}{72}m \bigl(-24\bar h^{{\rm R}t}_{(1)}{}^{b,cd} + 30\bar h^{{\rm R}t}_{(1)}{}^{b,c} a{}^{d} 
							+ 36\mathcal{E}{}^{cd} \bar h^{{\rm R}t}_{(1)}{}^{b} - 81 a{}^{b} a{}^{c} \bar h^{{\rm R}t}_{(1)}{}^{d}\bigr) \hat{n}{}^{a}{}_{bcd} 
							+ \tfrac{2}{9} m \mathcal{B}{}^{bc} \epsilon{}_{bd}{}^{j} \bar h^{{\rm R}}_{(1)}{}^{di} \hat{n}{}^{a}{}_{cij}\nonumber\\ 
						&\quad - \tfrac{1}{3} m \mathcal{B}{}^{bc} \epsilon{}^{a}{}_{b}{}^{j} \bar h^{{\rm R}}_{(1)}{}^{di} \hat{n}{}_{cdij} 
							+ \tfrac{1}{504}m \hat{n}{}_{bc} \bigl(144 \bar h^{{\rm R}t}_{(1)}{}^{a,bc} - 720 \bar h^{{\rm R}t}_{(1)}{}^{b,ac} 
							- 180 \bar h^{{\rm R}t}_{(1)}{}^{b,c} a{}^{a} - 684 \bar h^{{\rm R}t}_{(1)}{}^{a,b} a{}^{c} 
							+ 828 \bar h^{{\rm R}t}_{(1)}{}^{b,a} a{}^{c}\nonumber\\
						&\quad - 720 \mathcal{E}{}^{bc} \bar h^{{\rm R}t}_{(1)}{}^{a} + 1368 a{}^{b} a{}^{c} \bar h^{{\rm R}t}_{(1)}{}^{a}
						 	- 208 \mathcal{E}{}^{ac} \bar h^{{\rm R}t}_{(1)}{}^{b} + 972a{}^{a} a{}^{b} \bar h^{{\rm R}t}_{(1)}{}^{c} 
							+ 504 \bar h^{{\rm R}}_{(1)}{}^{bc} a{}^{a}{}_{,t}\bigr) + \tfrac{1}{10}m^2a{}^{b} \hat{n}{}^{a}{}_{b}{}^{c} a{}_{c}{}_{,t}\nonumber\\
						&\quad - \tfrac{1}{2} m \hat{n}{}_{b}{}^{c} \bigl(\bar h^{{\rm R}}_{(1)}{}^{ab} a{}_{c}{}_{,t} 
							- 3a{}^{b}\bar h^{{\rm R}}_{(1)}{}^{a}{}_{,ct}\bigr)+m\hat{n}{}^{bc}\bigl(-\tfrac{1}{2}\bar h^{{\rm R}}_{(1)}{}^{a}{}_{b,ct}
							+ a{}^{a} \bar h^{{\rm R}}_{(1)}{}_{bc}{}_{,t}\bigr) 
							+ \tfrac{1}{6} m \hat{n}{}^{ac} \bigl(3 \bar h^{{\rm R}}_{(1)}{}^{b}{}_{b} a{}_{c}{}_{,t} 
							- 8 a{}^{b} \bar h^{{\rm R}}_{(1)}{}_{bc}{}_{,t}\bigr)\nonumber\\
						&\quad + \tfrac{1}{42} m \hat{n}{}^{a}{}_{b} \bigl(26 \bar h^{{\rm R}t}_{(1)}{}^{b,c}{}_{c} + 48 \bar h^{{\rm R}t}_{(1)}{}^{b,c} a{}_{c} 
							- 7 (7 \bar h^{{\rm R}}_{(1)}{}^{bc} a{}_{c}{}_{,t} + 3 a{}^{b} \bar h^{{\rm R}}_{(1)}{}^{c}{}_{c,t} 
							- 7 a{}^{b} \bar h^{{\rm R}tt}_{(1)}{}_{,t})\bigr) + \tfrac{1}{90}m \bigl(28\bar h^{{\rm R}t}_{(1)}{}^{a,b}{}_{b} 
							- 84\bar h^{{\rm R}t}_{(1)}{}^{b,a}{}_{b} \nonumber\\
						&\quad + 60\bar h^{{\rm R}t}_{(1)}{}^{b}{}_{,b} a{}^{a} 
							- 90\bar h^{{\rm R}t}_{(1)}{}^{a,b} a{}_{b} + 90\bar h^{{\rm R}t}_{(1)}{}^{b,a} a{}_{b} 
							- 96 \mathcal{B}{}^{bc} \epsilon{}^{a}{}_{cd} \bar h^{{\rm R}}_{(1)}{}_{b}{}^{d} - 78 a{}_{b} a{}^{b} \bar h^{{\rm R}t}_{(1)}{}^{a} 
							+ 354 a{}^{a} a{}^{b} \bar h^{{\rm R}t}_{(1)}{}_{b} - 154 \mathcal{E}{}^{a}{}_{b} \bar h^{{\rm R}t}_{(1)}{}^{b} \nonumber\\
						&\quad + 120 \bar h^{{\rm R}}_{(1)}{}^{ab}{}_{,bt} - 15 \bar h^{{\rm R}}_{(1)}{}^{b}{}_{b}{}^{,a}{}_{t} 
							- 45 \bar h^{{\rm R}tt}_{(1)}{}^{,a}{}_{t}+70\bar h^{{\rm R}tt}_{(1)}a{}^{a}{}_{,t}-20\bar h^{{\rm R}}_{(1)}{}^{ab} a{}_{b}{}_{,t} 
							- 70 a{}^{b} \bar h^{{\rm R}}_{(1)}{}^{a}{}_{b}{}_{,t} + 60 a{}^{a} \bar h^{{\rm R}}_{(1)}{}^{b}{}_{b}{}_{,t}
							+ 70 \bar h^{{\rm R}t}_{(1)}{}^{a}{}_{,tt} \nonumber\\
						&\quad + 140 a{}^{a} \bar h^{{\rm R}tt}_{(1)}{}_{,t}\bigr) 
							+ \tfrac{1}{12} m \hat{n}{}^{ab} \bigl(3 \bar h^{{\rm R}}_{(1)}{}^{c}{}_{c,b}{}_{t} - 9 \bar h^{{\rm R}tt}_{(1)}{}_{,b}{}_{t} 
							- 2 (10 \bar h^{{\rm R}tt}_{(1)} a{}_{b}{}_{,t} + \bar h^{{\rm R}t}_{(1)}{}_{b}{}_{,tt})\bigr)\bigg]
							+r\ln(r)m^2\Big[\tfrac{104}{15} a{}^{b} \hat{n}{}^{a} a{}_{b}{}_{,t} \nonumber\\
						&\quad	-\tfrac{1}{45} (61 \mathcal{B}{}_{c}{}^{d} \epsilon{}^{a}{}_{bd} + 200 \mathcal{B}{}_{b}{}^{d} \epsilon{}^{a}{}_{cd}
							- 11 \mathcal{B}{}^{ad} \epsilon{}_{bcd}) a{}^{b} \hat{n}{}^{c} - \tfrac{256}{15}a{}^{b} \hat{n}{}_{b} a{}^{a}{}_{,t} 
							- \tfrac{2}{15} \hat{n}{}^{b} (13 \dot{\mathcal{E}}{}^{a}{}_{b} + 68 a{}^{a} a{}_{b}{}_{,t})\Big]+O(r^2),
\end{align}
\begin{align}
\bar h^{{\rm IH}ab}_{(2)} &= -\frac{7m^2(\nhat^{ab}+\tfrac{1}{3}\delta^{ab})}{r^2}+\frac{1}{r}\bigg[\tfrac{31}{5} m^2 a^{(a} n^{b)} 
							-\tfrac{12}{5}\delta^{ab}m^2 a^in_i + 2m\bar{h}^{{\rm R}i(a}_{(1)}\hat{n}^{b)}{}_i 
							-m\delta^{ab}\bar{h}^{{\rm R}(1)ij}\hat{n}_{ij} - m \left(\bar{h}^{{\rm R}i}_{(1)}{}_{i} 
							+\bar{h}^{{\rm R}tt}_{(1)}\right) \hat{n}^{ab}\nonumber\\
						&\quad + \tfrac{14}{3} m^2 a_i\hat{n}^{abi} \bigg] + m\left(\tfrac{4}{5}a^{(a}\bar{h}^{{\rm R}b)}_{(1)}{}_{c}
							-\delta^{ab}\bar{h}^{{\rm R}t}_{(1)}{}_{c,t}-\tfrac{1}{5}\delta^{ab}\bar{h}^{{\rm R}d}_{(1)}{}_{c,d}
							-\tfrac{7}{10}\delta^{ab} \bar{h}^{{\rm R}d}_{(1)}{}_{d,c}+\tfrac{6}{5}\bar{h}^{{\rm R}ab}_{(1)}{}_{,c}
							-\tfrac{4}{5}\bar{h}^{{\rm R}(a}_{(1)}{}_c{}^{,b)}\right)n^c\nonumber\\
						&\quad +\tfrac{1}{5} m a^c \delta^{ab} \bar{h}^{{\rm R}}_{(1)cd}n^d +\tfrac{1}{5}m\left(6\bar{h}^{{\rm R}c(a}_{(1)}{}_{,c}
							-\bar{h}^{{\rm R}c}_{(1)}{}_c{}^{,(a}+4a_c \bar{h}^{{\rm R}c(a}_{(1)}+6\bar{h}^{{\rm R}c}_{(1)}{}_ca^{(a}
							-\bar{h}^{{\rm R}tt,(a}_{(1)}+4\bar{h}^{{\rm R}tt}_{(1)} a^{(a}\right)n^{b)} \nonumber\\
						&\quad+m\left(\tfrac{2}{15}a^c\bar{h}^{{\rm R}ab}_{(1)}-\tfrac{17}{15}a^c\bar{h}^{{\rm R}d}_{(1)}{}_d \delta^{ab}
							-\tfrac{11}{10}\bar{h}^{{\rm R}tt,c}_{(1)}\delta^{ab}-\tfrac{49}{15}\bar{h}^{{\rm R}tt}_{(1)}a^c\delta^{ab}\right)n_c 
							+m^2 \left(\tfrac{2}{3}\E^{c(a}- a^ca^{(a}\right)\nhat^{b)}{}_c -\tfrac{7}{2} m^2 a_ca^c\nhat^{ab}\nonumber\\
						&\quad + m^2 \delta^{ab} \left(a^ca^d-\tfrac{2}{3}\E^{cd}\right)\nhat_{cd}+ m\left(\bar{h}^{{\rm R}c(a}_{(1)}{}_{,d}
							-a^c\bar{h}^{{\rm R}(a}_{(1)}{}_d\right)\nhat^{b)d}{}_c+\tfrac{1}{2}m\left( a^c\bar{h}^{{\rm R}d}_{(1)}{}_d
							-\bar{h}^{{\rm R}tt,c}_{(1)}-\bar{h}^{{\rm R}tt}_{(1)}a^c\right)\nhat^{ab}{}_c\nonumber\\
						&\quad -\tfrac{1}{2}m\bar{h}^{{\rm R}c}_{(1)}{}_{c,d}\nhat^{abd}+\tfrac{1}{2}m\delta^{ab}\left(a^c\bar{h}^{{\rm R}di}_{(1)}
							-\bar{h}^{{\rm R}cd,i}_{(1)}\right)\nhat_{cdi}+m^2\left(\tfrac{7}{5}\E^{cd}-\tfrac{56}{15}a^ca^d\right)\nhat^{ab}{}_{cd}
							-\tfrac{4}{15} m^2\ln(r) (4 \mathcal{E}{}^{ab} + 6 a{}^{a} a{}^{b}\nonumber\\
						&\quad  + 3 a{}_{c} a{}^{c} \delta{}^{ab}) +r\bigg[\tfrac{1}{126}m \bigl(72\bar h^{{\rm R}c}_{(1)}{}_{c}{}^{,d} a{}^{(a} 
							+ 54 a{}^{c}\bar h^{{\rm R}(a}_{(1)}{}_{c}{}^{,|d|} - 30 a{}^{c}\bar h^{{\rm R}d(a}_{(1)}{}_{,c} 
							- 34 \mathcal{E}{}^{cd} \bar h^{{\rm R}(a}_{(1)}{}_{c} - 78a{}^{c} a{}^{d} \bar h^{{\rm R}(a}_{(1)}{}_{c} 
							+ 45 a{}_{c} a{}^{c} \bar h^{{\rm R}d(a}_{(1)} \nonumber\\
						&\quad - 14 \bar h^{{\rm R}d}_{(1)}{}_{c}\mathcal{E}^{c(a}- 294 a^{c} \bar h^{{\rm R}d}_{(1)}{}_{c}a^{(a}  
							+ 294 \mathcal{B}^{di} \bar h^{{\rm R}tc}_{(1)}\epsilon^{(a}{}_{ci}\bigr) \hat{n}^{b)}{}_{d} 
							+ \tfrac{1}{63}m \bigl(-41 \mathcal{B}_{c}{}^{d} \bar h^{{\rm R}tc}_{(1)}\epsilon^{(a}{}_{di} 
							+ 27 \epsilon_{cdi} \bar h^{{\rm R}tc}_{(1)}\mathcal{B}{}^{d(a} \bigr) \hat{n}{}^{b)i} \nonumber\\
						&\quad + \tfrac{1}{252}m \bigl(108 a{}^{(a}\bar h^{{\rm R}b)c,d}_{(1)} 
							- 144 \bar h^{{\rm R}ab,d}_{(1)}a{}^{c} + 108 \bar h^{{\rm R}d(a,b)}_{(1)} a{}^{c} 
							- 306 \bar h^{{\rm R}tt,cd}_{(1)} \delta{}^{ab} 
							- 603 \bar h^{{\rm R}tt,c}_{(1)}a^{d} \delta{}^{ab} - 88 \mathcal{E}{}^{cd} \bar h^{{\rm R}ab}_{(1)} \nonumber\\
						&\quad - 36 a{}^{c} a{}^{d} \bar h^{{\rm R}ab}_{(1)} - 88 \mathcal{E}^{c(a} \bar h^{{\rm R}b)d}_{(1)} 
							- 324 a{}^{c}a{}^{(a}\bar h^{{\rm R}b)d}_{(1)} 
							- 180 \mathcal{E}^{ab} \bar h^{{\rm R}cd}_{(1)} 
							+ 504 a{}^{a} a{}^{b} \bar h^{{\rm R}cd}_{(1)} - 122 \mathcal{E}^{cd} \delta{}^{ab} \bar h^{{\rm R}i}_{(1)}{}_{i} \nonumber\\
						&\quad	+ 162 a{}^{c} a{}^{d}\delta{}^{ab}\bar h^{{\rm R}i}_{(1)}{}_{i} 
							- 580\mathcal{E}{}^{cd}\delta{}^{ab} \bar h^{{\rm R}tt}_{(1)} 
							+ 66 a{}^{c} a{}^{d} \delta{}^{ab} \bar h^{{\rm R}tt}_{(1)}\bigr) \hat{n}{}_{cd} 
							+ \tfrac{1}{28}m \bigl(-5\bar h^{{\rm R}d}_{(1)}{}_{d}{}^{,i} 
							+ 2 \bar h^{{\rm R}di}_{(1)}{}_{,d}\bigr) a{}^{c} \delta{}^{ab} \hat{n}{}_{ci} \nonumber\\  
						&\quad + \tfrac{1}{252} m\delta{}^{ab} \bigl(18 \bar h^{{\rm R}d}_{(1)}{}_{c}{}^{,i} a{}^{c} 
							+ 30\bar h^{{\rm R}di}_{(1)}{}_{,c} a{}^{c} 
							+ 100 \mathcal{E}{}^{cd} \bar h^{{\rm R}i}_{(1)}{}_{c} + 156 a{}^{c} a{}^{d} \bar h^{{\rm R}i}_{(1)}{}_{c} 
							- 45 a{}_{c} a{}^{c} \bar h^{{\rm R}di}_{(1)}\bigr) \hat{n}{}_{di} \nonumber\\
						&\quad - \tfrac{40}{63}m\mathcal{B}{}^{d(a} \epsilon{}^{b)}{}_{ci} \bar h^{{\rm R}tc}_{(1)} 
							- \tfrac{95}{63} m \mathcal{B}{}^{di} \epsilon{}_{cij} \delta{}^{ab} \bar h^{{\rm R}tc}_{(1)}\hat{n}{}_{d}{}^{j} 
							- \tfrac{7}{810} m^2 a{}^{c} \bigl(287 \mathcal{E}{}_{c}{}^{d} - 621 a{}_{c} a{}^{d}\bigr) \hat{n}{}^{ab}{}_{d}
							- \tfrac{4}{27} m \mathcal{B}{}^{di} \bar h^{{\rm R}tc}_{(1)} \epsilon_{c}{}^{j(a}\hat{n}{}^{b)}{}_{dij}   \nonumber\\
						&\quad + \tfrac{1}{1620} m^2 \bigl(1737 \mathcal{E}{}^{cd(a} - 148 \mathcal{E}{}^{cd} a{}^{(a} + 6292a{}^{c} \mathcal{E}{}^{d(a}  
							- 3564 a{}^{c} a{}^{d}a{}^{(a} \bigr) \hat{n}{}^{b)}{}_{cd} 
							- \tfrac{3}{5} m^2 \dot{\mathcal{B}}{}^{cd} \epsilon_{c}{}^{i(a} \hat{n}{}^{b)}{}_{di} \nonumber\\
						&\quad - \tfrac{1}{405} m^2 \bigl(309 \mathcal{E}{}^{cdi} + a{}^{c} (257 \mathcal{E}{}^{di} 
							+ 153 a{}^{d} a{}^{i})\bigr) \delta{}^{ab} \hat{n}{}_{cdi} - \tfrac{1}{72} m \bigl(24 \bar h^{{\rm R}tt,cd}_{(1)} 
							+ 30 \bar h^{{\rm R}tt,c}_{(1)} a{}^{d} - 20 \mathcal{E}{}^{cd} \bar h^{{\rm R}i}_{(1)}{}_{i} 
							+ 27 a{}^{c} a{}^{d} \bar h^{{\rm R}i}_{(1)}{}_{i} \nonumber\\
						&\quad + 12 \mathcal{E}{}^{cd} \bar h^{{\rm R}tt}_{(1)} 
							+ 15 a{}^{c} a{}^{d} \bar h^{{\rm R}tt}_{(1)}\bigr) \hat{n}{}^{ab}{}_{cd} 
							+ \tfrac{1}{4} m \bar h^{{\rm R}d}_{(1)}{}_{d}{}^{,i} a{}^{c} \hat{n}{}^{ab}{}_{ci} 
							- \tfrac{1}{9}m \bigl(-3\bar h^{{\rm R}c}_{(1)}{}_{c}{}^{,di} 
							- 4 \mathcal{E}{}^{cd} \bar h^{{\rm R}}_{(1)}{}_{c}{}^{i}\bigr) \hat{n}{}^{ab}{}_{di} \nonumber\\
						&\quad + \tfrac{2}{27} m \mathcal{B}{}^{di} \epsilon{}_{cd}{}^{j} \bar h^{{\rm R}tc}_{(1)} \hat{n}{}^{ab}{}_{ij} 
							+ \tfrac{1}{36}m \bigl(24\bar h^{{\rm R}c(a,|di|}_{(1)} - 18 a{}^{c}\bar h^{{\rm R}d(a,|i|}_{(1)} 
							- 4\mathcal{E}{}^{cd} \bar h^{{\rm R}i(a}_{(1)} + 27 a{}^{c} a{}^{d} \bar h^{{\rm R}i(a}_{(1)} 
							+ 16\bar h^{{\rm R}di}_{(1)}\mathcal{E}{}^{c(a} \bigr) \hat{n}{}^{b)}{}_{cdi} \nonumber\\
						&\quad - \tfrac{32}{27} m \mathcal{B}{}^{di} \bar h^{{\rm R}tc}_{(1)} \epsilon_{d}{}^{j(a}\hat{n}{}^{b)}{}_{cij} 
							- \tfrac{1}{72}m \delta{}^{ab} \bigl(24 \bar h^{{\rm R}cd,ij}_{(1)} - 18\bar h^{{\rm R}di,j}_{(1)} a{}^{c} 
							+ 12\mathcal{E}{}^{cd}\bar h^{{\rm R}ij}_{(1)}+27a{}^{c}a{}^{d}\bar h^{{\rm R}ij}_{(1)}\bigr) \hat{n}{}_{cdij}\nonumber\\
						&\quad + \tfrac{1}{45}m^2\bigl(15 \mathcal{E}{}^{cdi} + 8 a{}^{c} (-13 \mathcal{E}{}^{di} 
							+ 18 a{}^{d} a{}^{i})\bigr)\hat{n}{}^{ab}{}_{cdi}+ \tfrac{8}{63}m\bigl(\mathcal{B}^{cd}\bar h^{{\rm R}t(a}_{(1)}\epsilon{}^{b)}{}_{di}
							+ \mathcal{B}{}^{d(a}\epsilon{}^{b)}{}_{di} \bar h^{{\rm R}tc}_{(1)}\bigr) \hat{n}{}_{c}{}^{i} \nonumber\\
						&\quad + \tfrac{1}{126}m \hat{n}{}^{(a}{}_{c} \bigl(18 \bar h^{{\rm R}tt,b)c}_{(1)} + 90a{}^{b)} \bar h^{{\rm R}tt,c}_{(1)}  
							- 72 \bar h^{{\rm R}b)d}_{(1)}{}_{,d} a{}^{c} + 9 \bar h^{{\rm R}|d|}_{(1)}{}_{d}{}^{,b)} a{}^{c} 
							+ 27 \bar h^{{\rm R}tt,b)}_{(1)} a{}^{c} + 154 \mathcal{E}{}^{b)c} \bar h^{{\rm R}d}_{(1)}{}_{d} 
							+ 36 a{}^{b)} a{}^{c} \bar h^{{\rm R}d}_{(1)}{}_{d} \nonumber\\
						&\quad + 212 \mathcal{E}{}^{b)c} \bar h^{{\rm R}tt}_{(1)} 
							+ 6 a{}^{b)} a{}^{c} \bar h^{{\rm R}tt}_{(1)} + 210 a{}^{b)}{}_{,t} \bar h^{{\rm R}tc}_{(1)} 
							+ 42 \bar h^{{\rm R}tb)}_{(1)}{}_{,t}a{}^{c}\bigr) 
							+ \tfrac{1}{6} m \delta{}^{ab} \hat{n}{}_{c}{}^{d} \bigl(-7 \bar h^{{\rm R}tc}_{(1)}a{}_{d}{}_{,t} 
							+ a{}^{c} \bar h^{{\rm R}td}_{(1)}{}_{,t}\bigr) \nonumber\\
						&\quad + \tfrac{1}{21} m \bigl(-26 \bar h^{{\rm R}(a}_{(1)}{}_{c}{}^{,|d|}{}_{d} + 46 \bar h^{{\rm R}d(a}_{(1)}{}_{,cd} 
							+ 28 \bar h^{{\rm R}}_{(1)}{}_{c}{}^{d,(a}{}_{d} - 25 \bar h^{{\rm R}}_{(1)}{}^{d}{}_{d}{}^{,(a}{}_{c} 
							+ 14 a{}_{c}{}_{,t}\bar h^{{\rm R}t(a}_{(1)}  + 28 \bar h^{{\rm R}t}_{(1)}{}_{c,t} a{}^{(a} 
							- 7 \bar h^{{\rm R}(a}_{(1)}{}_{c,tt}\bigr)\hat{n}{}^{b)c}  \nonumber\\
						&\quad - \tfrac{11}{30} m^2 \hat{n}{}^{abc} a{}_{c}{}_{,tt} 
							+ \tfrac{1}{42}m \hat{n}{}^{cd} \bigl(32 \bar h^{{\rm R}ab}_{(1)}{}_{,cd} 
							+ 8 \bar h^{{\rm R}(a}_{(1)}{}_{c}{}^{,b)}{}_{d} 
							- 28 \bar h^{{\rm R}}_{(1)}{}_{cd}{}^{,ab} + 26 \bar h^{{\rm R}}_{(1)}{}_{cd}{}^{,i}{}_{i} \delta{}^{ab} 
							- 50 \bar h^{{\rm R}}_{(1)}{}_{c}{}^{i}{}_{,di} \delta{}^{ab} \nonumber\\
						&\quad - 11 \bar h^{{\rm R}i}_{(1)}{}_{i,cd} \delta{}^{ab} 
							- 21 \delta{}^{ab} \bar h^{{\rm R}t}_{(1)}{}_{c,dt} + 7 \delta{}^{ab} \bar h^{{\rm R}}_{(1)}{}_{cd,tt}\bigr) 
							+ \tfrac{1}{252} m\hat{n}{}^{ab} \bigl(156 \bar h^{{\rm R}c}_{(1)}{}_{c}{}^{,d}{}_{d} - 168\bar h^{{\rm R}cd}_{(1)}{}_{,cd} 
							- 12\bar h^{{\rm R}tt,c}_{(1)}{}_{c} + 6\bar h^{{\rm R}tt,c}_{(1)} a{}_{c} \nonumber\\
						&\quad + 30\bar h^{{\rm R}d}_{(1)}{}_{d,c} a{}^{c} 
							+ 40\mathcal{E}{}^{cd} \bar h^{{\rm R}}_{(1)}{}_{cd} + 168 a{}^{c} a{}^{d} \bar h^{{\rm R}}_{(1)}{}_{cd} 
							- 45 a{}_{c} a{}^{c} \bar h^{{\rm R}d}_{(1)}{}_{d} + 171 a{}_{c} a{}^{c} \bar h^{{\rm R}tt}_{(1)} 
							- 336 \bar h^{{\rm R}tc}_{(1)}a_{c,t} - 168 a^{c} \bar h^{{\rm R}t}_{(1)}{}_{c,t} \nonumber\\
						&\quad + 42 \bar h^{{\rm R}c}_{(1)}{}_{c,tt} + 42 \bar h^{{\rm R}tt}_{(1)}{}_{,tt}\bigr) 
							+ \tfrac{1}{270}m \bigl(312\bar h^{{\rm R}ab,c}_{(1)}{}_{c} - 936 \bar h^{{\rm R}c(a,b)}_{(1)}{}_{c} 
							+ 468 \bar h^{{\rm R}tt}_{(1)}{}^{,ab} - 144 \bar h^{{\rm R}c(a}_{(1)}{}_{c} a{}^{b)} 
							- 216 \bar h^{{\rm R}c}_{(1)}{}_{c}{}^{,(a} a{}^{b)} \nonumber\\
						&\quad + 1260 \bar h^{{\rm R}tt,(a}_{(1)}a{}^{b)} - 564 \bar h^{{\rm R}ab}_{(1)}{}_{,c} a{}^{c} 
							+ 756 \bar h^{{\rm R}(a}_{(1)}{}_{c}{}^{,b)} a{}^{c}
							- 312 \bar h^{{\rm R}c}_{(1)}{}_{c}{}^{,d}{}_{d} \delta{}^{ab} + 468 \bar h^{{\rm R}cd}_{(1)}{}_{,cd} \delta{}^{ab} 
							- 156 \bar h^{{\rm R}tt,c}_{(1)}{}_{c} \delta{}^{ab} \nonumber\\
						&\quad - 435 \bar h^{{\rm R}tt,c}_{(1)} a{}_{c} \delta{}^{ab} 
							- 126 \bar h^{{\rm R}d}_{(1)}{}_{c,d} a{}^{c} \delta{}^{ab} + 69 \bar h^{{\rm R}d}_{(1)}{}_{d,c} a{}^{c}\delta{}^{ab} 
							+ 36 a{}_{c} a{}^{c} \bar h^{{\rm R}ab}_{(1)} + 356 \mathcal{E}^{c(a} \bar h^{{\rm R}b)}_{(1)}{}_{c} 
							- 1068 a{}^{c} a{}^{(a}\bar h^{{\rm R}b)}_{(1)}{}_{c} \nonumber\\
						&\quad + 356 \mathcal{E}{}^{cd} \delta{}^{ab} \bar h^{{\rm R}}_{(1)}{}_{cd} 
							- 486 a{}^{c} a{}^{d} \delta{}^{ab} \bar h^{{\rm R}}_{(1)}{}_{cd} - 352 m \mathcal{E}{}^{ab} \bar h^{{\rm R}c}_{(1)}{}_{c} 
							+ 144 a{}^{a} a{}^{b} \bar h^{{\rm R}c}_{(1)}{}_{c} - 126 a{}_{c} a{}^{c} \delta{}^{ab} \bar h^{{\rm R}d}_{(1)}{}_{d} 
							- 768 \mathcal{B}{}^{d(a} \epsilon{}^{b)}{}_{cd} \bar h^{{\rm R}tc}_{(1)}\nonumber\\
						&\quad + 184 \mathcal{E}{}^{ab} \bar h^{{\rm R}tt}_{(1)} 
							+ 1200 a{}^{a} a{}^{b} \bar h^{{\rm R}tt}_{(1)} - 1050 a{}_{c} a{}^{c} \delta{}^{ab} \bar h^{{\rm R}tt}_{(1)} 
							+ 540 \bar h^{{\rm R}t(a,b)}_{(1)}{}_{t} 
							+ 90 \delta{}^{ab} \bar h^{{\rm R}tc}_{(1)}{}_{,ct} + 1560 \bar h^{{\rm R}t(a}_{(1)} a{}^{b)}{}_{,t}\nonumber\\ 
						&\quad + 450 \delta{}^{ab} \bar h^{{\rm R}tc}_{(1)} a{}_{c}{}_{,t} 
							+ 960 \bar h^{{\rm R}t(a}_{(1)}{}_{,t}a{}^{b)} 
							- 180 a{}^{c} \delta{}^{ab} \bar h^{{\rm R}t}_{(1)}{}_{c,t} + 240 \bar h^{{\rm R}ab}_{(1)}{}_{,tt} 
							- 15 \delta{}^{ab} \bar h^{{\rm R}c}_{(1)}{}_{c,tt} + 195 \delta{}^{ab} \bar h^{{\rm R}tt}_{(1)}{}_{,tt}\bigr)\bigg]\nonumber\\
						&\quad +r\ln(r)m^2\Big[(\tfrac{104}{15} \mathcal{E}{}^{ab} a{}^{c} - \tfrac{176}{15} a{}^{a} a{}^{b} a{}^{c}) \hat{n}{}_{c} 
							+ \tfrac{16}{15} a{}_{c} a{}^{c} a{}^{d} \delta{}^{ab} \hat{n}{}_{d} 
							+ 2\hat{n}{}^{(a} (\tfrac{32}{15} a{}^{b)}{}_{,tt}-\tfrac{4}{15} \mathcal{E}{}^{b)}{}_{c} a{}^{c}
							+ \tfrac{8}{5} a{}^{b)} a_{c}a^{c}) \nonumber\\
						&\quad + \hat{n}{}^{c} (\tfrac{112}{15} \mathcal{E}{}^{(a}{}_{c} a{}^{b)}-\tfrac{4}{5} \mathcal{E}{}^{ab}{}_{c} 
							- \tfrac{12}{5} \dot{\mathcal{B}}{}^{d(a} \epsilon{}^{b)}{}_{cd} 
							- \tfrac{8}{15} \delta{}^{ab} a{}_{c}{}_{,tt})- \tfrac{16}{5}\E{}_{cd} a{}^{c} \delta{}^{ab} \hat{n}{}^{d} \Big]+O(r^2).
\end{align}
Indices within vertical bars are excluded from symmetrizations.
\end{widetext}

\bibliography{second-order-partI}
\end{document}